\documentclass[12pt]{article}
\usepackage{a4wide,amssymb,cite}
\pdfoutput=1

\usepackage{a4wide,amssymb,graphicx}
\usepackage{epsfig}
\usepackage[usenames,dvipsnames]{color}
\usepackage{slashed}
\usepackage{amssymb,cite,graphicx}
\usepackage{mathrsfs}
\usepackage{slashed}
\usepackage{amsmath,bm,bbm}
\usepackage{amsfonts}
\usepackage[titletoc,title]{appendix}
\usepackage[small]{caption}
\usepackage[margin=1in]{geometry}
\usepackage[multiple]{footmisc}
\usepackage{mathtools}
\usepackage{slashed}
\usepackage[nottoc]{tocbibind}
\usepackage{xcolor}
\usepackage{multirow}

\newcommand{\be}{\begin{equation}}
\newcommand{\ee}{\end{equation}}
\newcommand{\bea}{\begin{eqnarray}}
\newcommand{\eea}{\end{eqnarray}}

\def\circa#1{\,\raise.3ex\hbox{$#1$\kern-.75em\lower1ex\hbox{$\sim$}}\,}

\begin{document}

\begin{titlepage}

\rightline{CERN-TH-2020-041}

\begin{centering}
\vspace{1cm}
{\Large {\bf  Connecting between inflation and dark matter \vspace{.2cm}  \\ in models with gauged $Z_3$ symmetry}} \\

\vspace{1.5cm}

{\bf Soo-Min Choi$^{\dagger a,c}$,  Jinsu Kim$^{* b}$, Hyun Min Lee$^{\ddagger a,c}$ and Bin Zhu$^{^\# a,d}$}
\vspace{.5cm}

{\it $^a$Department of Physics, Chung-Ang University, Seoul 06974, Korea} 
\\ \vspace{0.2cm}
{\it $^b$Institute for Theoretical Physics, Georg-August University G\"ottingen,\\Friedrich-Hund-Platz 1, G\"ottingen, D-37077 Germany}
\\ \vspace{0.2cm}
{\it $^c$CERN, Theory department, 1211 Geneva 23, Switzerland}
\\ \vspace{0.2cm}
{\it $^d$School of Physics, Yantai University, Yantai 264005, China}

\end{centering}
\vspace{1.5cm}

\begin{abstract}
\noindent
We investigate the possibility of unifying the inflation and dark matter physics in the minimal model for a complex scalar dark matter with gauged $Z_3$ symmetry. The dark local $U(1)$ symmetry is broken spontaneously into the $Z_3$ symmetry by the dark Higgs mechanism. As compared to dark matter models with $Z_2$ parity, the dark matter cubic self-interaction restricts the inflation with non-minimal couplings to take place beyond the pure dark matter direction and plays an important role in the loop corrections for the inflationary predictions as well as determining the correct dark matter relic density. Considering either of $2\rightarrow 2$, $3\rightarrow 2$ or forbidden channels to be a dominant production mechanism for dark matter, we show the viable parameter space of the model that is consistent with the theoretical and phenomenological constraints combined from inflation and dark matter.

\end{abstract}

\vspace{2cm}

\begin{flushleft} 
$^\dagger$Email: soominchoi90@gmail.com \\
$^*$Email: jinsu.kim@theorie.physik.uni-goettingen.de  \\
$^\ddagger$Email: hminlee@cau.ac.kr \\
$^\#$Email: zhubin@mail.nankai.edu.cn
\end{flushleft}

\end{titlepage}

\section{Introduction}
\label{sec:intro}
There have been a lot of progresses in measuring the cosmological parameters of the universe with unprecedented precision, for instance, from the Cosmic Microwave Background (CMB) anisotropies. Thus, two pillars in the modern cosmology, inflation \cite{Akrami:2018odb} and dark matter \cite{planck}, have reinforced the motivations and efforts for searching for physics beyond the Standard Model (SM).
Inspired by the SM Higgs inflation \cite{Bezrukov:2007ep}, there have been recent attempts to integrate inflation in connection with low energy particle physics. Moreover, given that the nature of dark matter remains a mystery, the unification of inflation and dark matter provides an economical way of solving the problems in cosmology and particle physics at the same time \cite{DMwithInflation,Ko:2014eia}.
The merit of this kind consideration is that additional fields introduced for the consistency may also solve the problems in the SM, such as the vacuum instability of the SM Higgs potential.

Various candidates for dark matter are based on the $Z_2$ symmetry for guaranteeing the stability of dark matter and their production mechanism rely on the freeze-out processes in the so-called Weakly Interacting Massive Particles (WIMPs) paradigm.
Inclusion of discrete symmetries beyond the $Z_2$ symmetry leads to interesting and new avenues for dark matter physics, such as $2\rightarrow 2$ and $3\rightarrow 2$ semi-annihilation processes for the freeze-out. In particular, in the latter case, the freeze-out process would require dark matter to have large self-interactions but small couplings to the SM. This possibility, so-called Strongly Interacting Massive Particles (SIMPs) paradigm \cite{simp} has been realized in a simple extension with gauged $Z_3$ symmetry \cite{Choi:2015bya,Choi:2016tkj}, where there are a lot of interesting signatures at intensity beam experiments and astrophysical observations of galaxy rotation curves and cluster collisions. 

As inflation and dark matter physics involve largely different scales, it is important to include quantum corrections to match them by using the renormalization group (RG) equations.
In particular, it is notable that inflationary observables such as the spectral index $n_s$ and tensor-to-scalar ratio $r$ may be significantly affected by quantum effects; for instance, the predicted value for the inflation with non-minimal coupling changes the tensor-to-scalar ratio from $r \approx 0.003$ to a somewhat larger value by quantum effects \cite{Ko:2014eia}. In the original Higgs inflation, the tree-level unitarity is violated due to a large non-minimal coupling at the scale much lower than the Planck scale \cite{unitarity},  so there is a need of introducing new physics or degrees of freedom to solve the unitarity problem \cite{UVsolution,RGs,UV2,R2}. 
However, taking into account the running Higgs quartic coupling in Higgs inflation, we can tolerate a smaller value of the non-minimal coupling, as compared to the case without quantum effects, so the unitarity scale could be delayed to a higher scale at the expense of a fine-tuning in realizing the inflation near the inflection point of the potential \cite{Ko:2014eia}.

In this article, we study the dynamics of inflation and dark matter in the unified framework beyond the $Z_2$ symmetry, in particular, by adopting models with gauged $Z_3$ symmetry. Dark matter phenomenology in the same models has been investigated for the thermal freeze-out with WIMP \cite{Ko} and SIMP dark matter \cite{Choi:2015bya,Choi:2016tkj}. A generalization of the similar discussion was made for models with $Z_5$ discrete symmetry \cite{Choi:2016hid}. The focus of our work is to fill a gap between dark matter and inflation scales and make a consistent description of the early universe in a single framework beyond the $Z_2$ symmetry. The model with $Z_3$ symmetry consists of two complex scalar fields, one SM-singlet dark Higgs $\phi$ and one SM-singlet scalar dark matter $\chi$, beyond the SM. The SM-singlet dark Higgs $\phi$ is responsible for the spontaneous symmetry breaking of a dark $U(1)_{{\rm X}}$ into $Z_3$ symmetry, which ensures the stability of dark matter. 

We cover three distinct regimes of dark matter production with WIMP, SIMP, and forbidden dark matter.
We connect each of the cases to the inflationary regime for the first time. To do so, it is crucial to include nonzero non-minimal couplings to gravity for all the scalar fields, which are generated otherwise at the quantum level in curved spacetime.
We regard a slow-roll inflation as taking place along the flat direction in the system of two real scalar fields, one of which is frozen in the minimum of the potential during inflation. Thus, the dynamics of inflation is reduced down effectively to the case of a single-field inflation. The effective inflaton can be either the dark Higgs or a mixture of the dark Higgs and scalar dark matter \cite{Lebedev:2011aq}. We note, however, that it is impossible for the scalar dark matter $\chi$ solely to play the role of inflaton; namely the pure $\chi$ field direction is not at the minimum of the potential. This is one of the distinctive features of the $Z_3$ models in comparison to the $Z_2$ models where scalar dark matter can be the only inflaton. 
We connect between dark matter and inflation scales by employing the RG-improved effective action and look for the consistent parameter space with the correlations between the following relevant observables: the relic density on one side, and perturbativity, unitarity and stability on the other side.  

The paper is organized as follows.
We begin with a description of our setup in Sec.~\ref{sec:model}, with a complete model Lagrangian and the particle content.
In Sec.~\ref{sec:effLagInf}, we first discuss the details on the inflationary dynamics at the classical level and include the RG effects in the later part. In this section, inflationary observables such as the spectral index and tensor-to-scalar ratio are computed.
We continue to discuss the general Boltzmann equation for dark matter and focus on three distinct limits of production mechanism of dark matter in Sec.~\ref{sec:DMinf}, and then impose the consistency conditions for connecting between inflation and dark matter scales, such as stability, perturbativity and unitarity as well as various constraints from low-energy experiments. 
Finally, conclusions are drawn in Sec.~\ref{sec:conc}.
There are two appendices providing the formulas for the RG equations and dark matter annihilation and scattering cross sections.

\section{The model}
\label{sec:model}

We consider an extension of the SM with a local dark $U(1)_{{\rm X}}$, including one complex scalar $\phi$ for dark Higgs mechanism and one complex scalar $\chi$ for dark matter. Then, the Lagrangian of our model \cite{Choi:2016tkj} is given by
\begin{align}
\mathcal{L}
=
\mathcal{L}_{{\rm NM}}
+\mathcal{L}_{{\rm SM}}
+\mathcal{L}_{{\rm X}}
+\mathcal{L}_{{\rm portal}}\,,
\label{eqn:Lag}
\end{align}
where $\mathcal{L}_{{\rm SM}}$ is the SM Lagrangian, 
\begin{align}
\mathcal{L}_{{\rm NM}} =
\frac{M_{{\rm P}}^{2}}{2}\left(
1
+ 2\xi_{\chi}\frac{|\chi|^{2}}{M_{{\rm P}}^{2}}
+ 2\xi_{\phi}\frac{|\phi|^{2}}{M_{{\rm P}}^{2}}
+ 2\xi_{H}\frac{|H|^{2}}{M_{{\rm P}}^{2}}
\right)\mathcal{R}
\,,
\end{align}
describes the non-minimal couplings of the scalar fields to gravity, and
\begin{align}
\mathcal{L}_{{\rm X}}
&=
-\frac{1}{4} V_{\mu\nu}  V^{\mu\nu}+
(D_{\mu}\chi)^{\dagger}
(D^{\mu}\chi)
+(D_{\mu}\phi)^{\dagger}
(D^{\mu}\phi)
+m_{\phi}^{2}|\phi|^{2}
-m_{\chi}^{2}|\chi|^{2}
\nonumber \\
&\quad
-\lambda_{\phi}|\phi|^{4}
-\lambda_{\chi}|\chi|^{4}
-\lambda_{\phi\chi}|\phi|^{2}|\chi|^{2}
-\zeta\phi^{\dagger}\chi^{3}
-\zeta\phi (\chi^{\dagger})^{3}
\,,\\
\mathcal{L}_{{\rm portal}}
&=
-\lambda_{\phi H}|\phi|^{2}|H|^{2}
-\lambda_{\chi H}|\chi|^{2}|H|^{2} -\frac{1}{2} \sin\xi\, B_{\mu\nu} V^{\mu\nu}
\,.
\end{align}
The action is then given by $S = \int d^{4}x \sqrt{-g} \mathcal{L}$.
The SM Higgs doublet is denoted by $H$, $\phi$ is an SM-singlet complex field dubbed the dark Higgs, and $\chi$ is another SM-singlet complex field.
The covariant derivatives are given by
\begin{align}
D_{\mu}\phi =
(\partial_{\mu}
-iq_{\phi}g_{X}V_{\mu})\phi
\,,\qquad
D_{\mu}\chi =
(\partial_{\mu}
-iq_{\chi}g_{X}V_{\mu})\chi
\,,
\end{align}
with $g_{X}$ being the gauge coupling of the dark $U(1)_{{\rm X}}$ group.
Here $V_{\mu\nu}=\partial_\mu V_\nu -\partial_\nu V_\mu$ is the field strength tensor for $U(1)_{{\rm X}}$, and $\sin\xi$ is the gauge kinetic mixing between the SM hypercharge and $U(1)_{{\rm X}}$. 
Note that $\zeta=\sqrt{2}\kappa/3!$ as compared to Ref.~\cite{Choi:2016tkj}.

We consider the complex scalar $\chi$ as a dark matter candidate, having a charge $q_{\chi}=1$ under the dark $U(1)_{{\rm X}}$ symmetry.
Another field $\phi$, which has a charge $q_{\phi}=3$, takes a vacuum expectation value (VEV) by $\langle \phi \rangle=v_\phi/\sqrt{2}$, and thus, being responsible for the spontaneous symmetry breaking of $U(1)_{{\rm X}}$ into $Z_{3}$.
Therefore the remaining discrete $Z_{3}$ symmetry guarantees the dark matter stability.
Moreover, the $U(1)_{{\rm X}}$ gauge boson receives mass, $m_{Z'}=3 g_X v_\phi$, and the dark Higgs mass becomes, $m_{h'}=\sqrt{2\lambda_\phi}\, v_\phi$, in the limit of a vanishing $\lambda_{\phi H}$; see Ref.~\cite{Choi:2016tkj} for details.
We note that when the dark Higgs $\phi$ carries $q_\phi=+5$, the $U(1)_{{\rm X}}$ gauge symmetry is broken down to $Z_5$, for which single or multi-component dark matter scenarios were also discussed \cite{Choi:2016hid}.

\subsection{The inflation regime}

For our discussion on the inflation dynamics in the next section, we further expand the Lagrangian \eqref{eqn:Lag} by choosing the unitary gauges for the SM Higgs doublet and the dark Higgs field as $H^{T} = (0, h)/\sqrt{2}$ and $\phi = \varphi/\sqrt{2}$, and taking the scalar dark matter to be $\chi = \psi e^{i\gamma}/\sqrt{2}$.
The scalar part of the Lagrangian \eqref{eqn:Lag} then becomes
\begin{align}
\mathcal{L}_{{\rm NM}}
&=
\frac{M_{{\rm P}}^{2}}{2}\left(
1
+ \xi_{\chi}\frac{\psi^{2}}{M_{{\rm P}}^{2}}
+ \xi_{\phi}\frac{\varphi^{2}}{M_{{\rm P}}^{2}}
+ \xi_{H}\frac{h^{2}}{M_{{\rm P}}^{2}}
\right)\mathcal{R}
\,,\\
\mathcal{L}_{{\rm SM}}
&\supset
\frac{1}{2}(\partial_{\mu}h)^{2}
+\frac{1}{2}m_{H}^{2}h^{2}
-\frac{1}{4}\lambda_{H}h^{4}
\,,\\
\mathcal{L}_{{X}}
&=
\frac{1}{2}(\partial_{\mu}\varphi)^{2}
+\frac{1}{2}(\partial_{\mu}\psi)^{2}
+\frac{1}{2}\psi^{2}(\partial_{\mu}\gamma)^{2}
+\frac{1}{2}m_{\phi}^{2}\varphi^{2}
-\frac{1}{2}m_{\chi}^{2}\psi^{2} \nonumber \\
&\quad-\frac{1}{4}\lambda_{\phi}\varphi^{4}
-\frac{1}{4}\lambda_{\chi}\psi^{4}
-\frac{1}{4}\lambda_{\phi\chi}\varphi^{2}\psi^{2}
-\frac{1}{2}\zeta\varphi\psi^{3}\cos(3\gamma)
\,,\\
\mathcal{L}_{{\rm portal}}
&=
-\frac{1}{4}\lambda_{\phi H}\varphi^{2}h^{2}
-\frac{1}{4}\lambda_{\chi H}\psi^{2} h^{2}
\,.
\end{align}
We assume that the SM Higgs field rolls down rapidly to its VEV during inflation, due to the initial condition or the effective mass larger than the Hubble scale, so we set $h=0$.
The relevant Lagrangian for inflation is then given by
\begin{align}
\mathcal{L}_{{\rm inf}} &=
\frac{M_{{\rm P}}^{2}}{2}\left(
1 
+\xi_\phi \frac{\varphi^2}{M_{{\rm P}}^2}
+\xi_\chi \frac{\psi^2}{M_{{\rm P}}^2}
\right)\mathcal{R}
+\frac{1}{2}(\partial_{\mu}\varphi)^{2}
+\frac{1}{2}(\partial_{\mu}\psi)^{2}
+\frac{1}{2}\psi^{2}(\partial_{\mu}\gamma)^{2}
-V(\varphi,\psi,\gamma)\,, \label{Jinflation}
\end{align}
where the scalar potential reads
\begin{align}
V(\varphi,\psi,\gamma) &=
-\frac{1}{2}
m_{\phi}^{2}
\varphi^{2}
+\frac{1}{2}
m_{\chi}^{2}
\psi^{2}
+\frac{1}{4}\lambda_{\phi}\varphi^{4}
+\frac{1}{4}\lambda_{\chi}\psi^{4}
+\frac{1}{4}\lambda_{\phi\chi}\varphi^{2}\psi^{2}
+\frac{1}{2}\zeta\varphi\psi^{3}\cos(3\gamma)
\,.
\end{align}

For the inflationary period, we are interested in the large-field limit, $\xi_\phi \varphi^2 / M_{{\rm P}}^2+\xi_\chi \psi^2 /M_{{\rm P}}^2 \gg 1$, so we can drop the scalar mass terms.
After minimizing the scalar potential for the phase $\gamma$ of scalar dark matter for $\varphi\neq 0,\psi\neq 0$, we obtain the Jordan-frame potential as
\begin{align}
V_{{\rm inf}}(\varphi,\psi) &=
\frac{1}{4}\lambda_{\phi}\varphi^{4}
+\frac{1}{4}\lambda_{\chi}\psi^{4}
+\frac{1}{4}\lambda_{\phi\chi}\varphi^{2}\psi^{2}
-\frac{1}{2}\zeta\varphi\psi^{3}
\,.
\end{align}
We will present the detailed discussion on the decoupling of the phase $\gamma$ or the scalar dark matter $\chi$ for the general inflationary trajectories in Section 3.3. 

As a consequence, the Jordan-frame action in Eq.~\eqref{eqn:Lag} relevant for inflation becomes
\begin{align}
S_{{\rm inf}} &= \int d^{4}x \, \sqrt{-g} \, \bigg\{
\frac{M_{{\rm P}}^{2}}{2}\bigg(
1
+\xi_\phi \frac{\varphi^2}{M_{{\rm P}}^2}
+\xi_\chi \frac{\psi^2}{M_{{\rm P}}^2}
\bigg)\mathcal{R}
+\frac{1}{2}(\partial_{\mu}\varphi)^{2}
+\frac{1}{2}(\partial_{\mu}\psi)^{2} \nonumber \\
&\quad
-\frac{1}{4}\lambda_{\phi}\varphi^{4}
-\frac{1}{4}\lambda_{\chi}\psi^{4}
-\frac{1}{4}\lambda_{\phi\chi}\varphi^{2}\psi^{2}
+\frac{1}{2}\zeta\varphi\psi^{3}
\bigg\}
\,.
\end{align}
Then, the effective theory for inflation is a system of two real scalar fields with quartic couplings and non-minimal couplings, which has been extensively studied in Ref.~\cite{Lebedev:2011aq}. For the analysis of inflationary observables, it is convenient to work in the Einstein frame where the gravity action takes the standard form of Einstein-Hilbert action. Making a Weyl transformation to the Einstein frame by
\begin{align}
\tilde{g}_{\mu\nu} = \Omega^{2} g_{\mu\nu}
\end{align}
with 
\begin{align}
\Omega^2 = 1 + \xi_\phi \frac{\varphi^2}{M_{{\rm P}}^2}
+\xi_\chi \frac{\psi^2}{M_{{\rm P}}^2}
\,,
\end{align}
we obtain the following Einstein-frame action:
\begin{align}
S_{{\rm inf}} &= \int d^{4}x \, \sqrt{-\tilde{g}}\,\bigg\{
\frac{M_{{\rm P}}^{2}}{2}\tilde{\mathcal{R}}
+\frac{3}{4}M_{{\rm P}}^{2}\tilde{g}^{\mu\nu}
\partial_{\mu}\ln\Omega^2
\partial_{\nu}\ln\Omega^2 \nonumber \\
&\quad+\frac{1}{2\Omega^2}\tilde{g}^{\mu\nu}
\partial_{\mu}\varphi\partial_{\nu}\varphi
+\frac{1}{2\Omega^2}\tilde{g}^{\mu\nu}
\partial_{\mu}\psi\partial_{\nu}\psi
-U
\bigg\}
\,,
\end{align}
where the tilde indicates the quantities associated with the Einstein-frame metric $\tilde{g}_{\mu\nu}$, and the scalar potential in the Einstein frame is given by $U = V_{{\rm inf}}/\Omega^4$.

We will discuss the details of the inflation regime in Sec.~\ref{sec:effLagInf}.
The dimensionless parameters in the inflation regime are matched to low energy parameters for dark matter in the same section, thanks to the RG equations given in Appendix A.

\subsection{The dark matter regime}

For the discussion of dark matter annihilations, expanding the SM and dark Higgs fields by $H^{T} = (0, v_{\rm ew}+h)/\sqrt{2}$ and $\phi = (v_\phi+h')/\sqrt{2}$, and diagonalizing the scalar and neutral gauge boson mass matrices, we can obtain various interaction terms for scalar dark matter $\chi$ \cite{Choi:2015bya,Choi:2016tkj}. We denote the CP-even scalars and dark gauge-like bosons by $h_{1,2}$ and $Z'$, respectively.

The interaction Lagrangian for dark matter with Higgs and $Z'$ portals at low energy is given~\cite{Choi:2015bya} by
\begin{eqnarray}
{\cal L}_{\rm int}&=& -(\lambda_{\phi\chi}v_\phi \cos\theta-\lambda_{\chi H} v_{\rm ew}\sin\theta) h_1 |\chi|^2-(\lambda_{\phi\chi} v_\phi \sin\theta +\lambda_{\chi H} v_{\rm ew} \cos\theta) h_2 |\chi|^2
\nonumber \\
&&-(\lambda_\phi v_\phi \cos^3\theta-\lambda_H v_{\rm ew} \sin^3\theta) h^3_1 -3\sin\theta\cos\theta(\lambda_\phi v_\phi \cos\theta+\lambda_H v_{\rm ew} \sin\theta ) h^2_1 h_2
\nonumber \\
&&-3\sin\theta\cos\theta(\lambda_\phi v_\phi \sin\theta+\lambda_H v_{\rm ew} \cos\theta) h_1 h^2_2 
-(\lambda_\phi v_\phi \sin^3\theta+\lambda_H v_{\rm ew} \cos^3\theta) h^3_1
\nonumber \\
&&-\frac{1}{\sqrt{2}}\,\zeta\, ( \cos\theta\, h_1+\sin\theta\, h_2) \chi^3+{\rm h.c.} -\lambda_\chi |\chi|^4+\cdots
\nonumber \\
&&+g_X Z'_\mu J^\mu_D +e\, \varepsilon Z'_\mu J^\mu_{\rm EM}
\,,
\end{eqnarray}
where $\theta$ is the Higgs mixing angle, and $J^\mu_D, J^\mu_{\rm EM}$ are dark and electromagnetic currents, respectively, given by $J^\mu_D=-iq_\chi (\chi \partial^\mu \chi^*-\chi^*\partial^\mu \chi)$ and $J^\mu_{\rm EM}=Q_f {\bar f}\gamma^\mu f-iQ_W  (W^\nu\partial^\mu W^\dagger_\nu-W^\dagger_\mu \partial^\mu W_\nu)$, and $\varepsilon\simeq \cos\theta_W\,\xi$ for $\xi\ll 1$.
Here, the ellipsis in the fourth line corresponds to the quartic couplings for CP-even scalars, which are reduced to $-\frac{1}{4}\lambda_\phi h^4_1-\frac{1}{4}\lambda_H h^4_2$ in the limit of a vanishing Higgs mixing angle.
Henceforth, we denote $h_1\approx h'$ for a small Higgs mixing angle.

As a consequence, dark matter can annihilate into a pair of the SM particles through Higgs and $Z'$ portal couplings for nonzero $\theta$ and $\varepsilon$. Moreover, dark matter can also annihilate into the dark sector, such as $\chi\chi\rightarrow \chi^* A'$, $\chi\chi^*\rightarrow AA$ with $A=h', Z'$. When the self-interaction of dark matter is sizable, dark matter can also annihilate by $3\rightarrow 2$ processes, such as $\chi\chi\chi\rightarrow \chi\chi^\ast$, $\chi\chi\chi^\ast\rightarrow \chi^\ast\chi^\ast$, and $\chi\chi\chi^\ast\rightarrow \chi A$ with $A=h', Z'$.
The details on the Boltzmann equation for determining the relic density, $n_{\rm DM}=n_\chi+n_{\chi*}$, are presented in Appendix B.
For the later discussion, we assume that dark Higgs $h'$ and dark photon $Z'$ are in thermal equilibrium with the SM thermal bath, so we can take $n_{h'}=n^{\rm eq}_{h'}$ and $n_{Z'}=n^{\rm eq}_{Z'}$.
Using the expressions for all the dark matter annihilation cross sections summarized in Appendix B, we will determine the dark matter relic density  in Sec.~\ref{sec:DMinf}.

\section{Inflationary dynamics}
\label{sec:effLagInf}

In this section, we discuss the details of the dynamics of inflation and constraints in our model.
We also discuss the justification of decoupling the extra scalar degree(s) of freedom for the effective theory of inflation.

\subsection{Inflation in large-field limit}
Closely following the discussion in Ref.~\cite{Lebedev:2011aq} and making the field redefinitions with
\begin{align}
\sigma &\equiv
\sqrt{\frac{3}{2}} M_{{\rm P}} \ln \Omega^2
\,,\quad
\tau \equiv \frac{\varphi}{\psi}
\,,
\end{align}
we rewrite the scalar kinetic terms and potential in the Einstein frame, respectively, as
\begin{align}
\mathcal{L}_{{\rm kin}} &=
\frac{1}{2}
\frac{M_{{\rm P}}^{2}(\xi_{\chi}^{2}+\xi_{\phi}^{2}\tau^{2})}
{(\xi_{\chi} + \xi_{\phi}\tau^{2})^{3}}\left(
1 - e^{-\sqrt{\frac{2}{3}}\frac{\sigma}{M_{{\rm P}}}}
\right)
(\tilde{\partial}\tau)^2
\nonumber \\
&\quad +\frac{1}{2}\left(
\frac{e^{\sqrt{\frac{2}{3}}\frac{\sigma}{M_{{\rm P}}}}(1+6\xi_{\chi}+(1+6\xi_{\phi})\tau^{2})-6(\xi_{\chi}+\xi_{\phi}\tau^{2})}
{6(\xi_{\chi}+\xi_{\phi}\tau^{2})(e^{\sqrt{\frac{2}{3}}\frac{\sigma}{M_{{\rm P}}}}-1)}\right)
(\tilde{\partial}\sigma)^2
\nonumber\\
&\quad
+\frac{M_{{\rm P}}(\xi_{\chi}-\xi_{\phi})\tau}
{\sqrt{6}(\xi_{\chi}+\xi_{\phi}\tau^{2})^{2}}
(\tilde{\partial}\sigma)(\tilde{\partial}\tau)
\,,
\end{align}
and
\begin{align}
U =
U_0(\tau)\left(
1 - e^{-\sqrt{\frac{2}{3}}\frac{\sigma}{M_{{\rm P}}}}
\right)^{2}
\,,
\end{align}
with
\begin{align}
U_0(\tau) \equiv
\frac{M_{{\rm P}}^{4}(
\lambda_{\chi} - 2\zeta\tau
+\lambda_{\phi\chi}\tau^{2}+\lambda_{\phi}\tau^{4}
)}
{4(\xi_\chi+ \xi_\phi \tau^2)^2}
\,.
\label{eqn:vac}
\end{align}

In the large-field limit for inflation, the kinetic terms become approximated to
\begin{align}
\mathcal{L}_{{\rm kin}} &\simeq 
\frac{1}{2}(\tilde{\partial}\sigma)^{2}
+\frac{1}{2}\frac{M_{{\rm P}}^{2}(\xi_{\chi}^{2}+\xi_{\phi}^{2}\tau^{2})}{(\xi_{\chi}+\xi_{\phi}\tau^{2})^{3}}
(\tilde{\partial}\tau)^{2}
+\frac{M_{{\rm P}}(\xi_{\chi}-\xi_{\phi})\tau}{\sqrt{6}(\xi_{\chi}+\xi_{\phi}\tau^{2})^{2}}
(\tilde{\partial}\sigma)(\tilde{\partial}\tau)
\nonumber \\
& \simeq 
\frac{1}{2}(\tilde{\partial}\sigma)^{2}
+\frac{1}{2}(\tilde{\partial}\tau_c)^{2}
+(\tilde{\partial}\sigma)(\tilde{\partial}\tau_c)
\frac{(\xi_{\chi}-\xi_{\phi})\tau}
{\sqrt{6}\sqrt{\xi_{\chi}^{2}+\xi_{\phi}^{2}\tau^{2}}
\sqrt{\xi_{\chi}+\xi_{\phi}\tau^{2}}}
\,,
\end{align}
where, in the second line, we used the canonical scalar field $\tau_c$ with
\begin{align}
\left(\frac{d\tau_c}{d\tau}\right)^2
=
\frac{M_{{\rm P}}^{2}(\xi_{\chi}^{2}+\xi_{\phi}^{2}\tau^{2})}{(\xi_{\chi}+\xi_{\phi}\tau^{2})^{3}}
\,.
\label{eqn:dtaucdtau}
\end{align}
Here, we find that the kinetic mixing term is suppressed by $1/\sqrt{\xi_{\chi}+\xi_{\phi}\tau^{2}}$ with large non-minimal couplings.
As a result, it is sufficient to consider the effective action in the later discussion on inflation in the following form,
\begin{align}
S_{{\rm inf}} = \int d^{4}x \, \sqrt{-\tilde{g}} \, \bigg\{
\frac{M_{{\rm P}}^{2}}{2}\tilde{\mathcal{R}}
+\frac{1}{2}(\tilde{\partial}\sigma)^{2}
+\frac{1}{2}(\tilde{\partial}\tau_c)^{2}
-U
\bigg\}\,.
\end{align}

We note that there is no symmetry exchanging between $\varphi$ and $\psi$, leading to $\tau\rightarrow 1/\tau$,  for a nonzero value $\zeta$ and unequal values of non-minimal couplings. Therefore, we will show in the next subsection that the ratio of the fields, $\tau$, is stabilized at a nonzero value by the potential $U_0(\tau)$ in Eq.~(\ref{eqn:vac}). We will take the non-minimal couplings to equal values in some benchmark models for dark matter in the later discussion, but there is no symmetry associated with $\tau\rightarrow 1/\tau$ due to the cubic coupling for dark matter, $\zeta$. In the limit of a vanishing $\zeta$, the above effective Lagrangian coincides with the one with self-quartic and mixing quartic couplings only, known in the literature \cite{Lebedev:2011aq}.

\subsection{Inflationary minima}

In order to guarantee the stability of a single-field inflation with the $\sigma$ field, we impose the following conditions on the minimum of the potential with respect to $\tau=\tau_m$, 
\begin{align}
U\bigg\vert_{\tau=\tau_m} \geq 0
\,,\quad
\frac{\partial U}{\partial\tau_c}\bigg\vert_{\tau=\tau_m}
=0
\,,\quad
\frac{\partial^2 U}{\partial\tau_c^2}\bigg\vert_{\tau=\tau_m} \geq 0
\,.
\label{eqn:infcond}
\end{align}
Once $\tau_m$ is fixed by the minimization of the potential, the vacuum energy during inflation is given accordingly by $U_0(\tau_m)$ with Eq.~\eqref{eqn:vac}.
For the forthcoming discussion on the inflationary minima, we note that the first and second derivatives of the inflation potential with respect to $\tau_c$ are given, respectively, by
\begin{align}
\frac{\partial U}{\partial \tau_c}
&\approx
\frac{M_{{\rm P}}^{3}}{2(\xi_{\chi}+\xi_{\phi}\tau^{2})^{3/2}\sqrt{\xi_{\chi}^{2}+\xi_{\phi}^{2}\tau^{2}}}\bigg[
-\zeta \xi_{\chi}
+(\lambda_{\phi\chi}\xi_{\chi}-2\lambda_{\chi}\xi_{\phi})\tau \nonumber \\
&\quad+3\zeta\xi_{\phi}\tau^{2}
+(2\lambda_{\phi}\xi_{\chi}-\lambda_{\phi\chi}\xi_{\phi})\tau^{3}
\bigg]
\label{eqn:pot1stderiv}
\end{align}
and
\begin{align}
\frac{\partial^{2}U}{\partial\tau_c^2}
&\approx
\frac{M_{{\rm P}}^{2}}
{2(\xi_\chi + \xi_\phi \tau^2)(\xi_\chi^2 + \xi_\phi^2 \tau^2)^{2}}
\bigg\{
2\lambda_{\chi}\xi_{\phi}\left(
3\xi_{\phi}^{3}\tau^4+2\xi_\phi\xi_\chi^2\tau^2
-\xi_\chi^3
\right) \nonumber \\
&\quad +\lambda_{\phi\chi}\left(
\xi_\phi^4\tau^6
-5\xi_\phi^3\xi_\chi\tau^4
-5\xi_\phi\xi_\chi^3\tau^2
+\xi_\chi^4
\right)
\nonumber\\
&\quad
+\tau\left[
2\lambda_{\phi}\xi_\chi\tau\Big(
3\xi_{\chi}^{3}+2\xi_\phi^2\xi_\chi\tau^2
-\xi_\phi^3\tau^4
\right) \nonumber \\
&\qquad +\zeta\xi_\phi\Big(
9\xi_{\chi}^{3}+\xi_\phi\xi_\chi^2
+\xi_\phi\xi_\chi(
7\xi_\phi-3\xi_\chi)\tau^2
-6\xi_\phi^3\tau^4
\Big)
\Big]
\bigg\}
\,.
\label{eqn:pot2ndderiv}
\end{align}

\subsubsection{Dark matter inflaton}

From the first derivative of the potential in Eq.~\eqref{eqn:pot1stderiv}, it is easy to see that $\tau =\varphi/\psi= 0$ would not be an extremum for $\zeta \xi_\chi\neq 0$, due to the linear potential term. Thus, the inflation with pure dark matter ($\psi$) is not possible for $\zeta\neq 0$. (We note that $\xi_\chi = 0$ is not a fixed point under RG; see Appendix A.)
This shows a clear distinction from the dark matter scenarios with $Z_2$ symmetry where the inflation with pure dark matter becomes a stable minimum.
Since the case with $Z_2$ symmetry has been extensively discussed in the literature, we focus on the case with $\zeta \neq 0$, so we do not pursue any longer the inflation scenario with pure dark matter in the later discussion.

\subsubsection{Dark Higgs inflaton}

We find that $\tau=\varphi/\psi= \infty$, corresponding to the inflation with pure dark Higgs, is always an extremum of the potential, independent of $\zeta$.
From the second derivative of the potential \eqref{eqn:pot2ndderiv}, we find that
\begin{align}
\frac{\partial^{2}U}{\partial\tau_{c}^{2}}
=
\frac{M_{{\rm P}}^{2}}{2\xi_{\phi}^{2}}
\left(
\lambda_{\phi\chi}\xi_{\phi} - 2\lambda_{\phi}\xi_{\chi}
\right)\,
\end{align}
at $\tau = \infty$.
Therefore, $\tau = \infty$ is the minimum if
\begin{align}
\lambda_{\phi\chi}\xi_{\phi} - 2\lambda_{\phi}\xi_{\chi}
>0
\,,
\label{eqn:dHinfcond1}
\end{align}
or it is an inflection point if
\begin{align}
\lambda_{\phi\chi}\xi_{\phi} - 2\lambda_{\phi}\xi_{\chi}
=0
\,.
\label{eqn:dHinfcond2}
\end{align}

\subsubsection{Mixed inflaton}

In order to obtain the inflationary minimum with a finite $\tau$, one needs to solve the cubic equation, the solutions to which we do not present here, due to lengthy expressions. The general solution is known, determined by the discriminant $D$, given by
\begin{align}
\frac{D}{4} &= 
27\zeta^4 \xi_\phi^3 \xi_\chi
+\left(\lambda_{\phi\chi}\xi_\phi
-2\lambda_\phi \xi_\chi\right)
\left(\lambda_{\phi\chi}\xi_{\chi}
-2\lambda_\chi \xi_\phi \right)^{3}  \nonumber \\
&\quad+9\zeta^{2}\left[
\lambda_\chi^2 \xi_\phi^4
+\lambda_{\phi\chi}^{2}\xi_\phi^2 \xi_\chi^2
-3\lambda_{\phi}^{2}\xi_{\chi}^{4}
+2\lambda_{\chi}\xi_{\phi}^{2}\xi_{\chi}\left(
3\lambda_{\phi}\xi_{\chi}-2\lambda_{\phi\chi}\xi_{\chi}
\right)
\right]
\,.
\end{align}
If $D>0$, there exist three real values for $\tau$; if $D<0$ there exists only one real value for $\tau$. 
Thus, there exists at least one minimum for a finite $\tau$. 
Taking $\tau>0$ and using the second derivative of the potential in Eq.~\eqref{eqn:pot2ndderiv}, we can find the value for $\tau_m$ that satisfies the inflation conditions \eqref{eqn:infcond}.

For a finite value of $\tau$, the inflaton is a mixture of the dark Higgs ($\varphi$) and dark matter ($\psi$). Then, taking $\tau = \tau_m$ to be constant along the inflation direction, we can write down the Jordan-frame action as follows:
\begin{align}
S &= \int d^{4}x \, \sqrt{-g} \, \bigg\{
\frac{M_{{\rm P}}^{2}}{2}\left[
1+\left(
\xi_\chi + \xi_\phi \tau_m^{2}
\right)\frac{\psi^{2}}{M_{{\rm P}}^{2}}
\right]\mathcal{R}
\nonumber \\
&\quad
+\frac{1+\tau_m^{2}}{2}g^{\mu\nu}
\partial_{\mu}\psi\partial_{\nu}\psi
-\frac{1}{4}(\lambda_{\chi}-2\zeta\tau_m
+\lambda_{\phi\chi}\tau_m^2+\lambda_\phi\tau_m^4)\psi^4
\bigg\}
\,.
\end{align}
Redefining $\Phi \equiv \sqrt{1+\tau_m^2}\,\psi$, the above action becomes
\begin{align}
S = \int d^{4}x \, \sqrt{-g} \, \left\{
\frac{M_{{\rm P}}^{2}}{2}\left(
1+\xi_\Phi\frac{\Phi^{2}}{M_{{\rm P}}^{2}}
\right)\mathcal{R}
+\frac{1}{2}g^{\mu\nu}
\partial_{\mu}\Phi\partial_{\nu}\Phi
-\frac{1}{4}\lambda_\Phi\Phi^4
\right\}
\,,
\end{align}
where
\begin{align}
\xi_\Phi \equiv
\frac{\xi_\chi + \xi_\phi \tau_m^{2}}
{1+\tau_m^2}
\,,\qquad
\lambda_\Phi \equiv
\frac{\lambda_{\chi}-2\zeta\tau_m
+\lambda_{\phi\chi}\tau_m^2+\lambda_\phi\tau_m^4}
{(1+\tau_m^2)^2}
\,.
\label{eqn:xilambdaPhi}
\end{align}
Therefore, we obtain a single-field inflation for $\Phi$ with the effective parameters $\xi_\Phi$ and $\lambda_\Phi$. 

Unlike the inflation along the single field direction, that is, dark Higgs inflation, the effective inflaton quartic coupling can be very small due to the cancellation with the $\zeta$ parameter in the definition of $\lambda_\Phi$ in Eq.~\eqref{eqn:xilambdaPhi}.
Then, fixing the combination $\lambda_\Phi / \xi_\Phi^2$ by the Planck normalization, we can take a small effective non-minimal coupling $\xi_\Phi$, as will be shown shortly.
Consequently, the violation of tree-level unitarity can occur at a higher scale than the case with a large non-minimal coupling.

\subsection{Decoupling of the extra degree(s) of freedom from dark matter}

In the previous subsections, we chose  $\Omega^2\gg 1$ for a slow-roll inflation during which the field values of $\varphi$ and $\psi$ are almost constant: 1) dark Higgs inflaton with $\varphi\neq 0$ and $\psi=0$; 2) mixed inflaton with $\varphi\sim \psi\neq 0$. 

First, for the case with $\varphi\neq 0$ and $\psi\neq 0$, which is called the mixed inflaton, ignoring the mass terms, the part of the Lagrangian  in Eq.~(\ref{Jinflation}) for the phase $\gamma$ during inflation is given in Einstein frame by
\bea
{\cal L}_{\gamma} = \frac{1}{2\Omega^2}\,\psi^2(\partial_\mu\gamma)^2 - \frac{1}{2\Omega^4}\,\zeta\varphi \psi^3 \cos(3\gamma). \label{Lag-gamma}
\eea
Then, for $\xi_\phi\sim\xi_\chi$ and $\varphi\sim \psi$, we get $\Omega^2\sim 2\xi_\phi \varphi^2$. Here, we set $M_P=1$.
In this case, we simplify the above Lagrangian for the canonical field, $\gamma\sim \sqrt{2\xi_\phi}\,{\tilde\gamma}$, in the following,
\bea
{\cal L}_{\tilde\gamma} \simeq \frac{1}{2} (\partial_\mu{\tilde\gamma})^2 -\frac{\zeta}{8\xi^2_\phi}\, \cos\Big(3\sqrt{2\xi_\phi} {\tilde\gamma} \Big).
\eea
Therefore, for $\zeta>0$, expanding the phase as $3\sqrt{2\xi_\phi}\, {\tilde\gamma}=\pi+3\sqrt{2\xi_\phi}\, \delta{\tilde\gamma}$, we can identify the squared mass of the perturbation $ \delta{\tilde\gamma}$ as $m^2_{ \delta{\tilde\gamma}}\sim \frac{9\zeta}{4\xi_\phi}$, which is much greater than the squared Hubble parameter, $H^2=\frac{\lambda_{\Phi}}{4\xi^2_{\Phi}}$ with $\xi_{\Phi}\sim \xi_\phi\gg1$ and a sizable $\zeta$.
As a result, we can safely ignore the dynamics of the phase of the scalar dark matter during inflation.

Next, for $\varphi\neq 0$ and $\psi=0$, the radial coordinate representation for the scalar dark matter, $\chi=\psi\,e^{i\gamma}/\sqrt{2}$, is not a valid description. In this case, instead we can take $\chi=(\psi+i b)/\sqrt{2}$ in the Cartesian representation with $b$ being a pseudo-scalar field.
Then, ignoring the mass terms, the part of the Lagrangian in Eq.~(\ref{Jinflation}) for the scalar dark matter including the mixing quartic coupling is given by
\bea
{\cal L}_{\psi,b} =\frac{1}{2\Omega^2}\Big((\partial_\mu\psi)^2+(\partial_\mu b)^2\Big) - \frac{1}{4\Omega^4}\,\lambda_{\phi\chi} (\psi^2+b^2) \varphi^2.
\eea
Then, taking $\Omega^2\simeq \xi_\phi \varphi^2$  during inflation and canonically normalizing the dark scalars by $\psi=\sqrt{\xi_\phi}\langle\varphi\rangle\,{\tilde\psi}$ and $b=\sqrt{\xi_\phi}\langle\varphi\rangle\,{\tilde b}$ with $\langle\varphi\rangle$ being an almost constant field value during inflation, we can rewrite the above Lagrangian as follows,
\bea
{\cal L}_{{\tilde\psi},{\tilde b}} \simeq \frac{1}{2} (\partial_\mu{\tilde\psi})^2+\frac{1}{2} (\partial_\mu{\tilde b})^2 -\frac{\lambda_{\phi\chi}}{4\xi_\phi}\, ({\tilde\psi}^2+{\tilde b}^2).
\eea
Consequently, we obtain the squared masses for dark scalars as $m^2_{\tilde\psi}=m^2_{\tilde b}\sim \frac{\lambda_{\phi\chi}}{2\xi_\phi}$, which is again much larger than the squared Hubble parameter, $H^2=\frac{\lambda_\phi}{4\xi^2_\phi}$ with $ \xi_\phi\gg1$ and a sizable $\lambda_{\phi\chi}$.
As a result, we can ignore the dynamics of dark scalars in $\chi$ during inflation.

In summary, independent of the inflationary vacua, the phase of dark matter or the full complex scalar dark matter can be decoupled during inflation, thus justifying our analysis focusing on the two real scalar fields, $\varphi$ and $\psi$, in the previous subsections.
Moreover, the results show that there is no isocurvature perturbation generated along the complex scalar dark matter in any inflation minimum.

\subsection{Inflationary observables}
For a given potential, it is straightforward to compute inflationary observables such as the scalar power spectrum, the spectral index, and the tensor-to-scalar ratio, in terms of the inflation vacuum energy and slow-roll parameters.

In order to link inflation to dark matter physics, it is essential to include the RG effects to account for the difference in energy scales. We follow the procedures outlined in Ref.~\cite{Ko:2014eia}.
Quantizing the theory in the Jordan frame and considering the RG-improved effective action, we get the leading effective action~\cite{Sher:1988mj} as
\begin{align}
\Gamma = \int d^{4}x \, \sqrt{-g} \, \left[
\frac{M_{{\rm P}}^{2}}{2}\Omega^2\mathcal{R}
+\frac{1}{2}g^{\mu\nu}G^2\partial_\mu \Phi
\partial_\nu \Phi
-V_{{\rm eff}}
\right]
\,,
\end{align}
where $\Phi$ is the inflaton and
\begin{align}
\Omega^{2}(t) &=
1+\xi_{\Phi}(t)G^{2}(t)\frac{\Phi^{2}(t)}{M_{{\rm P}}^{2}}
\,,\ \\
V_{{\rm eff}}(t) &=
\frac{\lambda_{\Phi}(t)}{4}G^{4}(t)\Phi^{4}(t)
\,, \\
G(t) &= \exp\left(
-\int^{t}
dt'\,
\frac{\gamma_\Phi}{1+\gamma_{\Phi}}
\right)
\,.
\end{align}
Here, $t = \ln(\mu/\mu_0)$ and $\mu$ is the renormalization scale, and we choose $\mu=\Phi$ and $\mu_0 = M_Z$.
We present the RG equations determining the running couplings and anomalous dimensions in our model in Appendix A.
We identify the inflaton as $\Phi = \varphi$ in the dark Higgs inflation and $\Phi = \sqrt{1+\tau_m^2}\,\psi$ in the mixed inflation.

Inflationary observables are computed in the Einstein frame. The relevant Einstein-frame potential is given by
\begin{align}
U_{{\rm eff}} =\frac{V_{\rm eff}}{\Omega^4}=
\frac{\lambda_{\Phi}(t)G^{4}(t)\Phi^{4}(t)}
{4[1+\xi_{\Phi}(t)G^{2}(t)\Phi^{2}(t)/M_{{\rm P}}^{2}]^{2}}
\,.
\end{align}
The slow-roll parameters are then obtained from the standard definitions:
\begin{align}
\epsilon = \frac{M_{{\rm P}}^{2}}{2}\left(
\frac{U_{{\rm eff}}^{\prime}}{U_{{\rm eff}}}
\right)^{2}
\,,\quad
\eta = M_{{\rm P}}^{2}\frac{U_{{\rm eff}}^{\prime\prime}}{U_{{\rm eff}}}
\,,\quad
\kappa^2 = M_{{\rm P}}^{4}\frac{U_{{\rm eff}}^{\prime}U_{{\rm eff}}^{\prime\prime\prime}}{U_{{\rm eff}}^{2}}
\,,
\end{align}
where the prime represents the derivative with respect to the canonically normalized field $\Psi$ which is related to the Jordan-frame field $\Phi$ by
\begin{align}
\frac{d\Psi}{d\Phi} = 
\sqrt{\frac{G^2}{\Omega^{2}}
+\frac{3M_{{\rm P}}^{2}}{2\Omega^4}\left(
\frac{d\Omega^2}{d\Phi}    
\right)^2}
\,.
\end{align}
In the classical limit of $G \rightarrow 1$, we have
\begin{align}
\frac{d\Psi}{d\Phi} = 
\frac{\sqrt{1+(1+6\xi_{\Phi})\xi_{\Phi}\Phi^2 / M_{{\rm P}}^{2}}}
{1+\xi_\Phi \Phi^2 / M_{{\rm P}}^{2}}
\,.
\end{align}
From the slow-roll parameters, we compute the spectral index $n_s$ and the tensor-to-scalar ratio $r$ at the horizon exit, as follows \cite{Stewart:1993bc,Liddle:1994dx,Leach:2002ar},
\begin{align}
n_s &\approx 
1-6\epsilon + 2\eta - \frac{2}{3}(5+36c)\epsilon^2 + 2(-1+8c)\epsilon\eta
+\frac{2}{3}\eta^2 + \left( \frac{2}{3} - 2c\right)\kappa^2
\,,\\
r &\approx
16\epsilon\left[
1 + \left( -\frac{4}{3} + 4c \right)\epsilon + \left( \frac{2}{3} - 2c \right) \eta
\right]
\,,
\end{align}
where we have computed the quantities up to the second order in the slow-roll parameters and $c = \gamma + \ln 2 - 2$ with $\gamma \approx 0.5772$ being the Euler-Mascheroni constant.
The number of e-foldings, $N$, can be also obtained from
\begin{align}
N = \frac{1}{M_{{\rm P}}^{2}}
\int_{\Psi_e}^{\Psi_*}
d\Psi\,
\frac{U_{{\rm eff}}}{U_{{\rm eff}}^{\prime}}
\,,
\end{align}
where $\Psi_e$ ($\Psi_*$) is the canonically normalized field value at the end of inflation for $\epsilon \simeq 1$ (at the horizon exit).
For typical reheating scenarios, we take $N=60$ at the horizon exit.

Let us briefly discuss the inflation dynamics classically.
From $\epsilon \simeq 1$ we find that the Jordan-frame inflaton field at the end of inflation $\Phi_e$ is given by $\Phi_e \simeq M_{{\rm P}}(4/3)^{1/4}/\sqrt{\xi_\Phi}$.
We may then express $\Phi$ in terms of $N$ as
\begin{align}
\Phi(N) \approx 
\sqrt{\frac{4N}{3\xi_\Phi}}M_{{\rm P}}
\,.
\label{eqn:fldtoN}
\end{align}
Then, the magnitude of the scalar power spectrum is given by
\begin{align}
A_s = \frac{U_{\rm eff}}{24\pi^2 M_{{\rm P}}^{4} \epsilon}
\approx \left(
\frac{N^{2}}{72\pi^2}
\right)\frac{\lambda_\Phi}{\xi_\Phi^2}
\,.
\end{align}
As mentioned earlier, the non-minimal coupling parameter $\xi_\Phi$ is chosen to satisfy the magnitude of the scalar power spectrum at the horizon exit, $A_s \approx 2.1\times 10^{-9}$ \cite{Akrami:2018odb}. 

In the large-field limit, the slow-roll parameters, $\epsilon$ and $\eta$, are expressed in terms of the number of e-foldings, 
\begin{align}
\epsilon \approx \frac{3}{4N^2} \,,
\quad
\eta \approx -\frac{1}{N} \,.
\end{align}
At the first order in the slow-roll parameters, the spectral index and the tensor-to-scalar ratio are then given by
\begin{align}
n_s \approx 1  - \frac{2}{N} - \frac{9}{2N^2}
\,,\quad
r \approx \frac{12}{N^2}
\,.
\end{align}
Therefore, we recover the well-known results for inflation with non-minimal coupling at the classical level, which are $n_s \approx 0.9654$ and $r \approx 0.003$ for $N=60$, being consistent with Planck data \cite{Akrami:2018odb}.

We are now in a position to discuss quantum corrections for inflationary observables. 
To discuss effects of runnings of the parameters qualitatively, let us consider the running of the effective quartic coupling $\lambda_\Phi$, ignoring the running of the non-minimal coupling for now. The slow-roll parameters, $\epsilon$ and $\eta$, in the large-field limit are then corrected \cite{RGs,Ko:2014eia} to
\begin{align}
\epsilon \approx
\frac{4M_{{\rm P}}^{4}}
{3\xi_\Phi^2\Phi^4}\left[
1 + \frac{\xi_\Phi \Phi^2}{4M_{{\rm P}}^{2}}\left(
\frac{\beta_{\lambda_{\Phi}}}{\lambda_{\Phi}}
\right)
\right]^2\,,\qquad
\eta \approx 
-\frac{4M_{{\rm P}}^{2}}{3\xi_{\Phi}\Phi^2}\left(
1 - \frac{\beta_{\lambda_{\Phi}}}{2\lambda_{\Phi}}
\right)\,,
\end{align}
where $\beta_{\lambda_{\Phi}} \equiv d\lambda_{\Phi}/dt$ is the beta function of the effective quartic coupling.
Here we have ignored $d\beta_{\lambda_\Phi}/dt$ and $\beta^2_{\lambda_\Phi}$ terms.

One may relate the inflation field value $\Phi$ with the number of e-foldings $N$.
In the presence of quantum corrections, the number of e-foldings becomes
\begin{align}
N \approx 
\frac{3\xi_\Phi}{2M_{{\rm P}}^2}
\int^{\Phi_*}_{\Phi_e} d\Phi \,
\Phi\left[
1 - \frac{\xi\Phi^2}{4M_{{\rm P}}^{2}}\left(
\frac{\beta_{\lambda_\Phi}}{\lambda_\Phi}
\right)
\right]
\,,
\end{align}
from which we obtain
\begin{align}
\frac{\Phi(N)}{M_{{\rm P}}} \approx 
\sqrt{\frac{4N}{3\xi_\Phi}\left[
1 + \frac{N}{6}\left(
\frac{\beta_{\lambda_\Phi}}{\lambda_\Phi}
\right)
\right]}\,.
\end{align}
Therefore, if $\beta_{\lambda_\Phi}/\lambda_\Phi$ is positive, the inflation field value at the horizon exit becomes larger than the classical value.
In terms of $N$, the slow-roll parameters are given by
\begin{align}
\epsilon \approx
\frac{3}{4N^2}+\frac{1}{4N}\left(
\frac{\beta_{\lambda_\Phi}}{\lambda_\Phi}
\right)\,,
\quad
\eta \approx
-\frac{1}{N}
+\left(\frac{1}{6}+\frac{1}{2N}\right)\left(
\frac{\beta_{\lambda_\Phi}}{\lambda_\Phi}
\right)
\,.
\end{align}

The spectral index $n_s$ and tensor-to-scalar ratio $r$ are, up to the first order in the slow-roll parameters, given by $r \approx 16\epsilon$ and $n_s \approx 1-6\epsilon+2\eta$. We thus have
\begin{align}
r &\approx 
\frac{12}{N^2}+\frac{4}{N}\left(
\frac{\beta_{\lambda_\Phi}}{\lambda_\Phi}
\right)
\,,\\
n_s &\approx 
1 - \frac{2}{N} - \frac{9}{2N^2}
+\frac{\beta_{\lambda_\Phi}}{3\lambda_\Phi}
\,.
\end{align}
The quantum effect may become important in two ways: (i) $\beta_{\lambda_\Phi}$ is large and/or (ii) $\lambda_\Phi$ is small. 
In particular, in the mixed inflation, increasing the $\zeta$ parameter with the other parameters being fixed, $\lambda_{\Phi}$ becomes smaller, shifting the tensor-to-scalar ratio and spectral index to a larger value, provided that $\beta_{\lambda_\Phi}$ is positive.

We remark that the reheating process can be important for making the inflationary predictions more precise, due to the fact that the total number of e-foldings depends on the equation of state of the inflaton and the reheating temperature during reheating \cite{reheating}. In the later discussion on dark matter, we take dimensionless parameters of our model to be sizable for dark matter with freeze-out processes, as far as they are allowed by perturbativity and unitarity, so we expect the reheating process to be almost instantaneous and the reheating temperature to be high enough.
Moreover, there are uncertainties in quantifying the preheating dynamics.  However, the details of preheating or reheating dynamics do not affect the later discussion on thermal dark matter in the next section, because dark matter is quickly thermalized after reheating and the dark matter abundance is determined by the freeze-out process at a low temperature, being insensitive to reheating.
Therefore, we do not elaborate on the reheating dynamics further and take the number of e-foldings to be about $N=60$ for illustration.

\subsection{Consistency conditions from inflation}

We discuss the theoretical and phenomenological consistency conditions to be satisfied for inflation physics in our model.
The constraints for inflation at high scales are the following:
\begin{itemize}
\item Dark Higgs inflation \vspace{0.2cm} \\
The stability of the inflaton potential requires
\begin{align}
\lambda_\phi &> 0,  \\
\lambda_{\phi\chi} &> 2\lambda_\phi \frac{\xi_\chi}{\xi_\phi}\,.
\end{align}

\item Mixed inflation \vspace{0.2cm} \\
The stability of the inflaton potential at a finite value for $\tau=\tau_m$ satisfying the extremum condition \eqref{eqn:pot1stderiv} requires
\begin{align}
\lambda_\Phi&=
\frac{1}{(1+\tau_m^2)^2}\big(
\lambda_\chi-2\zeta\,\tau_m+\lambda_{\phi\chi}\tau^2_m +\lambda_\phi \tau^4_m
\big)
>0
\,, \\
0 &=
\left(
\lambda_{\phi\chi}
-2\lambda_\phi
\frac{\xi_\chi}{\xi_\phi}
\right)\tau_m^3
-3\zeta\tau_m^2 
+\left(
2\lambda_\chi
-\lambda_{\phi\chi}
\frac{\xi_\chi}{\xi_\phi}
\right)\tau_m
+\zeta\,\frac{\xi_\chi}{\xi_\phi}
\,.
\end{align}
We further require the second derivative of the potential \eqref{eqn:pot2ndderiv} to be positive.

\item Perturbativity \vspace{0.2cm}\\
We also require perturbativity on the SM and new quartic couplings,
\begin{align}
y_t < 4\pi\,,\quad
g < 4\pi\,,\quad 
g^\prime < 4\pi \,,\quad
g_3 < 4\pi \,,\quad
g_{X} < 4\pi \,,\quad 
\lambda_i < 4\pi \,,\quad 
\zeta < 4\pi \,,
\end{align}
where $i=\phi,$ $\chi$, $\phi\chi$, $\phi H$ and $\chi H$.

\item CMB normalization  \vspace{0.2cm} \\
The scalar power spectrum at the Planck pivot scale \cite{Akrami:2018odb} is given by
\begin{align}
\ln(10^{10} A_s) = 3.044 \pm 0.014
\quad
\text{(68\% C.L. Planck TT,TE,EE+lowE+lensing)}
\,.
\end{align}
Then, requiring $A_s \approx 2.1\times 10^{-9}$ at $N=60$, we obtain the relation between the effective quartic and non-minimal couplings as follows:
\begin{align}
\frac{\lambda_\Phi}{\xi_\Phi^2}
\approx 4.15 \times 10^{-10}
\,.
\end{align}
For instance, for $\lambda_\Phi \simeq 0.1$ ($10^{-7}$), the non-minimal coupling should take $\xi_\Phi \simeq 15500$ ($15.5$).

\item Spectral index and tensor-to-scalar rato \vspace{0.2cm} \\
The latest Planck data \cite{Akrami:2018odb} read
\begin{align}
n_s &= 0.9659 \pm 0.0041 \,&\text{(68\% C.L. Planck TT,TE,EE+lowEB+lensing)}
\,,\\
r &< 0.11 \,&\text{(95\% C.L. Planck TT,TE,EE+lowEB+lensing)}
\,.
\end{align}

\end{itemize}

\section{Connection between inflation and dark matter}
\label{sec:DMinf}

In this section, we first discuss three distinct regimes of dark matter production in our model with $Z_3$ symmetry: WIMP, SIMP, and forbidden dark matter scenarios \cite{Choi:2015bya,Choi:2016tkj}.
In the following, the relic density is calculated in each case and various theoretical and experimental constraints from dark matter are listed. 
Then, we explore the connection between inflation and dark matter physics and impose the consistency conditions for inflation and experimental constraints on the model. 

As explained in the previous section, we can take quantum corrections into account by considering the running of the couplings, which can be read from the RG equations summarized in Appendix A.
We choose the following input parameters,
\begin{align}
m_{h^\prime}\,,\quad
m_\chi\,,\quad
\theta\,,\quad
\lambda_\phi\,,\quad
\lambda_\chi\,,\quad
\lambda_{\phi\chi}\,,\quad
\lambda_{\chi H}\,,\quad
\zeta,\quad
g_X\,,\quad
\xi_H\,,\quad
\xi_\chi/\xi_\phi\,,
\end{align}
at the $Z$ boson mass scale.
The $Z'$ mass is also given by $m_{Z'}=3g_X m_{h'}/\sqrt{2\lambda_\phi}$ for the given set of parameters. Here, we ignore the RG effects of the gauge kinetic mixing $\sin\xi$ or $\varepsilon\simeq \cos\theta_W\, \xi$ for $\xi\ll 1$.
We again stress that one of the non-minimal coupling parameters are not a free parameter. Thus we consider the ratio of the non-minimal couplings $\xi_\chi / \xi_\phi$ together with the SM Higgs non-minimal coupling $\xi_H$. Since the SM Higgs does not participate in inflationary dynamics, we always choose $\xi_H = 0$ at the $Z$ boson mass scale. However, due to the quantum effects, $\xi_H$ takes a non-zero value at inflationary scale.

For a chosen set of input parameters at low-energy scale, we first analyze the dark matter phenomenology by imposing various experimental and theoretical constraints on dark matter.
We then run the input parameters with the RG equations up to the scale of tree-level unitarity $\mu \sim M_{{\rm P}}/\sqrt{\xi_\Phi}$, where $\xi_\Phi$ is the effective non-minimal coupling for the inflaton \eqref{eqn:xilambdaPhi}, which is $\xi_\Phi = \xi_\phi$ in dark Higgs inflation and $\xi_\Phi = (\xi_\chi + \xi_\phi \tau_m^2)/\sqrt{1+\tau_m^2}$ in mixed inflation.
We combine the stability and perturbativity constraints from inflation at high-energy scale  and various theoretical and experimental constraints from dark matter at low-energy scale, as discussed in Sec.~\ref{sec:effLagInf} and Sec.~\ref{sec:DMinf}.

In the following, we present the results of our numerical analyses, dividing the parameter space into three distinct cases, depending on the dominant production mechanisms for dark matter, WIMP, SIMP, and forbidden (FBDM) scenarios. We study dark matter and inflation constraints in each case, namely, the dependencies of the relic density, the spectral index and the tensor-to-scalar ratio on the $\zeta$ parameter, which is characteristic of the $Z_3$ model, distinguishing it from the $Z_2$ models.

\subsection{Consistency conditions from dark matter}

The theoretical and phenomenological constraints for dark matter at low energies are the following:
\begin{itemize}
\item Stability \vspace{0.2cm} \\
We impose the stability condition given by \cite{Choi:2016tkj}
\begin{align}
\lambda_{\chi} > -\frac{1}{2}\lambda_{\phi\chi}X_{{\rm min}}^{2}
+\frac{3}{2}|\zeta|X_{{\rm min}}
\,, \label{VSB}
\end{align}
where
\begin{align}
X_{{\rm min}} = \Bigg\{
\begin{array}{ll}
\left(
P+\sqrt{P^2 + Q^3}    
\right)^{1/3}
+\left(
P-\sqrt{P^2 + Q^3}    
\right)^{1/3}\,, & D>0 \\
2\sqrt{-Q}\cos\left(
\frac{1}{3}\cos^{-1}\left(
\frac{P}{\sqrt{-Q^3}}
\right)
\right)
\,, & D<0
\end{array}
\end{align}
with $D \equiv P^2 + Q^3$, $P \equiv |\zeta|/(4\lambda_\phi)$, and $Q \equiv \lambda_{\phi\chi}/(6\lambda_{\phi})$.
Here, writing the scalar potential with quartic couplings in the form, $V=|\chi|^4 f(X)$ where $X\equiv |\phi|/|\chi|$, we denoted $X_{\rm min}$ as a positive value satisfying $f'(X_{\rm min})=0$. Then, the global minimum condition, $f(X_{\rm min})>0$, gives rise to Eq.~(\ref{VSB}) \cite{Choi:2016tkj}. 
\item Unitarity \vspace{0.2cm}\\
We impose the unitarity conditions on the squared amplitudes for dark matter self-scattering with Eqs.~\eqref{M1} and \eqref{M2}:
\begin{align}
|\mathcal{M}_{\chi\chi}| < 8\pi \,,\quad
|\mathcal{M}_{\chi\chi^*}| < 8\pi \,.
\end{align}
We note that perturbativity at low scale is trivially satisfied once it is imposed at inflation scale.
\item Dark matter relic density \vspace{0.2cm}\\
We impose the correct relic density for dark matter at present, given by Planck data (TT,TE,EE$+$lowE$+$lensing) \cite{planck} as
\begin{eqnarray}
\Omega_{\rm DM} h^2=0.1200\pm 0.0012
\,. \nonumber 
\end{eqnarray}
\item Higgs decays \vspace{0.2cm}\\
In the numerical analyses  below, we choose the mixing angle between the SM and dark Higgs bosons such that the LHC constraints on Higgs visible and invisible decays \cite{Sirunyan:2018owy} are satisfied.
The current limit on the branching fraction of the Higgs invisible decay is ${\rm BR}(h\rightarrow {\rm inv})<0.19$ at $90\%$ C.L.
\item Direct detection bounds \vspace{0.2cm}\\
We consider Xenon10 \cite{Xenon10}, Xenon 1T \cite{Xenon1T} and SENSEI-100 1yr (expected)\cite{sensei100} on the spin-independent cross sections for DM-nucleon and DM-electron scatterings, which are summarized in Appendix B. The Xenon10 limit \cite{Xenon10} applies for $m_\chi=8.8 \,{\rm MeV}-3\,{\rm GeV}$, so do the Xenon 1T limits \cite{Xenon1T} for $m_\chi=6 \,{\rm GeV}-10 \,{\rm TeV}$.
The projected limit from SENSEI-100 1yr (expected) is relevant for $m_\chi=1\, {\rm MeV}-1 \,{\rm GeV}$ and $m_{Z'} = 3m_\chi$.
\item Indirect detection bounds \vspace{0.2cm}\\
The indirect bounds on dark matter annihilation channels are the following.
First, the $e^+e^-$ annihilation constraint in CMB is stringent. We thus used the result in Ref.~\cite{CMBemep} to constrain the $e^+e^-$ annihilation at $T=0.25$ eV for the mass range between $m_\chi=1$ MeV and 10 TeV. 
We also used the AMS results in Refs.~\cite{AMSWmWp,AMSggbb,AMSemep} to constrain the $W^+W^-$ annihilation for the mass range between $m_\chi=100$ GeV to 40 TeV, $\gamma\gamma,\ b\bar{b}$ annihilations for $m_\chi=10\,{\rm GeV}-10\,{\rm TeV}$ and the $e^+e^-$ annihilation for  $m_\chi=5 \,{\rm GeV}-300 \,{\rm GeV}$, respectively. In our model, the di-photon annihilation channel is induced by the combination of the SM loops and the Higgs portal interaction.
We also used the Fermi-LAT limit on the $W^+W^-$ annihilation from Ref.~\cite{FermiLATWmWpttbb} for $m_\chi=m_W-10\, {\rm TeV}$ and the Fermi-LAT limits on the $\tau\tau,\ b\bar{b}$ annihilations from Ref.~\cite{FermiLATWmWpttbb} for $m_\chi=m_b(m_\tau)-10\,{\rm TeV}$. Moreover, the Fermi-LAT limit on the $\gamma\gamma$ annihilation in Ref.~\cite{FermiLATgg} is imposed for  $m_\chi=0.2\, {\rm GeV}-1\,{\rm TeV}$.
Lastly, the HESS constraints on the $W^+W^-, \gamma\gamma$ annihilations are taken from Refs.~\cite{HESSWmWp,HESSgg} for $m_\chi=180 \,{\rm GeV} -67\, {\rm TeV}$ and $m_\chi=300 \,{\rm GeV}-60 \,{\rm TeV}$, respectively.

\item Dark matter self-scattering  \vspace{0.2cm} \\
In the case of SIMP or forbidden dark matter, the self-scattering cross section can be sizable, so it can solve the small-scale problems at galaxies or can be constrained by no hint for dark matter self-interactions at galaxy clusters. 
Thus, we require the self-interaction cross section to be small enough to evade the bounds from Bullet Cluster \cite{Bullet} and spherical halo shapes \cite{haloShapes}:
\begin{align}
\frac{\sigma_{{\rm self}}}{m_\chi} < 1
\,{\rm cm^2/g}
\,.
\end{align}
The solution to the small-scale problems at galaxies \cite{Yu} might favor $\sigma_{\rm self}/m_\chi\gtrsim 0.1\,{\rm cm^2/g}$, but we do not impose the lower bound in our work by having in mind existing alternative solutions such as baryonic feedback effects.
\end{itemize}

\subsection{WIMP scenarios}
\label{subsec:WIMPres}

WIMP dark matter annihilates into the SM particles through Higgs or $Z'$ portals, but the strong constraints for direct detection narrow down this case near the resonance regions at $m_\chi\simeq m_{h'}/2$ or $m_{Z'}/2$. The indirect detection and CMB bounds can also constrain the WIMP dark matter with sizable Higgs portal couplings for $m_\chi\lesssim 100\,{\rm GeV}$. 
Away from the resonance regions, we focus on the dark matter annihilations into the hidden sector, such as $\chi\chi^*\rightarrow AA$ and $\chi\chi\rightarrow \chi^* A$ with $A=h',Z'$, for which we are safe from the direct detection bounds. Moreover, for $m_\chi\gtrsim 100\,{\rm GeV}$, the indirect detection and CMB bounds are not as strong.

We take the standard $2\rightarrow 2$ annihilation channels to be dominant for determining the dark matter relic density, resulting in the Boltzmann equation \eqref{Boltzmann} in the following form,
\begin{equation}\begin{aligned}
\dot{n}_\text{DM} + 3Hn_\text{DM}\ =\ &-\frac{1}{2}\langle \sigma v \rangle_{\chi\chi^*\rightarrow f{\bar f}}\bigg(n_{\rm DM}^2-(n_\text{DM}^{\rm eq})^2 \bigg)  
\\
& 
-\frac{1}{2}\langle \sigma v \rangle_{\chi \chi \rightarrow \chi^\ast h'}\bigg(n_\text{DM}^2-n_{\rm DM}^\text{eq}\, n_\text{DM}  \bigg)  -\frac{1}{2}\langle \sigma v \rangle_{ \chi \chi^\ast  \rightarrow h' h'}\bigg(n_\text{DM}^2-(n_{\rm DM}^\text{eq})^2 \bigg) \\
& -\frac{1}{2}\langle \sigma v \rangle_{ \chi \chi \rightarrow  \chi^\ast Z'}\bigg(n_\text{DM}^2- n_{\rm DM}^\text{eq}\, n_\text{DM}  \bigg)  - \frac{1}{2}\langle \sigma v \rangle_{ \chi \chi^\ast   \rightarrow Z' Z'}\bigg(n_\text{DM}^2-(n_{\rm DM}^\text{eq})^2 \bigg)\\
\approx& -\langle \sigma v\rangle_{2\rightarrow 2}\, n^2_{\rm DM}
\,,
\label{Boltzmann1}
\end{aligned}\end{equation}
with
\begin{eqnarray}
(\sigma v)_{2\rightarrow 2}&\equiv&\frac{1}{2} (\sigma v )_{\chi\chi^*\rightarrow f{\bar f}}+\frac{1}{2}( \sigma v )_{\chi \chi \rightarrow \chi^\ast h'}+\frac{1}{2}(\sigma v )_{ \chi \chi^\ast  \rightarrow h' h'} \nonumber \\
&&+\frac{1}{2}( \sigma v )_{ \chi \chi \rightarrow  \chi^\ast Z'}+\frac{1}{2}( \sigma v )_{ \chi \chi^\ast   \rightarrow Z' Z'} \nonumber \\
&\equiv & a+ b v^2.
\end{eqnarray}
Then, the relic density for WIMP dark matter is given by
\begin{eqnarray}
\Omega_{\rm DM} h^2 = 5.20\times 10^{-10}\,{\rm GeV}^{-2} \bigg(\frac{10.75}{g_*} \bigg)^{1/2} \Big(\frac{x_f}{20} \Big) \bigg(a+ \frac{3b}{x_f} \bigg)^{-1}
\,,
\end{eqnarray}
where $x_f=m_{\rm DM}/T_f$ with $T_f$ being the freeze-out temperature.

\begin{figure}[tbp]
\centering
\includegraphics[width=0.45\linewidth]{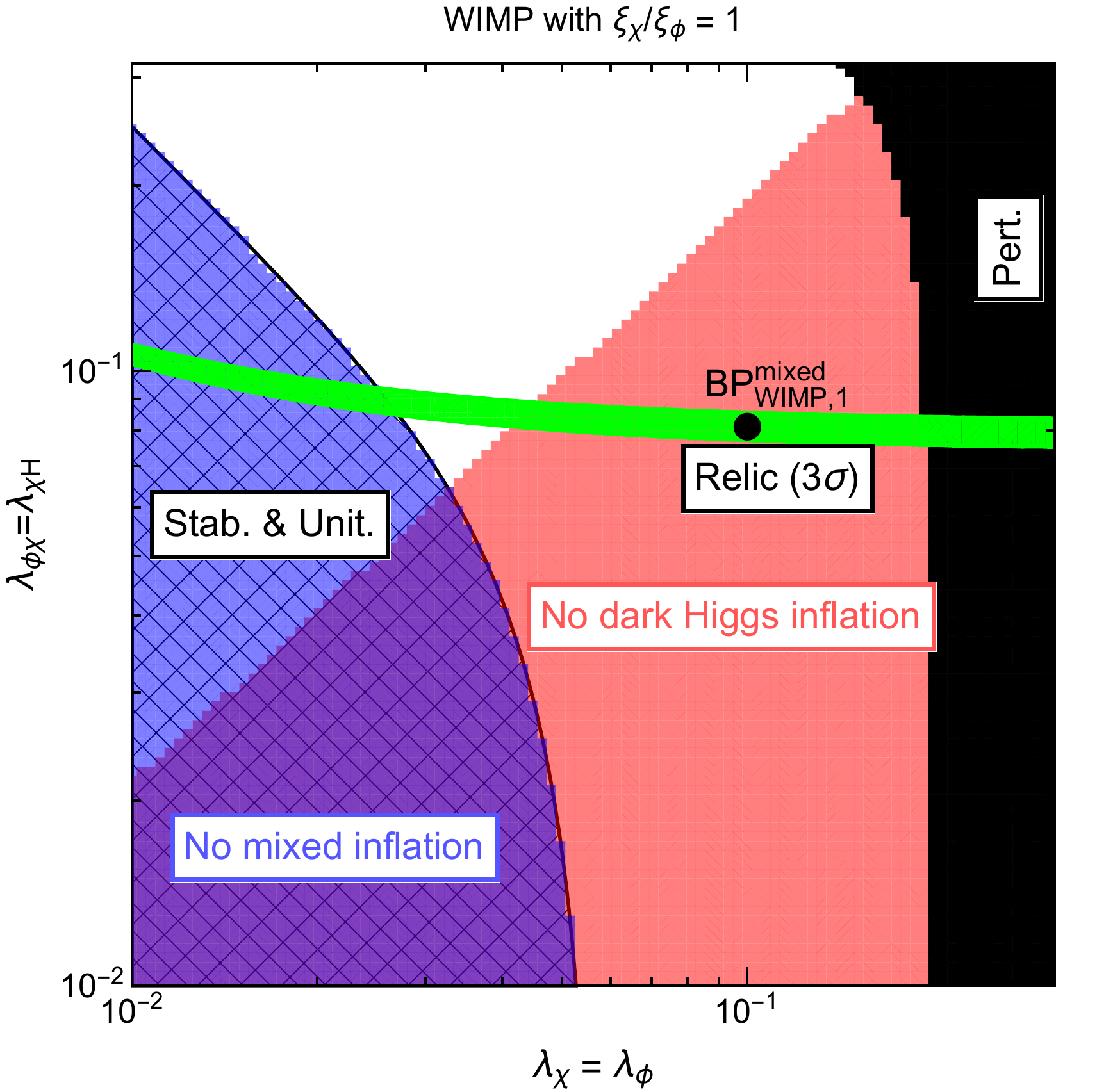} \,\,
\includegraphics[width=0.45\linewidth]{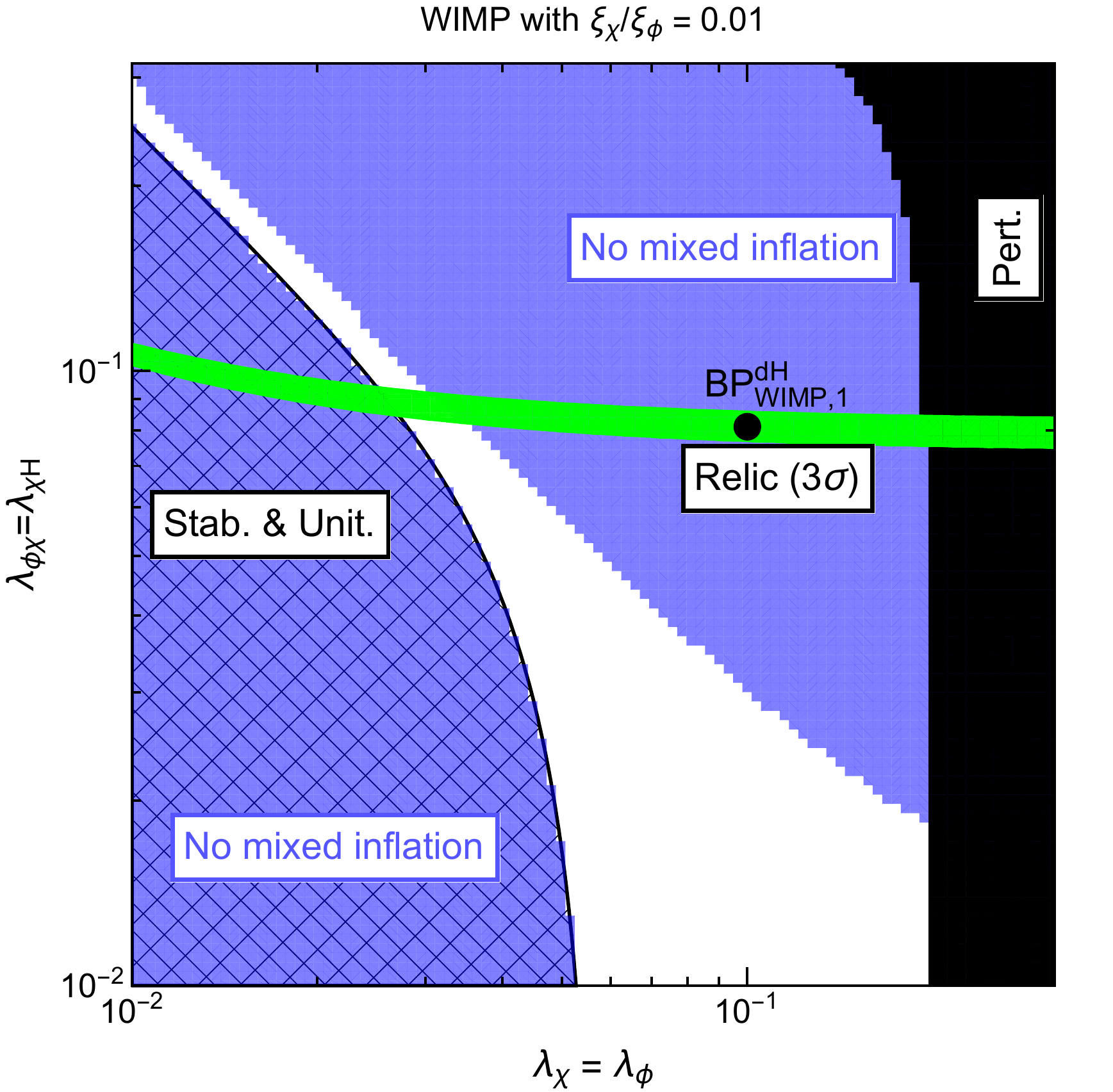}
\caption{Parameter space for WIMP dark matter scenario with $\xi_\chi/\xi_\phi=1$ (left) and $0.01$ (right) for the quartic self-couplings with $\lambda_\chi=\lambda_\phi$ versus the quartic mixing couplings with $\lambda_{\phi\chi}=\lambda_{\chi H}$. The parameters shown here are set at low-energy scale. The rest input parameters are chosen as $\zeta = 0.05$, $g_X =0.1, \, \sin\theta = 0.01$, $\varepsilon=10^{-4}$, $m_{h^\prime} = 100$ GeV and $m_\chi = 500$ GeV. Constraints from dark matter physics and cosmic inflation are shown. The green region represents the correct dark matter relic abundance within 3$\sigma$ range \cite{planck}. The hatched and black regions are disfavored due to stability, unitarity and perturbativity. The dark Higgs (mixed) inflation is not allowed in the red-colored (blue-colored) region. The black points are our benchmark points shown in Table \ref{tab:BPWIMPinput}.}
\label{fig:WIMP1}
\end{figure}
\begin{figure}[tbp]
\centering
\includegraphics[width=0.45\linewidth]{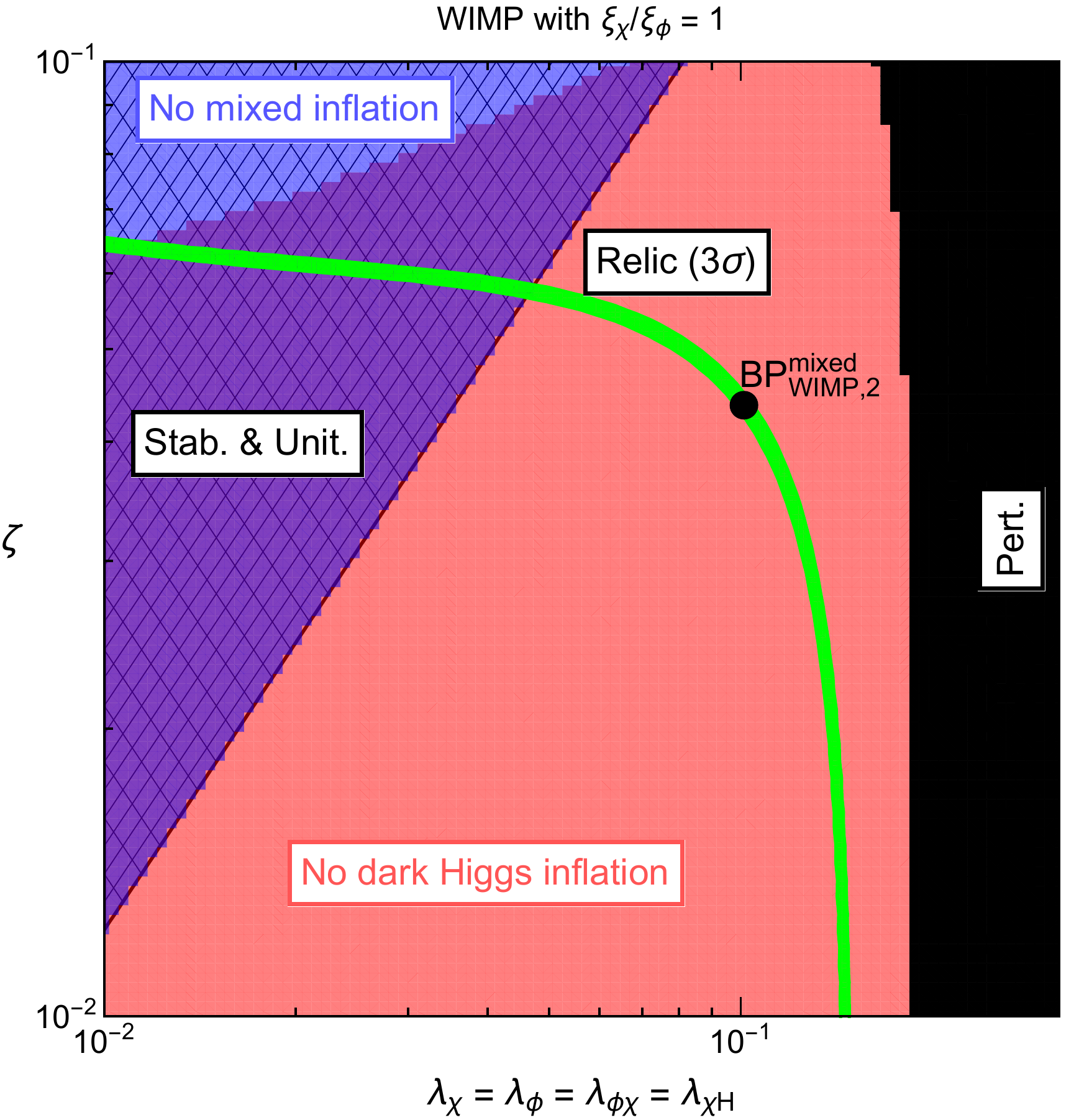}\,\,
\includegraphics[width=0.45\linewidth]{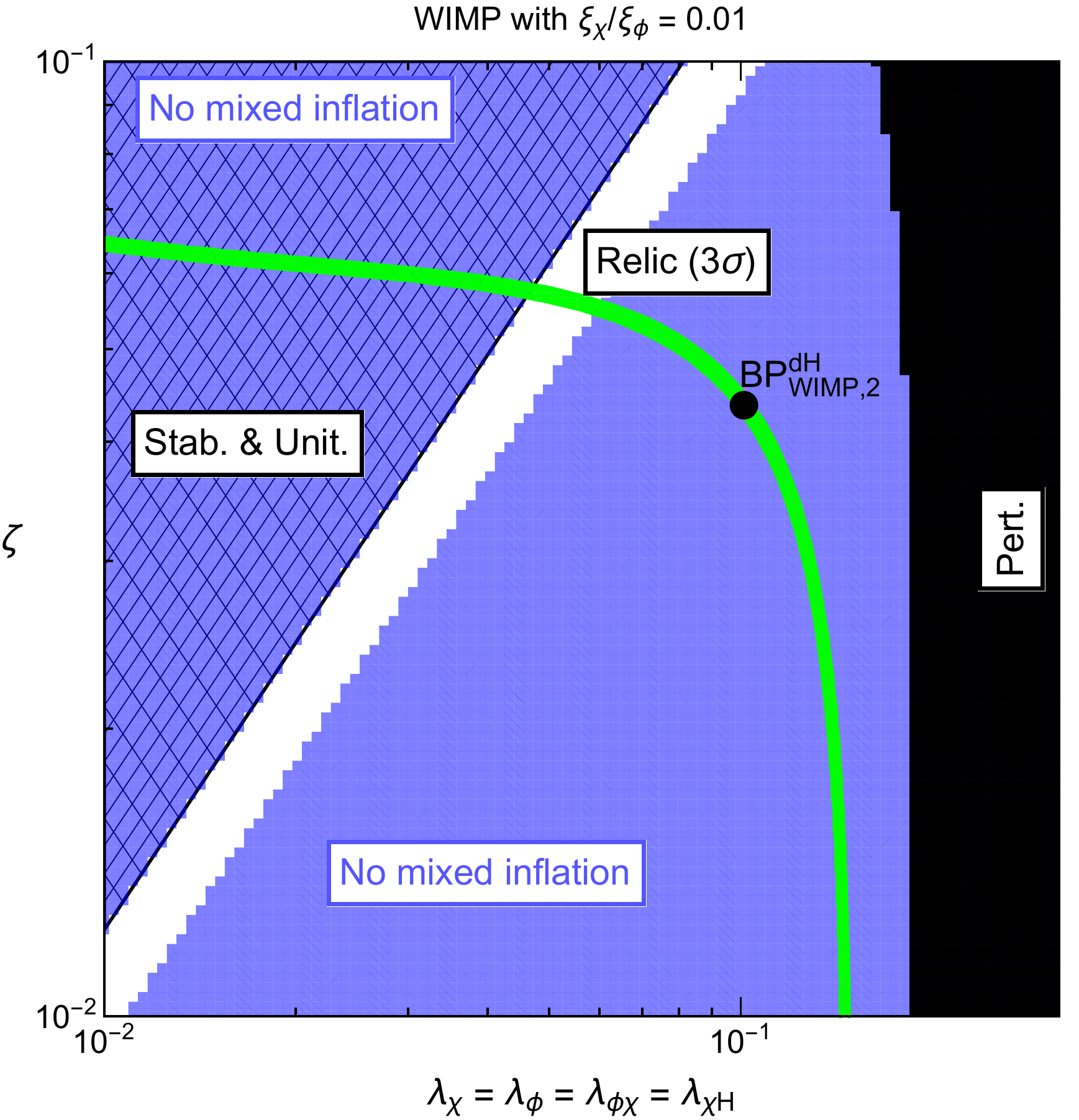}
\caption{Parameter space for WIMP dark matter scenario with $\xi_\chi/\xi_\phi=1$ (left) and $0.01$ (right) for the quartic couplings with $\lambda_\chi=\lambda_\phi=\lambda_{\phi\chi}=\lambda_{\chi H}$ versus the dark matter cubic coupling $\zeta$. The rest input parameters and color codes are as in Fig.~\ref{fig:WIMP1}, except that $\zeta$ is varying.}
\label{fig:WIMP2}
\end{figure}

For $m_{h'}\ll m_\chi$ and $m_{Z'}> m_\chi$, the dark matter annihilation cross section is dominated by $\chi\chi^*\rightarrow h'h'$ and $\chi\chi\rightarrow \chi^* h'$, which is approximated to
\begin{eqnarray}
\langle \sigma v\rangle_{2\rightarrow 2} \approx  \frac{\zeta^2 m^2_{Z'}}{768\pi  g^2_X m^4_\chi} \bigg(\frac{9g_X m_\chi}{m_{Z'}} -\frac{\lambda_{\phi\chi} m_{Z'}}{g_X m_\chi}\bigg)^2 + \frac{\lambda^2_{\phi\chi}}{128\pi m^2_\chi} \bigg(1-\frac{\lambda_{\phi\chi} m^2_{Z'}}{9 g^2_X m^2_\chi}+ \frac{\lambda_\phi m^2_{Z'}}{6g^2_Xm^2_\chi} \bigg)^2
\,.
\end{eqnarray}
Therefore, the resulting effective annihilation rate depends on the dark matter self-couplings, the cubic coupling $\zeta$ and the mixing quartic coupling $\lambda_{\phi\chi}$, as well as the resonance and dark matter masses. 
In a wide parameter space with reasonable values of the dimensionless parameters of weak strength in the model, we can make the dark matter physics consistent with the inflation regime.

\begin{table}[tbp]
\small
\begin{center}
\begin{tabular}{c|c|c|c|c|c|c}
\hline
Benchmark 
& \multirow{2}{*}{$\lambda_\chi=\lambda_\phi$}
& \multirow{2}{*}{$\lambda_{\phi\chi}=\lambda_{\chi H}$}
& \multirow{2}{*}{$\zeta$}
& \multirow{2}{*}{$\xi_\chi / \xi_\phi$}
& \multirow{2}{*}{$n_s$}
& \multirow{2}{*}{$r$}
\\
Points
& \multirow{2}{*}{} & \multirow{2}{*}{}
& \multirow{2}{*}{} & \multirow{2}{*}{}
& \multirow{2}{*}{} & \multirow{2}{*}{}
\\
\hline
\hline
\multirow{2}{*}{$\mathrm{BP^{mixed}_{WIMP,1}}$}
& \multirow{2}{*}{$0.1$}
& \multirow{2}{*}{$0.08$}
& \multirow{2}{*}{$0.05$}
& \multirow{2}{*}{$1$}
& \multirow{2}{*}{0.9702}
& \multirow{2}{*}{0.00321}
\\
&&&&&&
\\
\hline
\multirow{2}{*}{$\mathrm{BP^{dH}_{WIMP,1}}$}
& \multirow{2}{*}{$0.1$}
& \multirow{2}{*}{$0.08$}
& \multirow{2}{*}{$0.05$}
& \multirow{2}{*}{$0.01$}
& \multirow{2}{*}{0.9691}
& \multirow{2}{*}{0.00296}
\\
&&&&&&
\\
\hline
\multirow{2}{*}{$\mathrm{BP^{mixed}_{WIMP,2}}$}
& \multirow{2}{*}{$0.1$}
& \multirow{2}{*}{$0.1$}
& \multirow{2}{*}{$0.04$}
& \multirow{2}{*}{$1$}
& \multirow{2}{*}{0.9695}
& \multirow{2}{*}{0.00306}
\\
&&&&&&
\\
\hline
\multirow{2}{*}{$\mathrm{BP^{dH}_{WIMP,2}}$}
& \multirow{2}{*}{$0.1$}
& \multirow{2}{*}{$0.1$}
& \multirow{2}{*}{$0.04$}
& \multirow{2}{*}{$0.01$}
& \multirow{2}{*}{0.9691}
& \multirow{2}{*}{0.00297}
\\
&&&&&&
\\
\hline
\end{tabular}
\caption{Input parameters for the benchmark points, denoted as black points in Figs.~\ref{fig:WIMP1} and \ref{fig:WIMP2}. In this case, $m_{Z'}$ is also fixed to $m_{Z'}=67\,{\rm GeV}$, in the upper two and lower two cases, respectively. The superscript represents the inflation type; for example, $\mathrm{BP^{mixed}_{WIMP,1}}$ ($\mathrm{BP^{dH}_{WIMP,1}}$) is the benchmark point where mixed (dark Higgs) inflation is allowed. }
\label{tab:BPWIMPinput}
\end{center}
\end{table}

In Figs.~\ref{fig:WIMP1} and \ref{fig:WIMP2}, we show various constraints coming from dark matter physics and inflation in the parameter space for the quartic self-couplings with $\lambda_\chi=\lambda_\phi$ versus the quartic mixing couplings with $\lambda_{\phi\chi}=\lambda_{\chi H}$ in the former and the quartic couplings with $\lambda_\chi = \lambda_\phi = \lambda_{\phi\chi} = \lambda_{\chi H}$ versus the dark matter cubic self-coupling $\zeta$ in the latter. Here, we chose $m_\chi = 500$ GeV and $m_{h^\prime} = 100$ GeV, and $m_{Z'}$ varies in the range of $60\,{\rm GeV}\lesssim m_{Z'}\lesssim 212\,{\rm GeV}$, depending on $\lambda_\phi$ in the plots.
We selected four benchmark points, $\mathrm{BP^{mixed}_{WIMP,1}}$, $\mathrm{BP^{dH}_{WIMP,1}}$, $\mathrm{BP^{mixed}_{WIMP,2}}$, and $\mathrm{BP^{dH}_{WIMP,2}}$, denoted as black points in Figs.~\ref{fig:WIMP1} and \ref{fig:WIMP2}. The input parameters for those benchmark points are summarized in Table \ref{tab:BPWIMPinput}.
Imposing the normalization of the scalar power spectrum from the latest Planck, we computed inflationary observables, including the spectral index $n_s$ and the tensor-to-scalar ratio $r$. The annihilation cross sections are dominated by WIMP processes in all the cases.

\begin{figure}[tbp]
\centering
\includegraphics[width=0.46\linewidth]{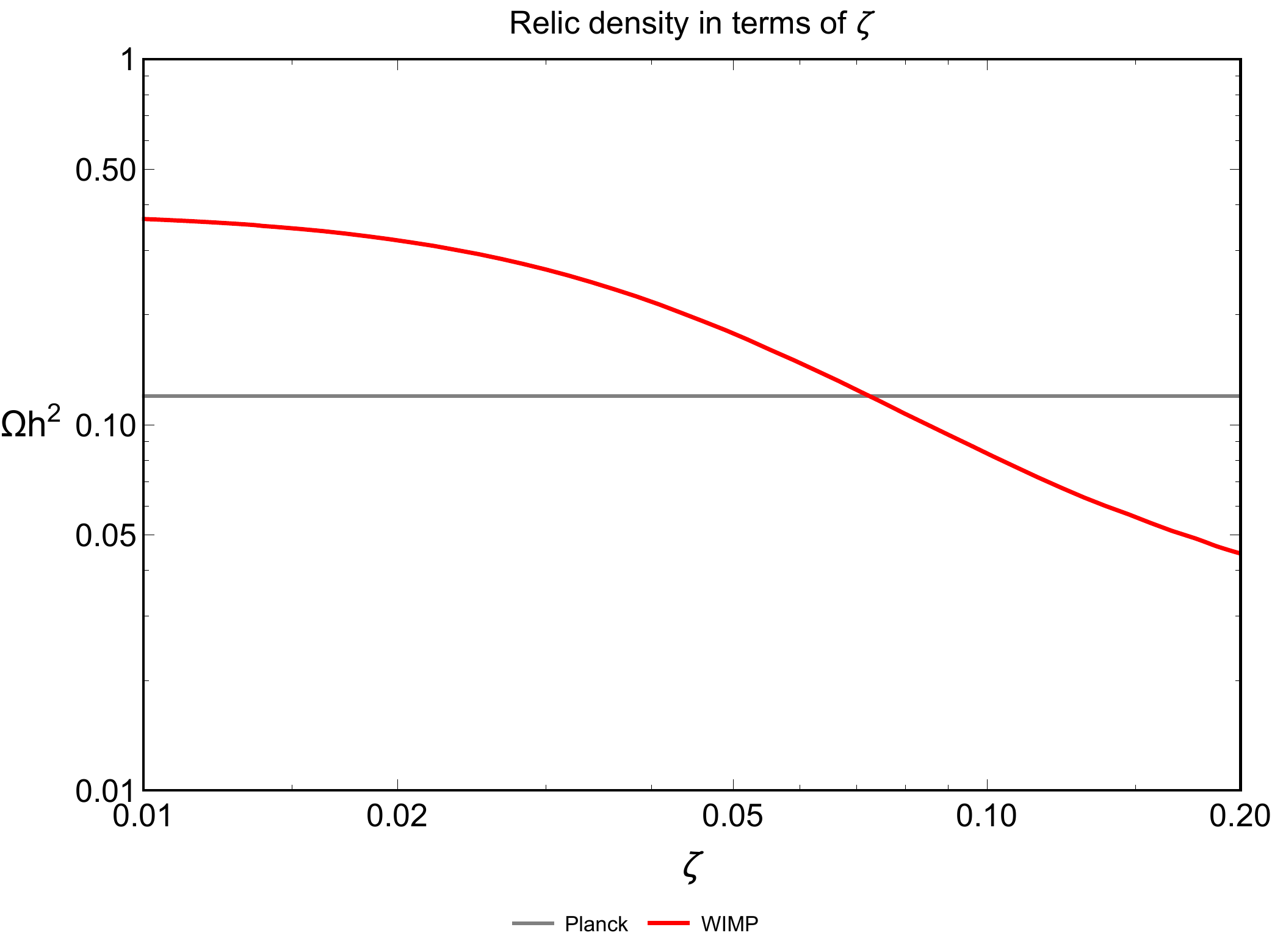}
\includegraphics[width=0.50\linewidth]{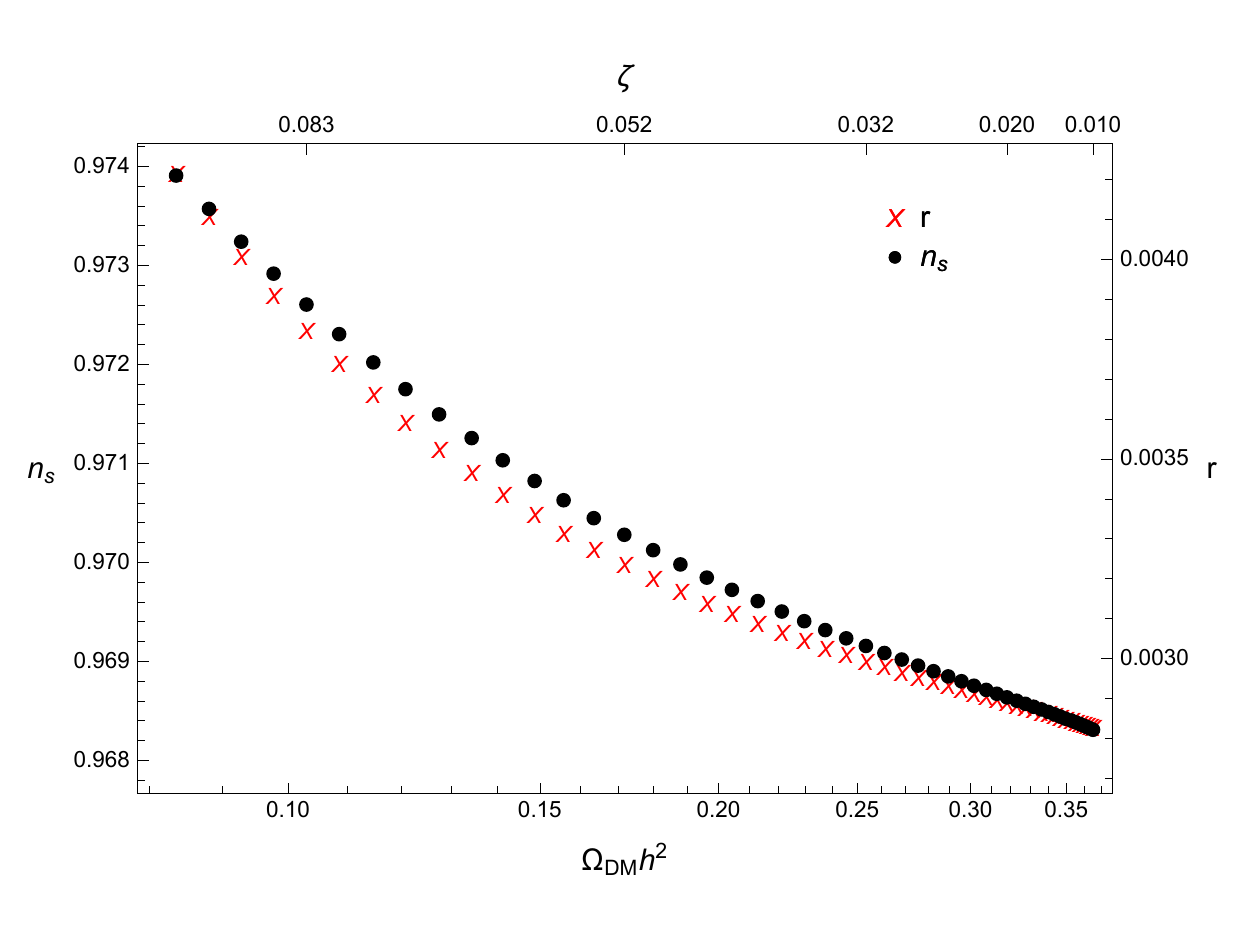}
\caption{
(Left) Relic density for WIMP dark matter as a function of $\zeta$. (Right) Relic density in the plane of spectral index and tensor-to-scalar ratio. The parameters are chosen as follows:
$\sin\theta = 0.01$, $\varepsilon = 10^{-4}$, $g_X = 0.1$, $m_{h^\prime}=100 \, {\rm GeV}$, $m_\chi=500 \, {\rm GeV}$, $\lambda_\phi = \lambda_\chi=\lambda_{\phi\chi} = \lambda_{\chi H} = 0.1$, thus  $m_{Z'}=67\,{\rm GeV}$.
The gray horizontal solid line on left represents the Planck data for the relic density.
}
\label{fig:WIMPRelDenZeta}
\end{figure}

Our model with discrete $Z_3$ symmetry is distinguished from models with $Z_2$ through the cubic self-coupling for scalar dark matter, namely, the $\zeta$ parameter. On the left of Fig.~\ref{fig:WIMPRelDenZeta}, we show the $\zeta$ dependence of the relic density by fixing the other parameters.
In comparison, on the right of Fig.~\ref{fig:WIMPRelDenZeta}, we present the relic density in the plot for the spectral index versus tensor-to-scalar ratio in order to see the correlation between dark matter and inflation constraints. 
We find that the spectral index and the tensor-to-scalar ratio vary, $n_s \approx 0.968-0.974$ and $r \approx 0.0028-0.0042$, depending on the $\zeta$ parameter. Thus, the spectral index can deviate sizably from the results of the classical non-minimal coupling inflation, but most of the parameter is still consistent with Planck within about $2\sigma$.

\subsection{SIMP scenarios}
\label{subsec:SIMPres}

For SIMP dark matter, the dark matter abundance is determined dominantly by the $3\rightarrow 2$ processes, which depend mostly on the dark matter self-interactions. In this case, the typical values of the dark matter mass are below GeV scale to be consistent with the correct relic density and the perturbativity bound imposed at the dark matter mass scale. But, the sizable self-couplings for dark matter would lead to the premature breakdown of perturbativity below the inflation scale. Thus, we need to rely on the enhancement of the $3\rightarrow 2$ processes at resonances, for instance, $m_{h'}\simeq 3m_\chi$ \cite{Choi:2016hid,Choi:2017mkk,Choi:2018iit}.

\begin{figure}[tbp]
\centering
\includegraphics[width=0.45\linewidth]{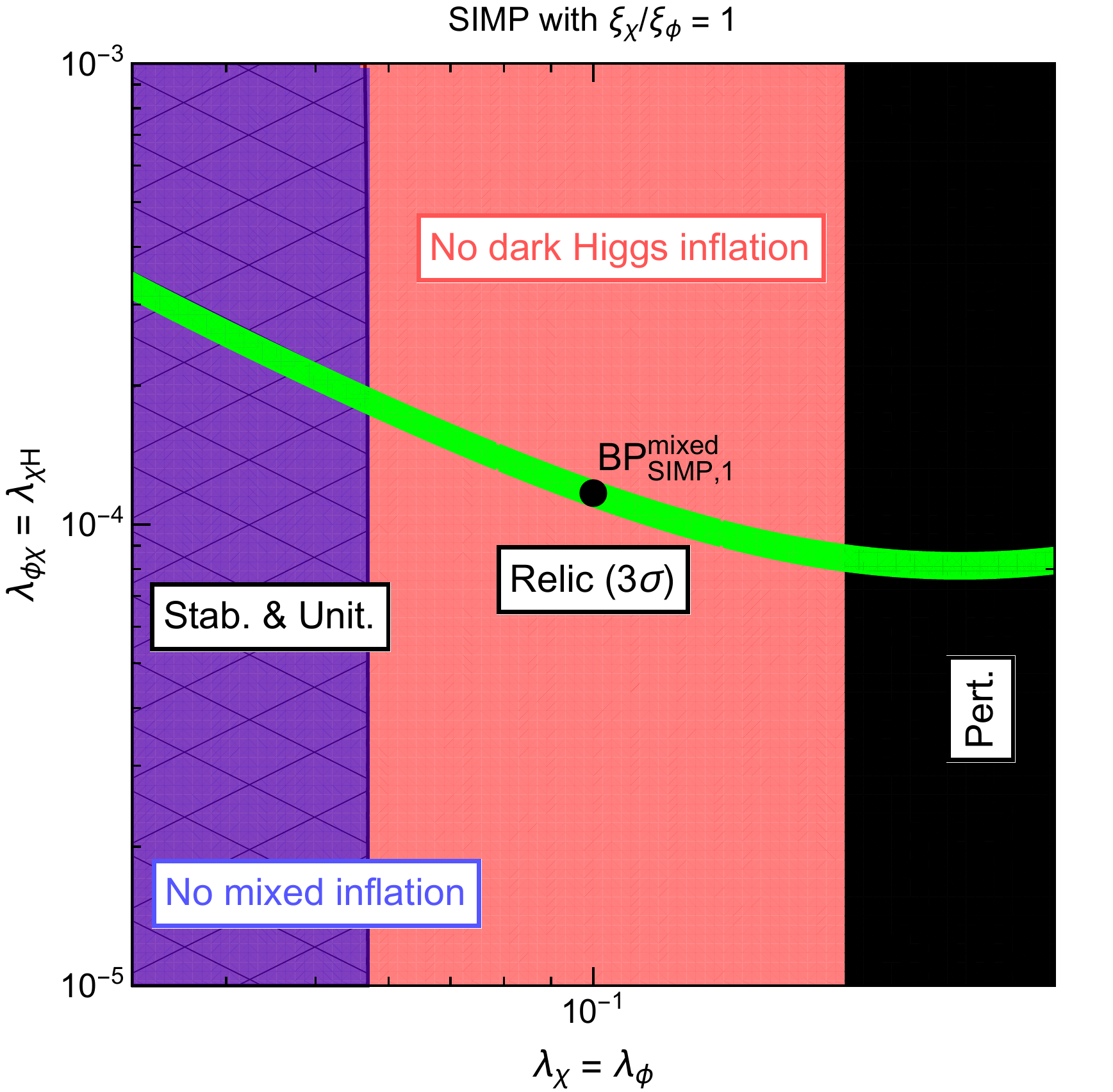}\,\;
\includegraphics[width=0.45\linewidth]{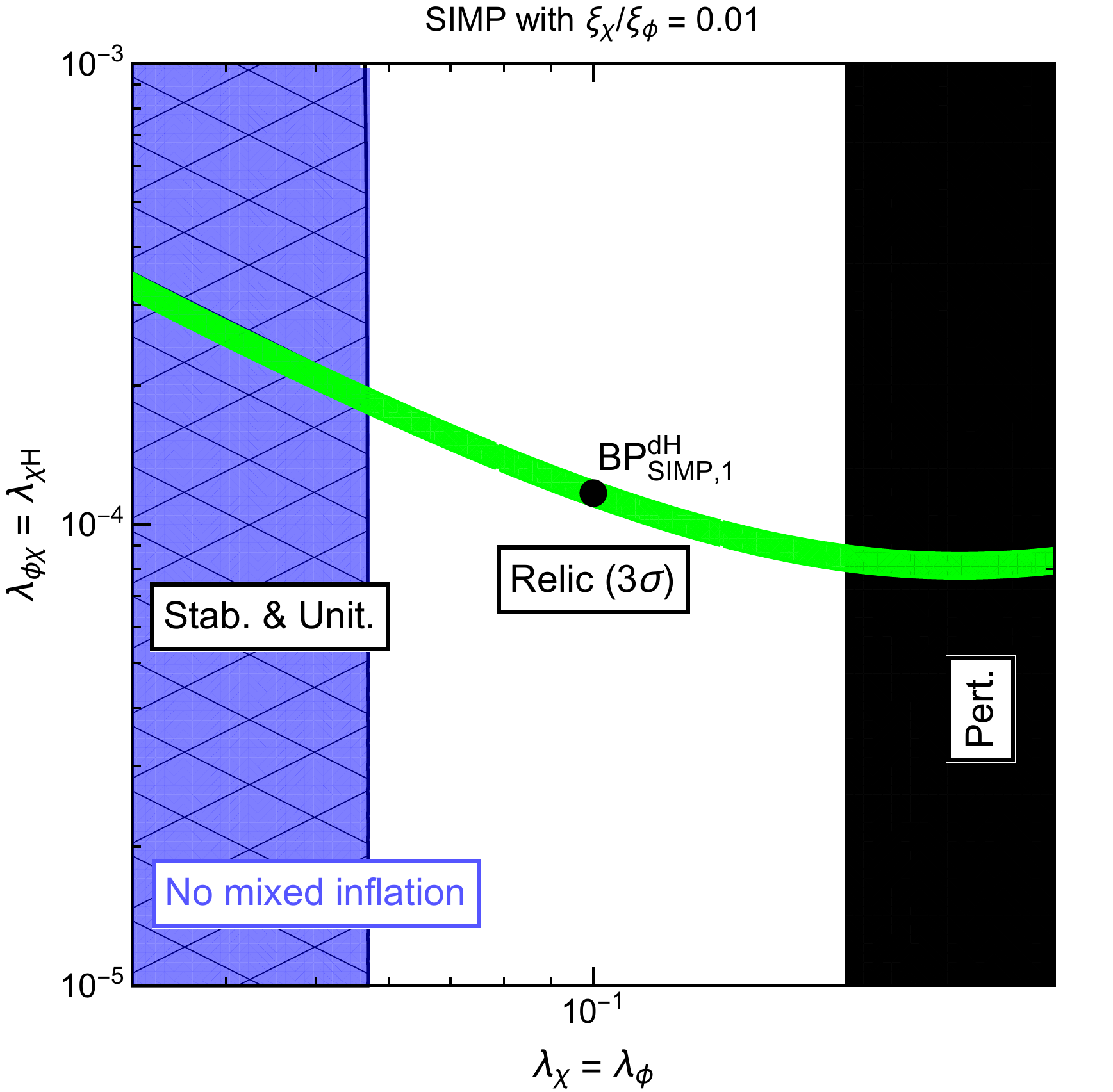}
\caption{Parameter space for SIMP dark matter scenario with $\xi_\chi/\xi_\phi=1$ (left) and $0.01$ (right), for $\lambda_\phi=\lambda_\chi$ versus $\lambda_{\phi\chi}=\lambda_{\chi H}$. The parameters shown here are set at low-energy scale. The rest input parameters are chosen as $g_X = 0.1$, $\sin\theta = 10^{-6}$, $\varepsilon=10^{-4}$, $m_{h^\prime} = 0.35$ GeV, $m_\chi = 0.1$ GeV and $\zeta = 0.05$. The hatched and black regions are disfavored by stability, unitarity and perturbativity. The dark Higgs (mixed) inflation is not allowed in the red-colored (blue-colored) region.  The green solid lines represent the correct dark matter relic abundance within 3$\sigma$ range \cite{planck}. The black points are our benchmark points shown in Table \ref{tab:BPSIMPinput}. }
\label{fig:SIMP1}
\end{figure}
\begin{figure}[tbp]
\centering
\includegraphics[width=0.45\linewidth]{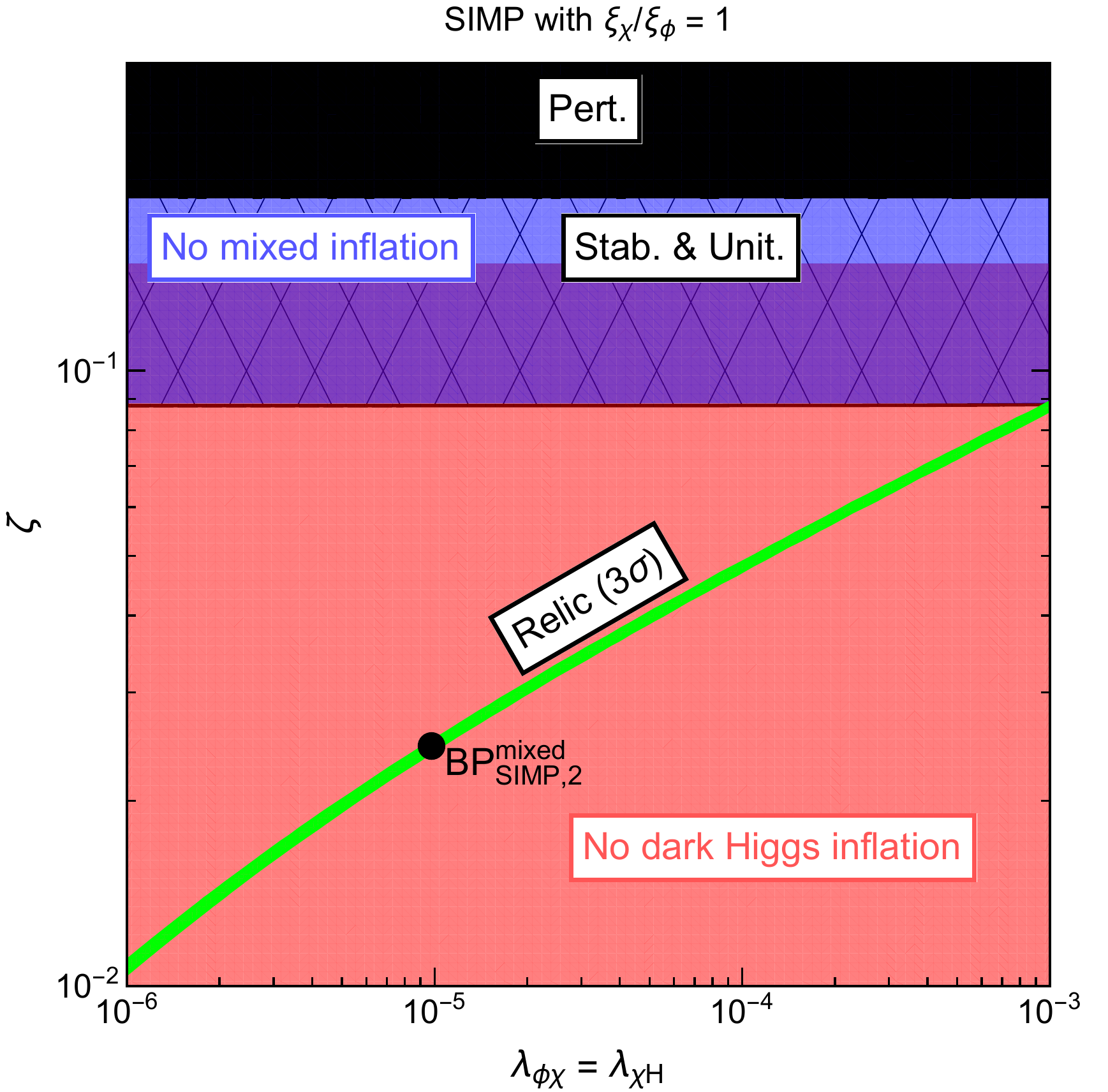}\,\;
\includegraphics[width=0.45\linewidth]{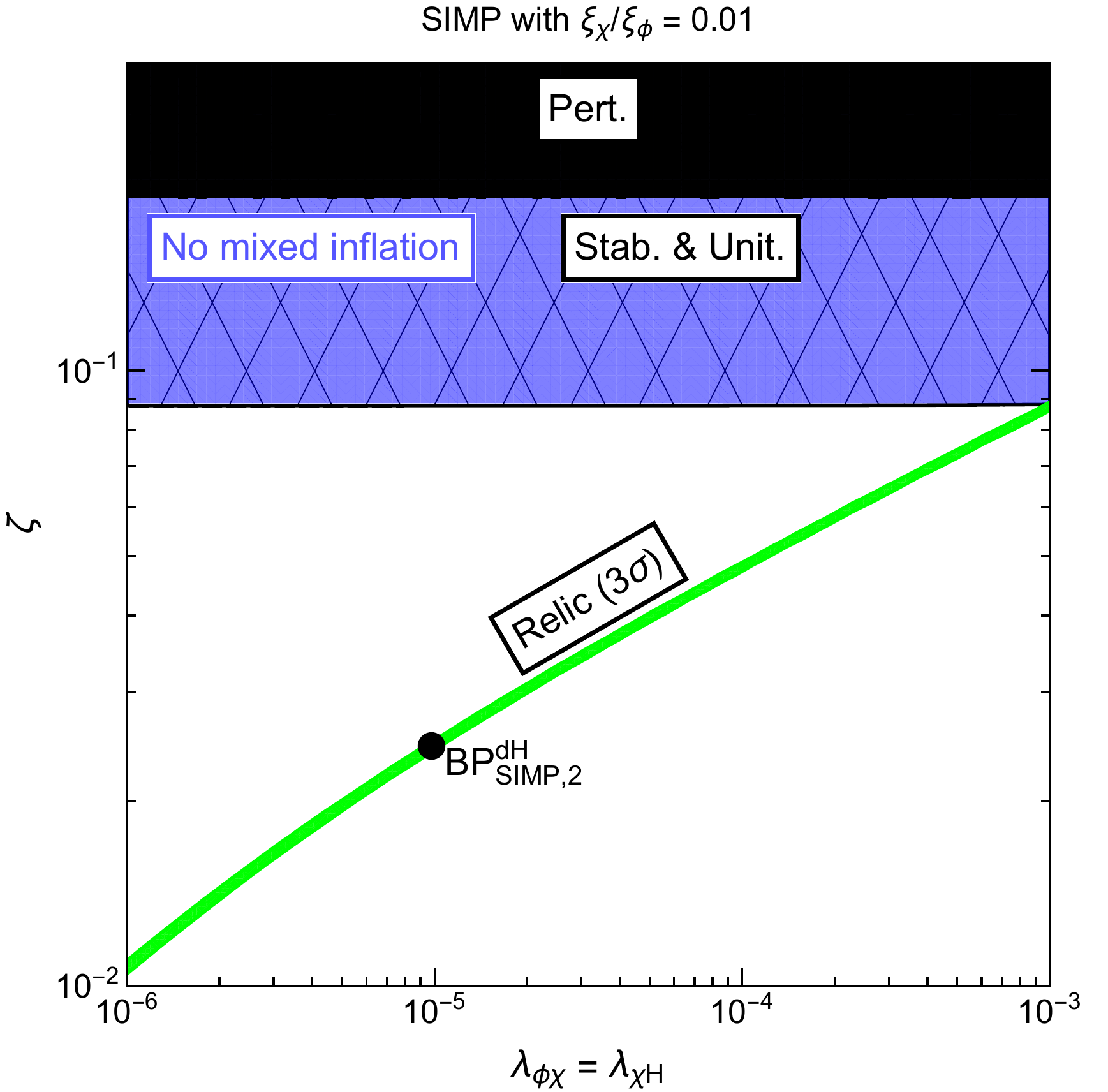}
\caption{Parameter space for SIMP dark matter scenario with $\xi_\chi/\xi_\phi=1$ (left) and $0.01$ (right), for $\lambda_{\phi\chi}=\lambda_{\chi H}$ versus $\zeta$. The input parameters and color codes are as in Fig.~\ref{fig:SIMP1}, except that $\lambda_{\chi} = \lambda_{\phi} = 0.1$ and $\zeta$ is varying. }
\label{fig:SIMP2}
\end{figure}

We consider the case where the $3\rightarrow 2$ annihilation processes are dominant for determining the dark matter relic density. In this case, the Boltzmann equation \eqref{Boltzmann} becomes
\begin{equation}\begin{aligned}
\dot{n}_\text{DM} + 3Hn_\text{DM}\ =\ & -\frac{1}{4}(\langle\sigma v^2\rangle_{\chi\chi\chi\rightarrow \chi\chi^\ast}+\langle\sigma v^2\rangle_{\chi\chi\chi^\ast\rightarrow \chi^\ast\chi^\ast})(n_\text{DM}^3-n_\text{DM}^\text{eq}n_\text{DM}^2)\\
& -\frac{1}{2}\langle\sigma v^2\rangle_{\chi\chi\chi^\ast\rightarrow \chi h'}\bigg( n_\text{DM}^3-(n_\text{DM}^\text{eq})^2 n_\text{DM}\bigg) \\
& -\frac{1}{2}\langle\sigma v^2\rangle_{\chi\chi\chi^\ast\rightarrow \chi Z'}\bigg( n_\text{DM}^3-(n_\text{DM}^\text{eq})^2 n_\text{DM} \bigg) \\
\approx &-\langle\sigma v^2\rangle_{ 3\rightarrow 2} \, n^3_{\rm DM} 
\,,
\label{Boltzmann2}
\end{aligned}\end{equation}
with
\begin{eqnarray}
\langle\sigma v^2\rangle_{ 3\rightarrow 2}&\equiv& \frac{1}{4}(\langle\sigma v^2\rangle_{\chi\chi\chi\rightarrow \chi\chi^\ast}+\langle\sigma v^2\rangle_{\chi\chi\chi^\ast\rightarrow \chi^\ast\chi^\ast})+\frac{1}{2}\langle\sigma v^2\rangle_{\chi\chi\chi^\ast\rightarrow \chi h'}+\frac{1}{2}\langle\sigma v^2\rangle_{\chi\chi\chi^\ast\rightarrow \chi Z'} \nonumber \\
&\equiv& \frac{\alpha^3_{\rm eff}}{m^5_\chi}.
\end{eqnarray}
Here, we note that $m_\chi<m_{h'}<2m_\chi$ or $m_\chi<m_{Z'}<2m_\chi$ in order for the assisted $3\rightarrow 2$ processes with $h'$ or $Z'$ to be kinematically open and for the hidden sector $2\rightarrow 2$ annihilations to be forbidden. 
Then, the relic density for SIMP dark matter \cite{Choi:2016tkj,Choi:2015bya} is given by
\begin{eqnarray}
\Omega_{\rm DM} h^2=1.41\times 10^{-8}\,{\rm GeV}^{-2} \bigg(\frac{10.75}{g_*} \bigg)^{3/4}  \Big(\frac{x_f}{20} \Big)^2 \bigg(\frac{M^{1/3}_{{\rm P}} m_\chi }{\alpha_{\rm eff}} \bigg)^{3/2}. 
\end{eqnarray}
In the presence of the velocity-dependence or resonance poles in the $3\rightarrow 2$ processes, special care should be taken for thermal averaging \cite{Choi:2017mkk,Choi:2018iit}.

For $m_{h'}\simeq 3m_\chi$, the averaged $3\rightarrow 2$ annihilation cross section for dark matter is dominated by the resonance term in $\chi\chi\chi\rightarrow \chi\chi^*$ \cite{Choi:2017mkk,Choi:2018iit}, approximated to
\begin{eqnarray}
\langle \sigma v^2\rangle \approx \frac{9\pi \zeta^2}{256m^5_\chi}\, \bigg(1+\frac{9\lambda^2_{\phi\chi}}{2\lambda_\phi} \bigg)^2\, \epsilon^2_R\,x^3\, e^{-\frac{3}{2} x\,\epsilon_R}\,\theta(\epsilon_R)
\,,
\end{eqnarray}
where  $x=m_\chi/T$ and
\begin{eqnarray}
\epsilon_R\equiv \frac{m^2_{h'}-9m^2_\chi}{9m^2_\chi}
\,.
\end{eqnarray}
Here, we assumed the narrow width approximation with $\Gamma_{h'}/m_{h'}\ll 1,\epsilon_R$, where the width of the dark Higgs is given by $\Gamma_{h'}=\frac{\lambda^2_{\phi\chi} v^2_\phi}{16\pi m_{h'}}\sqrt{1-\frac{4m^2_\chi}{m^2_{h'}}}$.
Then, we used the thermal average of $(\sigma v^2)=\frac{b_R\, \gamma_R}{(\epsilon_R-u^2)^2+\gamma^2_R}$ with $\gamma_R=m_{h'}\Gamma_{h'}/(9m^2_\chi)$, $u=\frac{1}{3} (v^2_1+v^2_2 +v^2_3)$ and $b_R$ being a velocity-independent coefficient: $\langle \sigma v^2\rangle=\frac{27}{16}\pi \epsilon^2_R x^3 \, e^{-\frac{3}{2} x\,\epsilon_R}\,\theta(\epsilon_R)$ from Ref.~\cite{Choi:2017mkk}.
Therefore, the resulting effective annihilation rate depends significantly on the dark matter cubic coupling $\zeta$ as well as the resonance and dark matter masses.

\begin{table}[h!]
\small
\begin{center}
\begin{tabular}{c|c|c|c|c|c|c}
\hline
Benchmark 
&  \multirow{2}{*}{$\lambda_\chi=\lambda_\phi$}
& \multirow{2}{*} {$\lambda_{\phi\chi}=\lambda_{\chi H}$}
& \multirow{2}{*}{$\zeta$}
& \multirow{2}{*}{$\xi_\chi / \xi_\phi$}
& \multirow{2}{*}{$n_s$}
& \multirow{2}{*}{$r$}
\\
Points
& 
& \multirow{2}{*}{}
& \multirow{2}{*}{} & \multirow{2}{*}{}
& \multirow{2}{*}{} & \multirow{2}{*}{}
\\
\hline
\hline
\multirow{2}{*}{$\mathrm{BP^{mixed}_{SIMP,1}}$}
&\multirow{2}{*}{$0.1$}
& \multirow{2}{*}{$1.2\times 10^{-4}$}
& \multirow{2}{*}{$0.05$}
& \multirow{2}{*}{$1$}
&\multirow{2}{*}{0.9709}
&\multirow{2}{*}{0.00337}
\\
&&&&&&
\\
\hline
\multirow{2}{*}{$\mathrm{BP^{dH}_{SIMP,1}}$}
& \multirow{2}{*}{0.1}
& \multirow{2}{*}{$1.2\times 10^{-4}$}
& \multirow{2}{*}{$0.05$}
& \multirow{2}{*}{$0.01$}
&\multirow{2}{*}{0.9688}
&\multirow{2}{*}{0.00292}
\\
&&&&&&
\\
\hline
\multirow{2}{*}{$\mathrm{BP^{mixed}_{SIMP,2}}$}
& \multirow{2}{*}{0.1}
& \multirow{2}{*}{$10^{-5}$}
& \multirow{2}{*}{$0.024$}
& \multirow{2}{*}{$1$}
&\multirow{2}{*}{0.9692}
&\multirow{2}{*}{0.00299}
\\
&&&&&&
\\
\hline
\multirow{2}{*}{$\mathrm{BP^{dH}_{SIMP,2}}$}
& \multirow{2}{*}{0.1}
& \multirow{2}{*}{$10^{-5}$}
& \multirow{2}{*}{$0.024$}
& \multirow{2}{*}{$0.01$}
&\multirow{2}{*}{0.9688}
&\multirow{2}{*}{0.00292}
\\
&&&&&&
\\
\hline
\end{tabular}
\caption{Input parameters for our benchmark points, denoted as black points in Figs.~\ref{fig:SIMP1} and \ref{fig:SIMP2}. In this case, $m_{Z'}$ is also fixed to $m_{Z'}=0.235\,{\rm GeV}$. The superscript represents the inflation type; for example, $\mathrm{BP^{mixed}_{SIMP,1}}$ ($\mathrm{BP^{dH}_{SIMP,1}}$) is the benchmark point where mixed (dark Higgs) inflation is allowed.}    
\label{tab:BPSIMPinput}
\end{center}
\end{table}
\begin{figure}[tbp]
\centering
\includegraphics[width=0.50\linewidth]{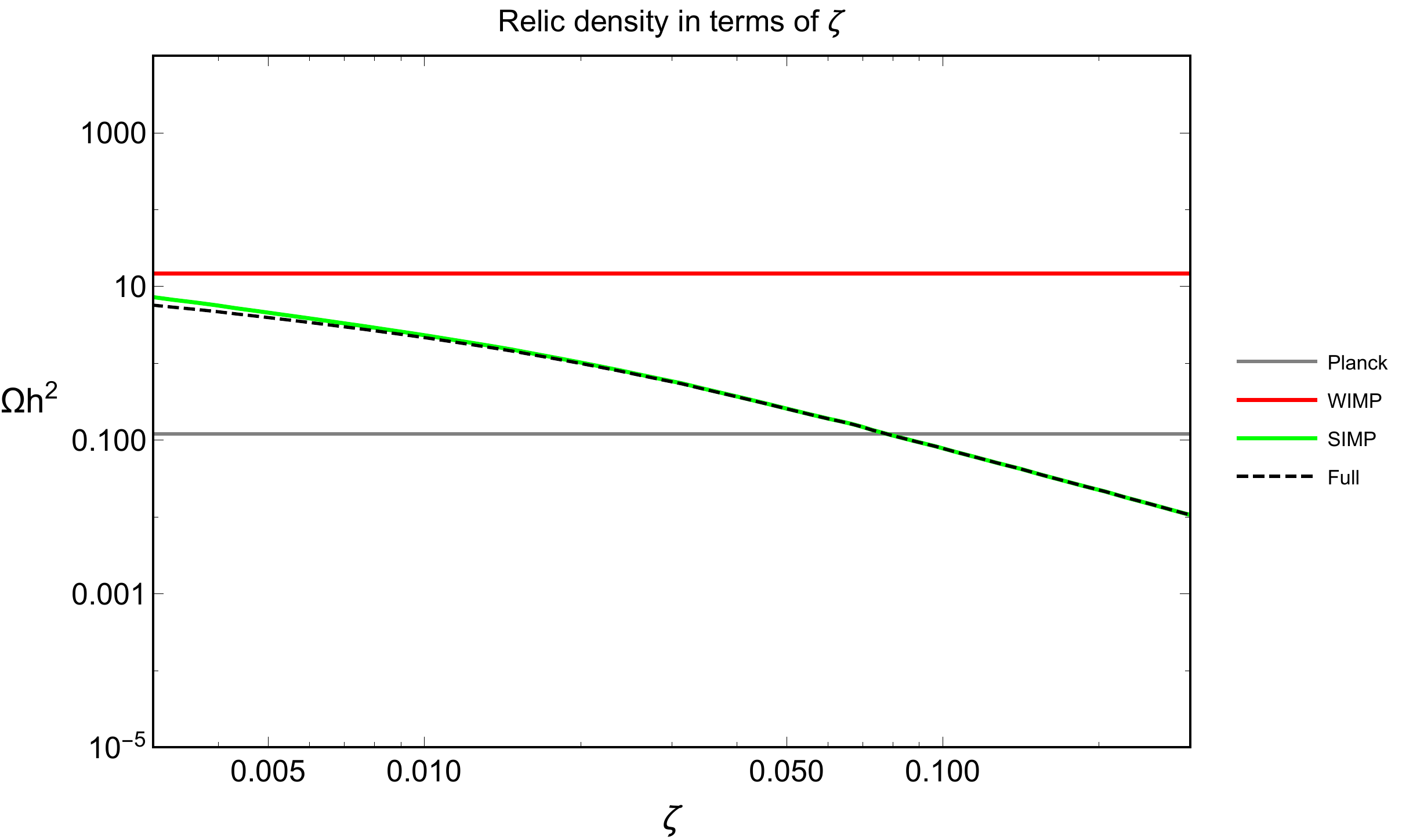}
\includegraphics[width=0.45\linewidth]{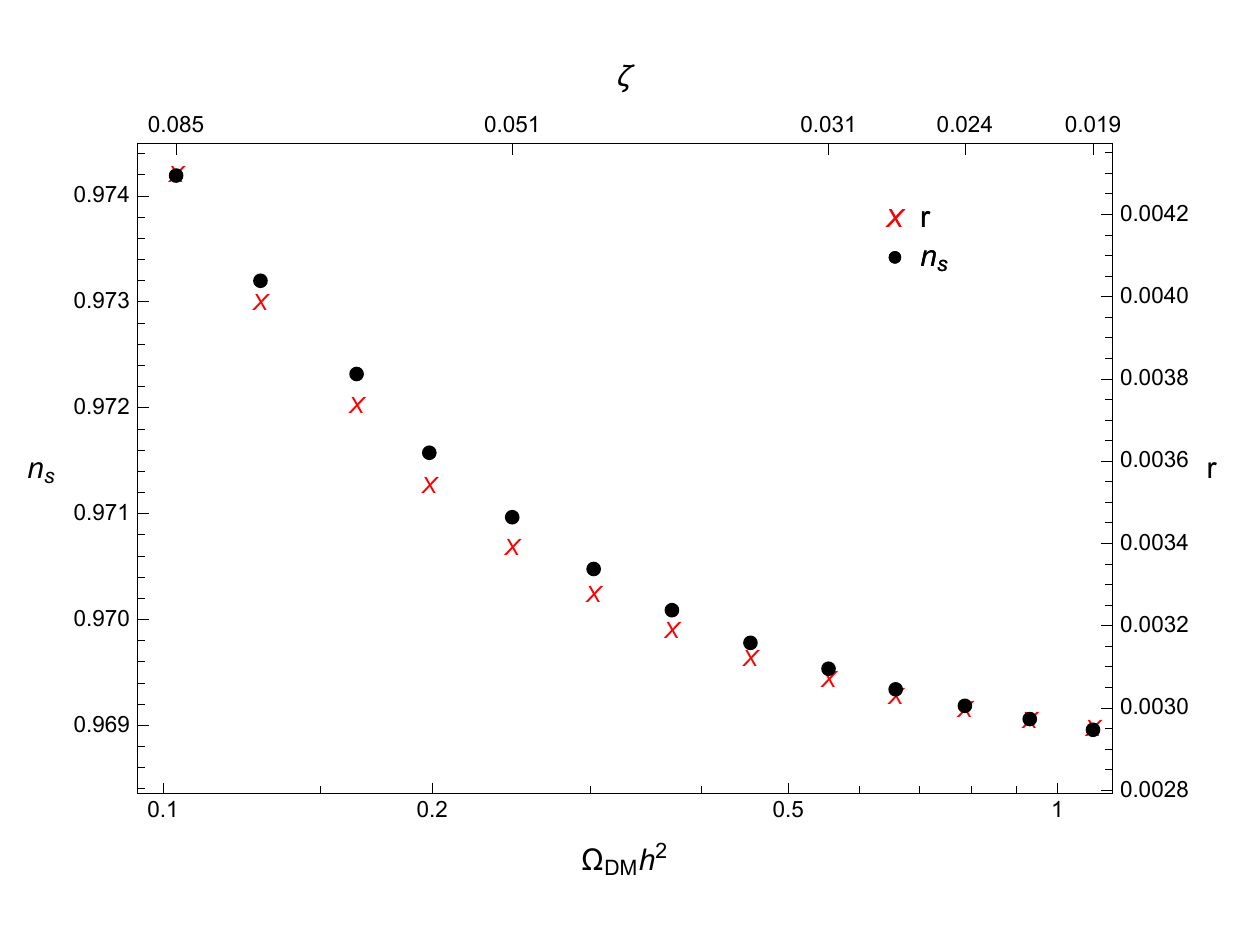}
\caption{
(Left) Relic densities for each dark matter components as a function of $\zeta$. (Right) Correlation between relic density, spectral index and tensor-to-scalar ratio. The parameters are chosen as follows:
$\sin\theta = 10^{-6}$, $\varepsilon = 10^{-4}$, $g_X = 0.1$, $m_{h^\prime}=0.35 \, {\rm GeV}$, $m_\chi=0.1 \, {\rm GeV}$, $\lambda_\phi = \lambda_\chi = 0.1$, $\lambda_{\phi\chi} =\lambda_{\chi H}= 10^{-4}$, thus $m_{Z'}=0.235\,{\rm GeV}$.
The grey horizontal solid line on left represents the Planck data for the relic density.
}
\label{fig:SIMPRelDenZeta}
\end{figure}

In Figs.~\ref{fig:SIMP1} and \ref{fig:SIMP2}, we show various constraints coming from dark matter physics and inflation in the quartic self-coupling with $\lambda_\chi=\lambda_\phi$ versus the mixing quartic couplings with $\lambda_{\phi\chi}=\lambda_{\chi H}$ in the former and the mixing quartic couplings with $\lambda_{\phi\chi}=\lambda_{\chi H}$ versus the dark matter cubic self-coupling $\zeta$ in the latter. Here, we chose the mass parameters near the resonance at $m_{h'}=3m_\chi$, for instance, $m_\chi = 0.1$ GeV and $m_{h^\prime} = 0.35$ GeV. Then, we vary $v_\phi$ to get the range, $0.209\,{\rm GeV}\lesssim m_{Z'}\lesssim 0.418\,{\rm GeV}$, depending on $\lambda_\phi$ for the fixed $g_X$ and $m_{h'}$.
We select four benchmark points in the above plots, $\mathrm{BP^{mixed}_{SIMP,1}}$, $\mathrm{BP^{dH}_{SIMP,1}}$, $\mathrm{BP^{mixed}_{SIMP,2}}$, and $\mathrm{BP^{dH}_{SIMP,2}}$, with the input parameters summarized in Table \ref{tab:BPSIMPinput}.
For the selected benchmark points, we computed inflationary observables, including the spectral index $n_s$ and the tensor-to-scalar ratio $r$. Note that the scalar power spectrum is normalized to satisfy the latest Planck result.

On the left of Fig.~\ref{fig:SIMPRelDenZeta}, we show the $\zeta$ dependence of the relic density by fixing the other parameters.
The relic densities that would have been determined by the WIMP or SIMP channels only are shown in red and green lines on the left of Fig.~\ref{fig:SIMPRelDenZeta}. The total relic density follows closely the one that would have been determined by SIMP channels only. In comparison, on the right of Fig.~\ref{fig:SIMPRelDenZeta}, we present the relic density in the plot for the spectral index and the tensor-to-scalar ratio in order to see the correlation between them.
We find that the spectral index and the tensor-to-scalar ratio vary, $n_s \approx 0.969-0.974$ and $r \approx 0.0029-0.0043$, depending on the $\zeta$ parameter. Thus, the spectral index can deviate sizably from the results of the classical non-minimal coupling inflation, but most of the parameter is still consistent with Planck within about $2\sigma$.

\subsection{Forbidden scenarios}
\label{subsec:FBDMres}

For the forbidden dark matter, the hidden sector annihilations are forbidden at zero temperature, but they are open at a finite temperature during the freeze-out \cite{DAgnolo:2015ujb,Choi:2016tkj}. In this case, due to the Boltzmann suppression with $e^{-\Delta\, x}$ for a large dark matter velocity where $\Delta=(m'-m_{\rm DM})/m_{\rm DM}$ with $m'$ being the mass of the hidden sector particle, the hidden sector annihilation cross section should be sizable, depending on the mass difference between dark matter and hidden sector particle masses, $m_{h'}$ or $m_{Z'}$, in our case \cite{Choi:2016tkj}. Importantly, for $T\ll \Delta$, the forbidden annihilation cross section is exponentially suppressed, so the CMB bound on the annihilation cross section for light dark matter does not apply.

For $m_\chi<m_{h'}, m_{Z'}$, we consider the $2\rightarrow 2$ forbidden annihilation processes to be dominant for determining the dark matter relic density. Then, we can approximate the Boltzmann equation \eqref{Boltzmann} as
\begin{equation}\begin{aligned}
\dot{n}_\text{DM} + 3Hn_\text{DM}\ =\ 
& \langle \sigma v \rangle_{\chi^\ast h' \rightarrow \chi \chi}\bigg( n_\text{DM} n_{h'} - \frac{n_{h'}^\text{eq}}{n_\text{DM}^\text{eq}}n_\text{DM}^2 \bigg) +2\langle \sigma v \rangle_{h' h'  \rightarrow \chi \chi^\ast}\bigg(n_{h'}^2-\frac{(n_{h'}^\text{eq})^2 }{(n_\text{DM}^\text{eq})^2}\,n_\text{DM}^2 \bigg) \\
& +\langle \sigma v \rangle_{\chi^\ast Z' \rightarrow \chi \chi}\bigg( n_\text{DM} n_{Z'} - \frac{n_{Z'}^\text{eq}}{n_\text{DM}^\text{eq}}n_\text{DM}^2 \bigg)  +2\langle \sigma v \rangle_{Z' Z'  \rightarrow \chi \chi^\ast}\bigg( n_{Z'}^2-\frac{(n_{Z'}^\text{eq})^2}{(n_\text{DM}^\text{eq})^2} \,n_\text{DM}^2\bigg) \\
\approx &- \langle \sigma v\rangle_{\rm FB}\,  n^2_{\rm DM}
\,,
\label{Boltzmann3}
\end{aligned}\end{equation}
with
\begin{eqnarray}
\langle \sigma v\rangle_{\rm FB}&\equiv&  \Big(\frac{n_{h'}^\text{eq}}{n_\text{DM}^\text{eq}}\Big)  \langle \sigma v \rangle_{\chi^\ast h' \rightarrow \chi \chi}+ \frac{2(n_{h'}^\text{eq})^2 }{(n_\text{DM}^\text{eq})^2}\, \langle \sigma v \rangle_{h' h'  \rightarrow \chi \chi^\ast} \nonumber \\
&& +\Big( \frac{n_{Z'}^\text{eq}}{n_\text{DM}^\text{eq}} \Big)\langle \sigma v \rangle_{\chi^\ast Z' \rightarrow \chi \chi} + \frac{2(n_{Z'}^\text{eq})^2}{(n_\text{DM}^\text{eq})^2}\,\langle \sigma v \rangle_{Z' Z'  \rightarrow \chi \chi^\ast} \nonumber \\
&\approx & \frac{1}{2} (1+\Delta_{h'})^{3/2} \,e^{-\Delta_{h'} x} \langle \sigma v \rangle_{\chi^\ast h' \rightarrow \chi \chi} + \frac{1}{2} (1+\Delta_{h'})^3 \,e^{-2\Delta_{h'} x}\langle \sigma v \rangle_{h' h'  \rightarrow \chi \chi^\ast}  \nonumber \\
&&+\frac{3}{2}  (1+\Delta_{Z'})^{3/2} \,e^{-\Delta_{Z'} x} \langle \sigma v \rangle_{\chi^\ast Z' \rightarrow \chi \chi} + \frac{9}{2} (1+\Delta_{Z'})^3 \,e^{-2\Delta_{Z'} x}\langle \sigma v \rangle_{Z' Z'  \rightarrow \chi \chi^\ast}\,. 
\end{eqnarray}
Here, the dark matter annihilation rates are computed by using the detailed balance conditions \cite{DAgnolo:2015ujb,Choi:2016tkj}.
Then, the general expression for the relic density for forbidden dark matter in our model \cite{Choi:2016tkj} is given by
\begin{eqnarray}
\Omega_{\rm DM} h^2= 5.20\times 10^{-10}\,{\rm GeV}^{-2}   \bigg(\frac{10.75}{g_*} \bigg)^{1/2} \Big(\frac{x_f}{20} \Big)\, \frac{e^{(\Delta_{Z'}+\Delta_{h'})x_f/2} gh}{e^{(\Delta_{Z'}-\Delta_{h'})x_f/2}g +e^{-(\Delta_{Z'}-\Delta_{h'})x_f/2}h }
\,,
\end{eqnarray}
where $\Delta_{h'}=(m_{h'}-m_\chi)/m_\chi$, $\Delta_{Z'}=(m_{Z'}-m_\chi)/m_\chi$, and
\begin{eqnarray}
h(\Delta_{h'},x_f) &\equiv&\bigg[ \frac{1}{2} c_1 (1+\Delta_{h'})^{3/2} \Big(1-\Delta_{h'} x_f\, e^{\Delta_{h'}x_f} \int^\infty_{\Delta_{h_1}x_f} dt\, t^{-1}\,e^{-t} \Big) \nonumber \\
&&+\frac{1}{2}c_2 (1+\Delta_{h'})^3 e^{-\Delta_{h'}x_f}  \Big(1-2\Delta_{h'} x_f\, e^{2\Delta_{h'}x_f}  \int^\infty_{2\Delta_{h'}x_f} dt\, t^{-1}\,e^{-t} \Big)\bigg]^{-1},  \\
g(\Delta_{Z'},x_f) &\equiv& \bigg[\frac{9d_1}{2x_f}\,(1+\Delta_{Z'})^{3/2}\Big(1-(\Delta_{Z'}x_f)^2\, e^{\Delta_{Z'}x_f} \int^\infty_{\Delta_{Z'}x_f} dt \,t^{-2} e^{-t}\Big) \nonumber \\
&&+\frac{9d_2}{2}\,(1+\Delta_{Z'})^3  e^{-\Delta_{Z'} x_f} \Big(1-2(\Delta_{Z'}x_f)\, e^{2\Delta_{Z'}x_f}  \int^\infty_{2\Delta_{Z'}x_f} dt \,t^{-1} e^{-t} \Big) \bigg]^{-1}
\,,
\end{eqnarray}
with $\langle \sigma v\rangle_{h'\chi^*\rightarrow \chi\chi}=c_1$, and $\langle \sigma v\rangle_{h'h'\rightarrow \chi\chi^*}=c_2$, $\langle \sigma v\rangle_{Z'\chi^*\rightarrow \chi\chi}=6d_1/x$, and $\langle \sigma v\rangle_{Z'Z'\rightarrow \chi\chi^*}=d_2$.

\begin{figure}[tbp]
\centering
\includegraphics[width=0.45\linewidth]{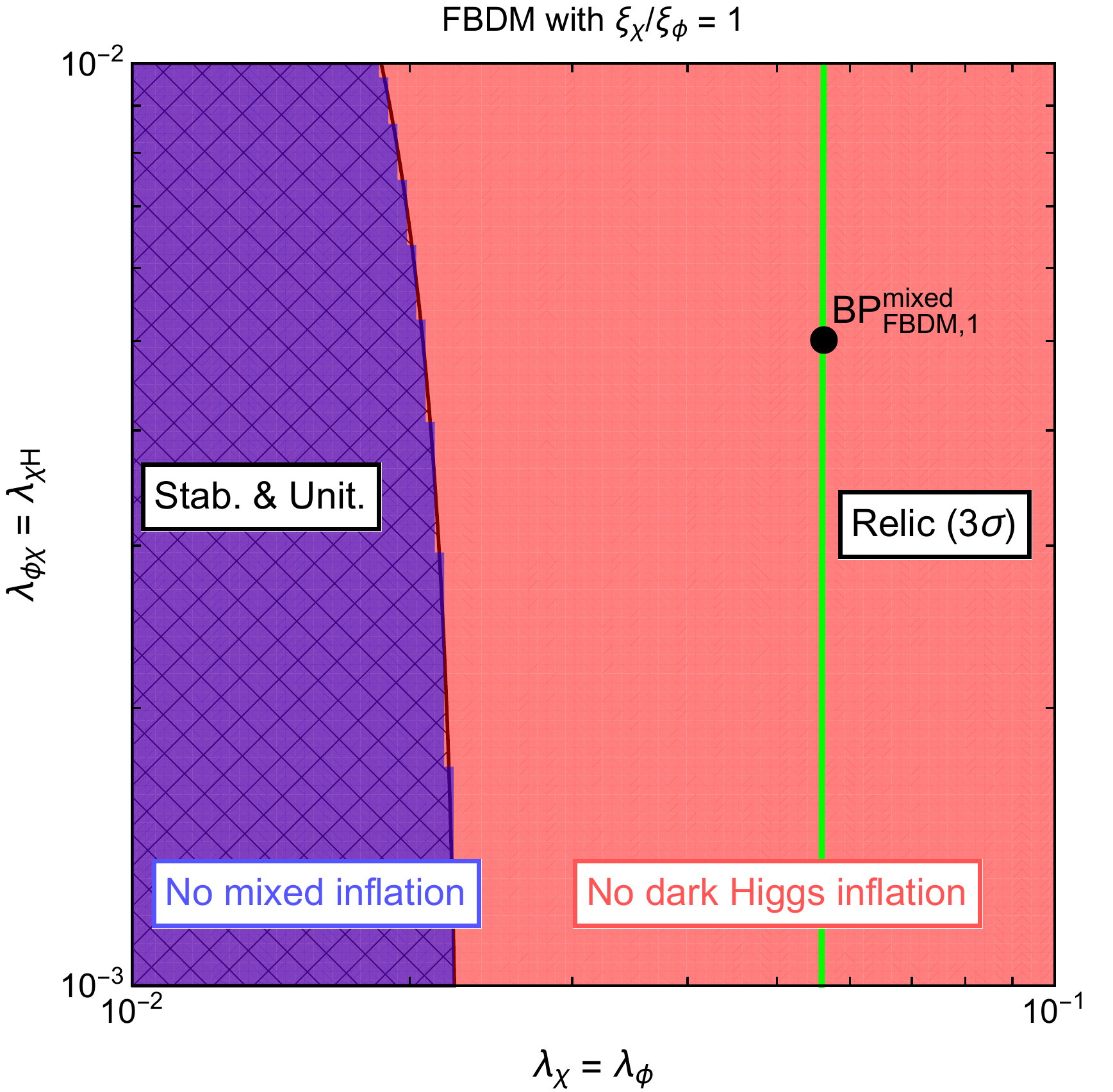}\,\;
\includegraphics[width=0.45\linewidth]{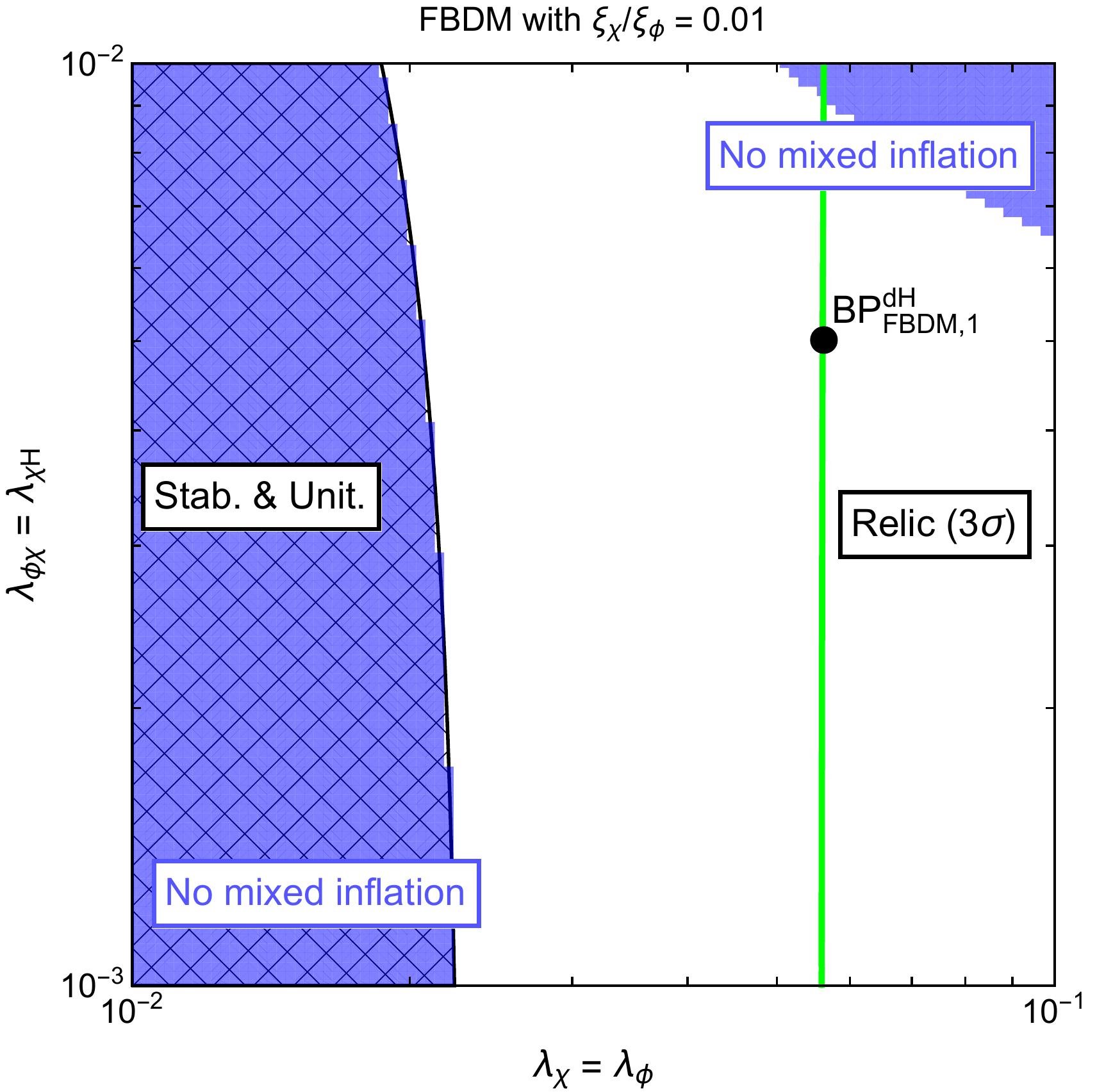}
\caption{Parameter space for forbidden dark matter scenario with $\xi_\chi/\xi_\phi=1$ (left) and $0.01$ (right) in $\lambda_\chi=\lambda_\phi$ versus $\lambda_{\phi\chi}=\lambda_{\chi H}$. The parameters shown here are set at low-energy scale. The rest input parameters are chosen as $g_X = 0.1$, $\sin\theta = 10^{-5}$, $\varepsilon=10^{-4}$, $m_{h^\prime} = 0.16$ GeV, $m_\chi = 0.1$ GeV, and $\zeta =0.02$. The hatched regions are disfavored due to the stability and unitarity. The dark Higgs (mixed) inflation is not allowed in the red-colored (blue-colored) region. The green solid lines represent the correct dark matter relic abundance within 3$\sigma$ range \cite{planck}. The black points are our benchmark points shown in Table \ref{tab:BPForbDMinput}. }
\label{fig:ForbDM1}
\end{figure}
\begin{figure}[tbp]
\centering
\includegraphics[width=0.45\linewidth]{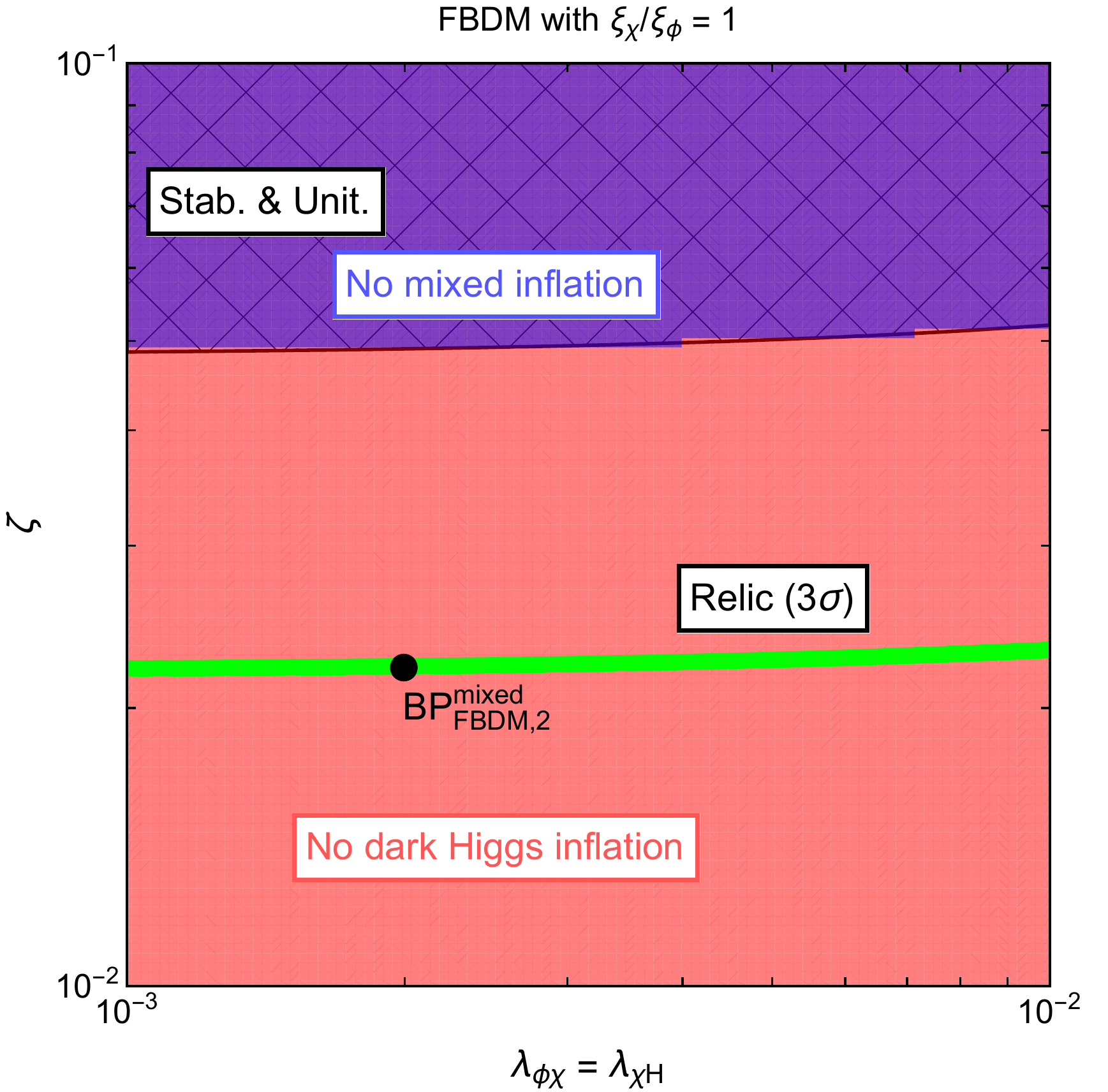}
\,\;\;\;\;\;\;\;\;
\includegraphics[width=0.45\linewidth]{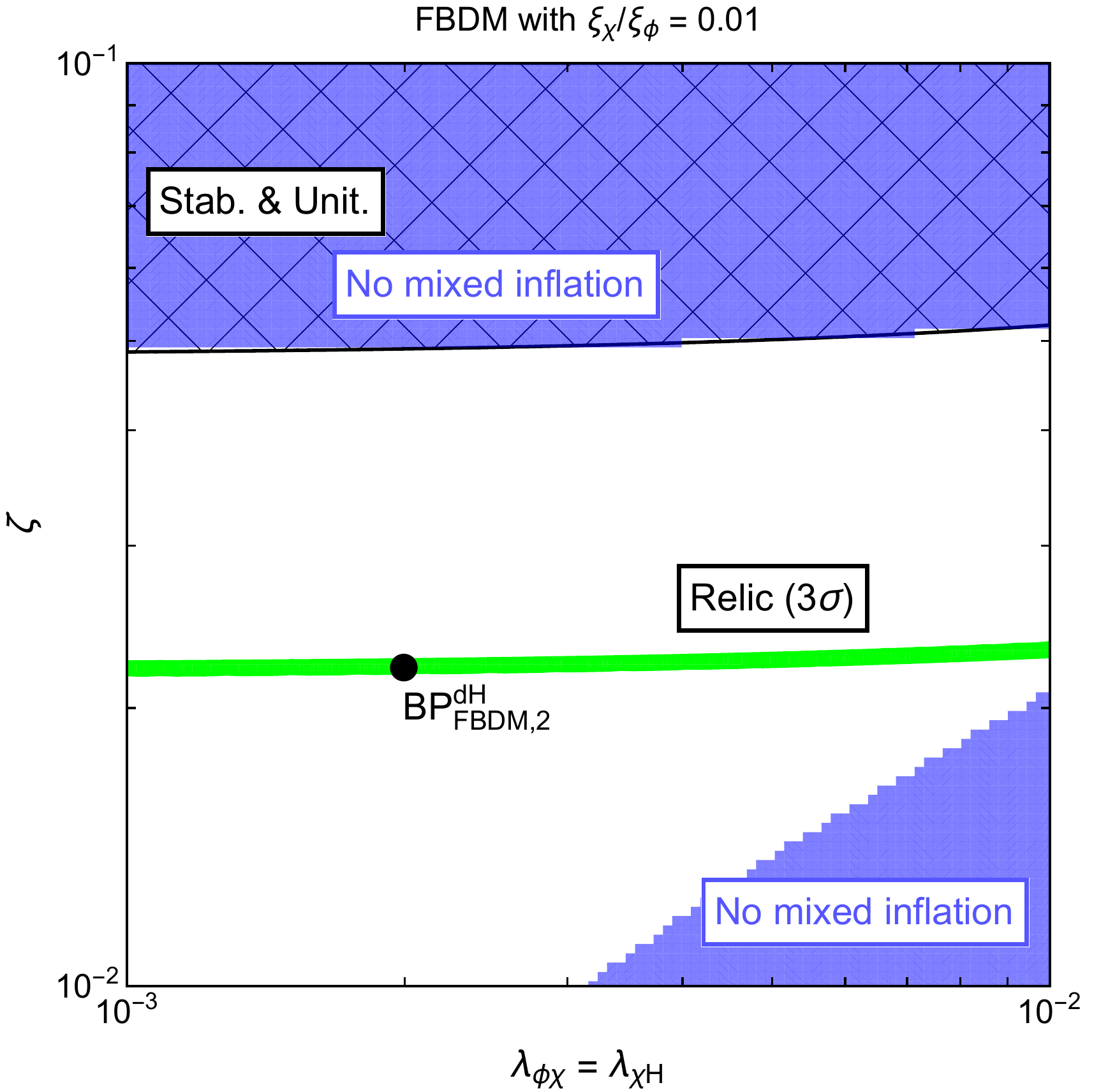}
\caption{Parameter space for forbidden dark matter scenario with $\xi_\chi/\xi_\phi=1$ (left) and $0.01$ (right) for $\lambda_{\phi\chi}=\lambda_{\chi H}$ versus $\zeta$. The parameters shown here are set at dark matter scale. The rest input parameters and color codes are as in Fig.~\ref{fig:ForbDM1}, except that $\lambda_\chi = \lambda_\phi = 0.055$ and $\zeta$ is varying.}
\label{fig:ForbDM2}
\end{figure}
\begin{table}[tbp]
\small
\begin{center}
\begin{tabular}{c|c|c|c|c|c|c}
\hline
Benchmark 
& \multirow{2}{*}{$\lambda_\chi=\lambda_\phi$}
& \multirow{2}{*}{$\lambda_{\phi\chi}=\lambda_{\chi H}$}
& \multirow{2}{*}{$\zeta$}
& \multirow{2}{*}{$\xi_\chi / \xi_\phi$}
& \multirow{2}{*}{$n_s$}
& \multirow{2}{*}{$r$}
\\
Points
& \multirow{2}{*}{} & \multirow{2}{*}{}
& \multirow{2}{*}{} & \multirow{2}{*}{}
& \multirow{2}{*}{} & \multirow{2}{*}{}
\\
\hline
\hline
\multirow{2}{*}{$\mathrm{BP^{mixed}_{FBDM,1}}$}
& \multirow{2}{*}{$0.056$}
& \multirow{2}{*}{$0.005$}
& \multirow{2}{*}{$0.02$}
& \multirow{2}{*}{$1$}
&\multirow{2}{*}{0.9700}
&\multirow{2}{*}{0.00316}
\\
&&&&&&
\\
\hline
\multirow{2}{*}{$\mathrm{BP^{dH}_{FBDM,1}}$}
& \multirow{2}{*}{$0.056$}
& \multirow{2}{*}{$0.005$}
& \multirow{2}{*}{$0.02$}
& \multirow{2}{*}{$0.01$}
&\multirow{2}{*}{0.9698}
&\multirow{2}{*}{0.00312}
\\
&&&&&&
\\
\hline
\multirow{2}{*}{$\mathrm{BP^{mixed}_{FBDM,2}}$}
& \multirow{2}{*}{$0.055$}
& \multirow{2}{*}{$0.002$}
& \multirow{2}{*}{$0.022$}
& \multirow{2}{*}{$1$}
&\multirow{2}{*}{0.9703}
&\multirow{2}{*}{0.00322}
\\
&&&&&&
\\
\hline
\multirow{2}{*}{$\mathrm{BP^{dH}_{FBDM,2}}$}
& \multirow{2}{*}{$0.055$}
& \multirow{2}{*}{$0.002$}
& \multirow{2}{*}{$0.022$}
& \multirow{2}{*}{$0.01$}
&\multirow{2}{*}{0.9699}
&\multirow{2}{*}{0.00313}
\\
&&&&&&
\\
\hline
\end{tabular}
\caption{Input parameters for the benchmark points. They correspond to the black points in Figs.~\ref{fig:ForbDM1} and \ref{fig:ForbDM2}. We chose $g_X=0.1,\, \sin\theta=10^{-5}$, $\varepsilon = 10^{-4}$, $m_{h'}=0.16\,{\rm GeV}$ and $m_\chi=0.1\,{\rm GeV}$. The superscript represents the inflation type; for example, $\mathrm{BP^{mixed}_{FBDM,1}}$ ($\mathrm{BP^{dH}_{FBDM,1}}$) is the benchmark point where the mixed (dark Higgs) inflation is allowed, etc.   }    
\label{tab:BPForbDMinput}
\end{center}
\end{table}

For $m_{h'}\gtrsim m_\chi$ and $m_{Z'}\gg m_{h'}$, the annihilation cross section for the $2\rightarrow 2$ forbidden channels is dominated by the one for $\chi\chi\rightarrow \chi^* h'$, approximated to
\begin{eqnarray}
\langle\sigma v\rangle_{\rm FB }&\approx& \frac{9\zeta^2}{256\pi m^2_\chi} \, \frac{\sqrt{\Delta_{h'} (1+\Delta_{h'})(4+\Delta_{h'})}}{(2+\Delta_{h'})(3+\Delta_{h'})^2}\,\bigg(2(3+\Delta_{h'}) -(1+\Delta_{h'}) (5+2\Delta_{h'})\frac{\lambda_{\phi\chi}}{\lambda_\phi}\bigg)^2\,e^{-\Delta_{h'}x} \nonumber \\
&\approx &\frac{\zeta^2\sqrt{\Delta_{h'}}}{256\pi m^2_\chi} \,\bigg(6-\frac{5\lambda_{\phi\chi}}{\lambda_\phi}\bigg)^2\, e^{-\Delta_{h'}x}
\,.
\end{eqnarray}
Therefore, the resulting effective annihilation rate depends on the dark matter cubic coupling $\zeta$ as well as the mass difference $\Delta_{h'}$ between the resonance and dark matter masses.

In Fig.~\ref{fig:ForbDM1}, we show various constraints coming from dark matter physics and inflation in the quartic self-coupling with $\lambda_\chi=\lambda_\phi$ versus the quartic mixing quartic couplings with $\lambda_{\phi\chi}=\lambda_{\chi H}$ in the former and the quartic mixing quartic couplings with $\lambda_{\phi\chi}=\lambda_{\chi H}$ versus the dark matter cubic self-coupling $\zeta$ in the latter. 
Here, we chose $\Delta_{h'}=(m_{h'}-m_\chi)/m_\chi=0.4$, for instance, $m_\chi = 0.1$ GeV and $m_{h^\prime} = 0.16$ GeV, and let $v_\phi$ vary to get the range, $0.107\,{\rm GeV}\lesssim m_{Z'}\lesssim 0.339\,{\rm GeV}$, depending on $\lambda_\phi$ for the fixed $g_X$ and $m_{h'}$.
We again selected four benchmark points, $\mathrm{BP^{mixed}_{FBDM,1}}$, $\mathrm{BP^{dH}_{FBDM,1}}$, $\mathrm{BP^{mixed}_{FBDM,2}}$, and $\mathrm{BP^{dH}_{FBDM,2}}$, with the input parameters shown in Table \ref{tab:BPForbDMinput}.
For the selected benchmark points, we computed inflationary observables, including the spectral index $n_s$ and the tensor-to-scalar ratio $r$.

\begin{figure}[tbp]
\centering
\includegraphics[width=0.50\linewidth]{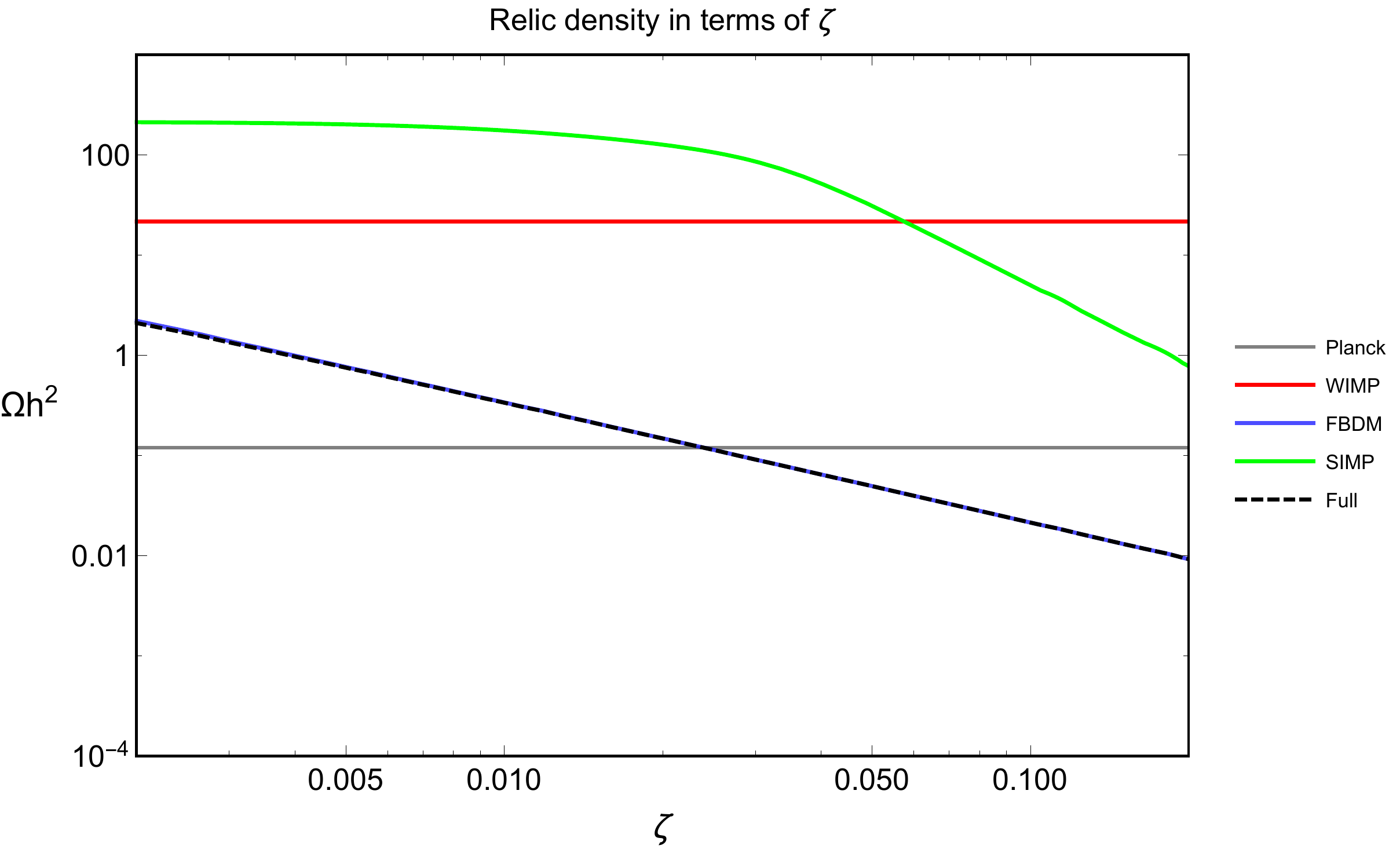}
\includegraphics[width=0.45\linewidth]{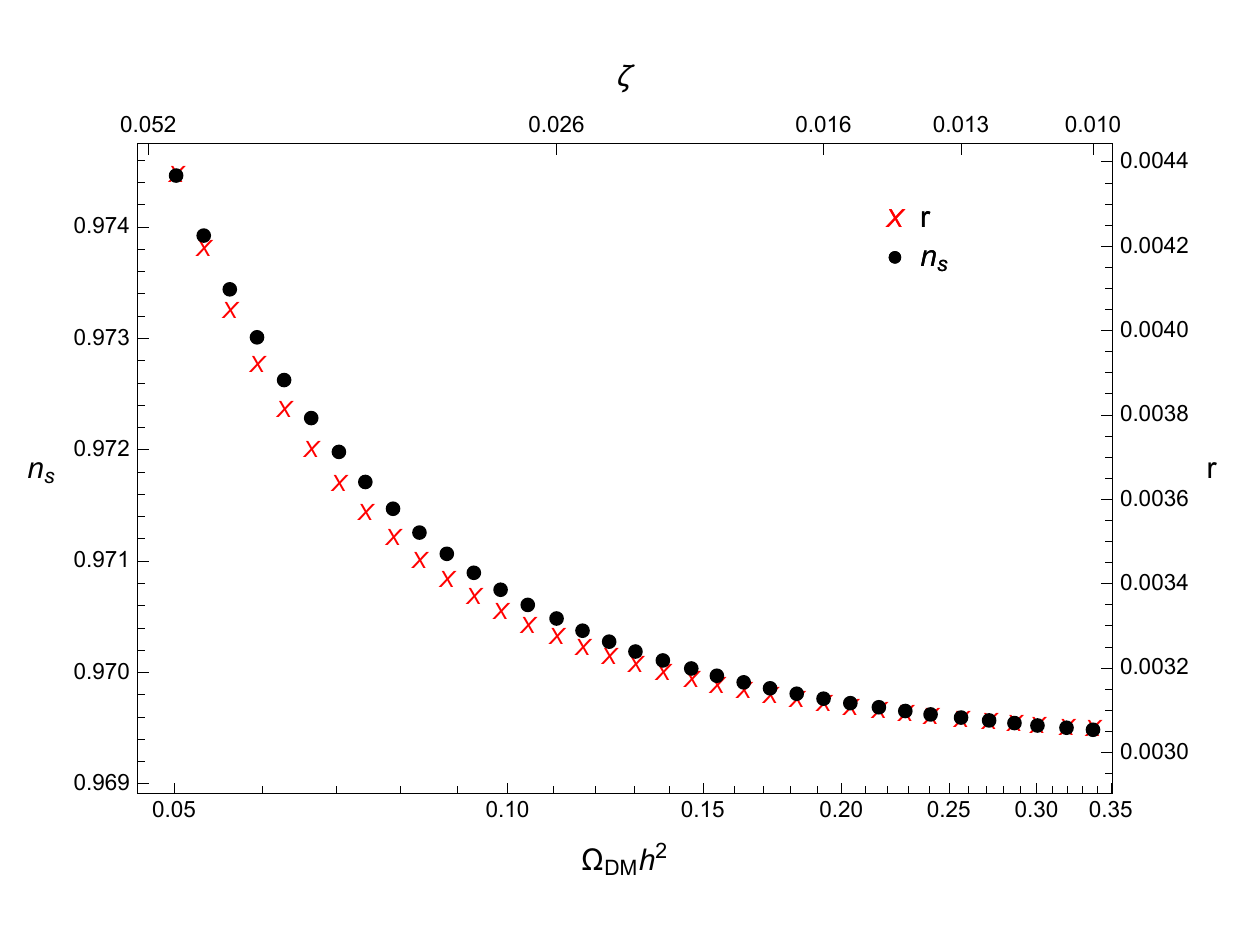}
\caption{
(Left) Relic densities for each dark matter components. (Right) Relic density in the plane of spectral index and tensor-to-scalar ratio. The parameters are chosen as follows:
$\sin\theta = 10^{-5}$, $\varepsilon = 10^{-4}$, $g_X = 0.1$, $m_{h^\prime}=0.16 \, {\rm GeV}$, $m_\chi=0.1 \, {\rm GeV}$, $\lambda_\phi = \lambda_\chi =0.055$, $\lambda_{\phi\chi} = \lambda_{\chi H} =0.006$, thus $m_{Z'}=0.14\,{\rm GeV}$.
The grey horizontal solid line on left represents the Planck data for the relic density.  
}
\label{fig:FBDMRelDenZeta}
\end{figure}

In order to see the $\zeta$ parameter dependence of the relic density and the inflationary predictions.
The relic densities corresponding to the WIMP, SIMP and forbidden components are shown in red, green and blue lines on the left of Fig.~\ref{fig:FBDMRelDenZeta}. The total relic density follows closely the one obtained for forbidden dark matter.  
In comparison, on the right of Fig.~\ref{fig:FBDMRelDenZeta}, we presented the relic density in the plot for the spectral index and the tensor-to-scalar ratio in order to see the correlation between them.
We find that the spectral index and the tensor-to-scalar ratio vary, $n_s \approx 0.969-0.974$ and $r \approx 0.0030-0.0044$, depending on the $\zeta$ parameter, similarly as in the WIMP and SIMP cases. 
Thus, the spectral index again can deviate sizably from the results of the classical non-minimal coupling inflation, but most of the parameter is still consistent with Planck within about $2\sigma$.

\section{Conclusions}
\label{sec:conc}

We have considered the minimal possibility that a complex scalar field beyond $Z_2$ parity can play roles of both the inflaton and dark matter at the same time. The discussion is focussed upon the models with gauged $Z_3$ symmetry, which is the remnant of a dark local $U(1)$ symmetry after the spontaneous breakdown. Thus, in this model, dark matter is communicated to the SM through double portals with the Higgs and $Z'$ bosons.  

The dark matter cubic self-interaction restricts the inflation with non-minimal couplings to be realized always as either the dark Higgs or the mixture of the dark Higgs and scalar dark matter, but not the pure dark matter direction. 
Moreover, the same dark matter cubic self-interaction opens additional channels for dark matter in the dark sector, such as allowed or forbidden $2\rightarrow 2$ (semi-)annihilations and the $3\rightarrow 2$ annihilations, thus it also plays an important role in determining the relic density while evading the stringent constraints from direct and indirect detection experiments. 

As a result, we searched for the parameter space for achieving the successful inflation well within the Planck $2\sigma$ band as well as the observably consistent dark matter with correct relic abundance, depending on the dominant production mechanism for dark matter. The parameter space in each case of dark matter productions is selected due to the theoretical and phenomenological bounds available from  inflation all the way to dark matter energy scales.

\section*{Acknowledgments}

The work of SMC and HML is supported in part by Basic Science Research Program through the National Research Foundation of Korea (NRF) funded by the Ministry of Education, Science and Technology (NRF-2019R1A2C2003738 and NRF-2018R1A4A1025334).  SMC is supported in part by CERN-CKC graduate student fellowship program. 
The work of JK is supported by Alexander von Humboldt Foundation.
The work of BZ is supported partially by Korea Research Fellowship Program through the National Research Foundation of Korea (NRF) funded by the Ministry of Science and ICT (2019H1D3A1A01070937).


\def\theequation{A.\arabic{equation}}
\setcounter{equation}{0}
\vskip0.8cm
\noindent
{\Large \bf Appendix A:  Renormalization Group Equations}
\label{apdx:RGEs}

It is essential to consider renormalization group equations of the relevant parameters to properly connect inflation scale with low-energy scale of dark matter physics.
Following Refs.~\cite{RGEs1} (see also Refs.~\cite{RGEs2}) and Ref.~\cite{Buchbinder:1992rb}, we obtain the renormalization group equations of the model as follows.
First, the beta functions for the gauge couplings are
\begin{align}
(4\pi)^2 \beta_{g_3}
&=
-7g_3^3
\,,\\
(4\pi)^2 \beta_g
&=
-\frac{19}{6}g^3
\,,\\
(4\pi)^2 \beta_{g^\prime}
&=
\frac{41}{6}g^{\prime 3}
\,,\\
(4\pi)^2 \beta_{g_X}
&=
\frac{9s_\phi + s_\chi}{3}g_X^3
\,.
\end{align}
The beta function for top Yukawa coupling is
\begin{align}
(4\pi)^2 \beta_{y_t}&=
y_{t}\left[
\frac{27}{6}y_{t}^{2}
-\left(
8g_{3}^{2}
+\frac{9}{4}g^{2}
+\frac{17}{12}g^{\prime 2}
\right)
\right]
\,,
\end{align}
and the beta functions for quartic couplings are
\begin{align}
(4\pi)^{2} \beta_{\lambda_H}
&=
24\lambda_{H}^{2}
-6y_{t}^{4}
-3\lambda_{H}\left(
g^{\prime 2} + 3g^{2} - 4y_{t}^{2}
\right)
+\frac{3}{8}\left[
\left(
g^{\prime 2}+g^{2}
\right)^{2} +2g^{4}
\right] \nonumber \\
&
\quad +\frac{1+s_\chi^2}{2}
\lambda_{\chi H}^{2}
+\frac{1+s_\phi^2}{2}
\lambda_{\phi H}^{2}
\,,
\end{align}
\begin{align}
(4\pi)^{2} \beta_{\lambda_{\phi H}}
&=
6\lambda_{\phi H}\left(
2\lambda_{H}
+y_{t}^{2}
\right)
-\frac{3}{2}\lambda_{\phi H}
\left(
g^{\prime 2} + 3g^{2}
+36g_{X}^{2}
\right)
+4s_\phi \lambda_{\phi H}^2 \nonumber \\
&
\quad +(1+s_\chi^2)\lambda_{\chi H}\lambda_{\phi \chi}
+2(1+3s_\phi^2)\lambda_\phi \lambda_{\phi H}
\,,\\
(4\pi)^{2} \beta_{\lambda_{\chi H}}
&=
6\lambda_{\chi H}\left(
2\lambda_{H}
+y_{t}^{2}
\right)
-\frac{3}{2}\lambda_{\chi H}\left(
g^{\prime 2}+3g^{2}
+4g_{X}^{2}
\right)
+4 s_\chi \lambda_{\chi H}^2 \nonumber \\
&
\quad+(1+s_\phi^2)\lambda_{\phi\chi}\lambda_{\phi H}
+2(1+3s_\chi^2)\lambda_\chi \lambda_{\chi H}
\,,
\end{align}
\begin{align}
(4\pi)^{2} \beta_{\lambda_{\phi}}
&=
2\lambda_{\phi H}^{2}
-108\lambda_{\phi}g_{X}^{2}
+486g_X^4
+2\left(
1 + 9s_\phi^2
\right)\lambda_\phi^2
+\frac{1+s_\chi^2}{2}\lambda_{\phi\chi}^2
\,,\\
(4\pi)^{2} \beta_{\lambda_{\chi}}
&=
2\lambda_{\chi H}^{2}
-12\lambda_{\chi}g_{X}^{2}
+6g_X^4
+2\left(
1 + 9s_\chi^2
\right)\lambda_\chi^2
+\frac{1+s_\phi^2}{2}\lambda_{\phi\chi}^2
+9\left(
1+ s_\phi s_\chi
\right)\zeta^2
\,,\\
(4\pi)^{2} \beta_{\lambda_{\phi\chi}}
&=
4\lambda_{\chi H}\lambda_{\phi H}
-60\lambda_{\phi\chi}g_{X}^{2}
+108g_X^4
+2(1+3s_\phi^2)\lambda_\phi \lambda_{\phi\chi}
+2(1+3s_\chi^2)\lambda_\chi \lambda_{\phi\chi}	\nonumber \\
&
\quad +4s_\phi s_\chi \lambda_{\phi\chi}^2
+18(1+s_\chi^2)\zeta^2\,, \\
(4\pi)^{2} \beta_{\zeta}
&=
6\left(
-1+ 3 s_\chi^2
\right)\lambda_\chi \zeta
+6 s_\phi s_\chi
\lambda_{\phi\chi} \zeta
-36g_X^2 \zeta
\,.
\end{align}
The beta functions for non-minimal couplings are
\begin{align}
(4\pi)^{2}  \beta_{\xi_H}&=
\left(
12\lambda_{H} + 6y_t^2 - \frac{3}{2}g^{\prime 2}
-\frac{9}{2}g^2
\right)
\left(
\xi_H + \frac{1}{6}
\right)
+(1+s_\phi)\lambda_{\phi H}
\left(
\xi_\phi + \frac{1}{6}
\right) \nonumber \\
&\quad +(1+s_\chi)\lambda_{\chi H}
\left(
\xi_\chi + \frac{1}{6}
\right)
\,,\\
(4\pi)^{2}\beta_{\xi_\phi} &=
4\lambda_{\phi H}\left(
\xi_H + \frac{1}{6}
\right)
+2\left[
(1+3s_\phi)\lambda_{\phi} - 27 g_X^2
\right]
\left(
\xi_\phi + \frac{1}{6}
\right) \nonumber \\
&\quad 
+(1+s_\chi)\lambda_{\phi\chi}
\left(
\xi_\chi + \frac{1}{6}
\right)
\,,\\
(4\pi)^{2}\beta_{\xi_\chi} &=
4\lambda_{\chi H}
\left(
\xi_H + \frac{1}{6}
\right)
+(1+s_\phi)\lambda_{\phi\chi}
\left(
\xi_\phi + \frac{1}{6}
\right) \nonumber \\
&\quad +2\left[
(1+3s_\chi)\lambda_{\chi} - 3g_X^2
\right]
\left(
\xi_\chi + \frac{1}{6}
\right)
\,.
\end{align}
The beta functions for mass parameters are given by
\begin{eqnarray}
(4\pi)^2\beta_{m_H^2} &=&
\bigg(
12\lambda_H+6y_t^2
-\frac{3g'^2}{2}-\frac{9g^2}{2}
\bigg)m_{H}^2
+(1+s_\phi) \lambda_{\phi H} m_\phi^2 \nonumber \\
&&-(1+s_\chi) \lambda_{\chi H} m_\chi^2\,, \\
(4\pi)^2\beta_{m_\phi^2} &=& 2\Big[
(1+3s_\phi) \lambda_\phi - 27g_X^2
\Big]m_\phi^2+
4\lambda_{\phi H} m_{H}^2
-(1+s_\chi) \lambda_{\phi\chi} m_\chi^2\,, \\
(4\pi)^2\beta_{m_\chi^2} &=& 2\Big[
(1+3s_\chi) \lambda_\chi - 3g_X^2
\Big]m_\chi^2
-4\lambda_{\chi H} m_{H}^2
-(1+s_\phi) \lambda_{\phi\chi} m_\phi^2
\,.
\end{eqnarray}
Finally the anomalous dimensions are given by
\begin{align}
(4\pi)^2 \gamma_{H} &=
-\frac{3}{4}g^{\prime 2} - \frac{9}{4}g^2 + 3y_t^2
\,,\\
(4\pi)^2 \gamma_{\phi} &=
-27 g_X^2
\,,\\
(4\pi)^2 \gamma_{\chi} &=
-3g_X^2
\,.
\end{align}
Here, $\beta_p \equiv dp/dt$ for a parameter $p$ and $t = \ln \mu/\mu_0$ with $\mu$ being the renormalization scale, $\mu_0=M_Z$, and $s_\chi$ and $s_\phi$ are the suppression factors that need to be taken into account when inflation occurs along the $\chi$ and/or $\phi$ field direction, or equivalently $\psi$ and $\varphi$ direction, respectively.
They are due to the fact that the presence of the non-minimal couplings modifies the propagators \cite{Salopek:1988qh,DeSimone:2008ei}. Explicitly, they are given by
\begin{align}
s_{\phi} = \frac{1+\xi_\phi \varphi^2/M_{{\rm P}}^{2}}
{1+(6\xi_{\phi} + 1)\xi_{\phi}\varphi^2 / M_{{\rm P}}^{2}}
\,,
\end{align}
and similarly for $s_\chi$. Note that $s_{\phi,\chi} \rightarrow 1$ for $\xi_{\phi,\chi} \rightarrow 0$.

\def\theequation{B.\arabic{equation}}

\setcounter{equation}{0}

\vskip0.8cm
\noindent
{\Large \bf Appendix B:  Dark matter annihilation/scattering cross sections}

We list the full Boltzmann equation for dark matter relic density and the necessary formulas for the dark matter annihilation and the dark matter-SM scattering as well as the dark matter self-scattering.

\subsection*{\underline{Boltzmann equation}}

The full Boltzmann equation governing the dark matter density, $n_{\rm DM}=n_\chi+n_{\chi*}$, is given \cite{Choi:2016tkj} by
\begin{equation}\begin{aligned}
\dot{n}_\text{DM} + 3Hn_\text{DM}\ =\ &-\frac{1}{2}\langle \sigma v \rangle_{\chi\chi^*\rightarrow f{\bar f}}\bigg(n_{\rm DM}^2-(n_\text{DM}^{\rm eq})^2 \bigg)  
\\
& -\frac{1}{4}\Big(\langle\sigma v^2\rangle_{\chi\chi\chi\rightarrow \chi\chi^\ast}+\langle\sigma v^2\rangle_{\chi\chi\chi^\ast\rightarrow \chi^\ast\chi^\ast}\Big)\bigg(n_\text{DM}^3-n_\text{DM}^\text{eq}n_\text{DM}^2\bigg)\\
& -\frac{1}{2}\langle\sigma v^2\rangle_{\chi\chi\chi^\ast\rightarrow \chi h'}\bigg( n_\text{DM}^3-\frac{(n_\text{DM}^\text{eq})^2}{n_{h'}^\text{eq}}n_\text{DM}n_{h'} \bigg) \\
& +\langle \sigma v \rangle_{\chi^\ast h' \rightarrow \chi \chi}\bigg( n_\text{DM} n_{h'} - \frac{n_{h'}^\text{eq}}{n_\text{DM}^\text{eq}}n_\text{DM}^2 \bigg) \\
& +2\langle \sigma v \rangle_{h' h'  \rightarrow \chi \chi^\ast}\bigg(n_{h'}^2-\frac{(n_{h'}^\text{eq})^2 }{(n_\text{DM}^\text{eq})^2}\,n_\text{DM}^2 \bigg) \\
& -\frac{1}{2}\langle\sigma v^2\rangle_{\chi\chi\chi^\ast\rightarrow \chi Z'}\bigg( n_\text{DM}^3-\frac{(n_\text{DM}^\text{eq})^2}{n_{Z'}^\text{eq}}n_\text{DM}n_{Z'} \bigg)\\
& +\langle \sigma v \rangle_{\chi^\ast Z' \rightarrow \chi \chi}\bigg( n_\text{DM} n_{Z'} - \frac{n_{Z'}^\text{eq}}{n_\text{DM}^\text{eq}}n_\text{DM}^2 \bigg) \\
& +2\langle \sigma v \rangle_{Z' Z'  \rightarrow \chi \chi^\ast}\bigg( n_{Z'}^2-\frac{(n_{Z'}^\text{eq})^2}{(n_\text{DM}^\text{eq})^2} \,n_\text{DM}^2\bigg).  \label{Boltzmann}
\end{aligned}\end{equation}
Here, we assumed that $n_\chi=n_{\chi^*}$ with no CP violation in the dark sector.

\subsection*{\underline{$2\rightarrow 2$ processes}}

The annihilation cross section of dark matter into a pair of SM fermions \cite{Choi:2015bya,Choi:2016tkj} is given by
\begin{eqnarray}
\langle\sigma v\rangle_{\chi\chi^*\rightarrow {\bar f}f} &=& \frac{\varepsilon^2e^2 g^2_X N_c}{\pi x}\, \frac{m^2_\chi+ \frac{1}{2} m^2_f}{(4m^2_\chi-m^2_{Z'})^2+m^2_{Z'} \Gamma^2_{Z'}}\, \sqrt{1-\frac{m^2_f}{m^2_\chi}} \nonumber \\
&& + \frac{N_c}{4\pi} \Big(\frac{m_f}{v_{\rm ew}}\Big)^2 \Big(1-\frac{m^2_f}{m^2_\chi}\Big)^{3/2}\bigg|\frac{y_{h_1 \chi^*\chi}}{4m^2_\chi-m^2_{h_1}} + \frac{y_{h_2 \chi^* \chi}}{4m^2_\chi-m^2_{h_2}} \bigg|^2
\,,
\end{eqnarray}
with $x\equiv m_\chi/T$,  $v_{\rm ew}=246\,{\rm GeV}$, $\varepsilon\simeq \cos\theta_W\xi$ for $\xi\ll 1$ and 
\begin{eqnarray}
y_{h_1 \chi^*\chi}&=&  -\sin\theta (\lambda_{\phi\chi} v_\phi \cos\theta -\lambda_{\chi H} v\sin\theta), \label{hportal1} \\
y_{h_2 \chi^* \chi}&=& \cos\theta (\lambda_{\phi\chi} v_\phi \sin\theta +\lambda_{\chi H} v\cos\theta) \label{hportal2}
\,,
\end{eqnarray}
where $h_{1,2}$ are dark Higgs-like and SM Higgs-like scalars \cite{Choi:2015bya,Choi:2016tkj}, respectively, and the mixing angle is given by
\begin{eqnarray}
\tan(2\theta) = \frac{\lambda_{\phi H} v_\phi v_{\rm ew}}{\lambda_H v^2_{\rm ew}-\lambda_\phi v^2_\phi}
\,.
\end{eqnarray}

The cross section for dark matter annihilating into $VV$  with $V=Z, W$ are given by
\begin{equation}
\begin{aligned}
\langle \sigma v\rangle_{\chi\chi^*\rightarrow VV}\ =\ &\frac{\delta_V e^4}{256\pi \sin^4{\theta_W}\cos^4{\theta_W}}\frac{v_{\rm ew}^2m_\chi^2}{m_V^4}\bigg(4-4\frac{m_V^2}{m_\chi^2}+3\frac{m_V^4}{m_\chi^4}\bigg)\\
&\times \sqrt{1-\frac{m_V^2}{m_\chi^2}}\bigg(\frac{y_{h_1\chi^*\chi}}{4m_\chi^2-m_{h_1}^2}+\frac{y_{h_2\chi^*\chi}}{4m_\chi^2-m_{h_2}^2}\bigg)^2\,,
\end{aligned}
\end{equation}
with $\delta_V=1$ $(2\cos^4\theta_W)$ for $V=Z$ ($W$).
The cross-section for dark matter annihilating into $\gamma\gamma$ is loop-induced, given by
\begin{equation}
\begin{aligned}
\langle \sigma v\rangle_{\chi\chi^*\rightarrow \gamma\gamma}\ =\ &\frac{e^4 m^2_\chi}{64\pi^5 v_{\rm ew}^2}\bigg|\sum_f N_c Q_f^2A_{1/2}(x_f)+A_1(x_w)\bigg|^2\bigg(\frac{y_{h_1 \chi^*\chi}}{4m^2_\chi-m^2_{h_1}} + \frac{y_{h_2 \chi^* \chi}}{4m^2_\chi-m^2_{h_2}} \bigg)^2
\,.
\end{aligned}
\end{equation}

The (semi-)annihilation cross sections for the allowed channels for $m_\chi>m_A$ with $A=h', Z'$ are
\begin{align}
\langle\sigma v\rangle_{\chi\chi\rightarrow \chi^* h'} &= \frac{3R^2}{128\pi m_\chi^2}\bigg( 1-\frac{m_{h'}^2}{9m_\chi^2}\bigg)^{1/2}\bigg(1-\frac{m_{h'}^2}{m_\chi^2}\bigg)^{1/2} \bigg(\frac{3m_\chi}{v_\phi}-\frac{\lambda_{\phi\chi}v_\phi(9m_\chi^2+m_{h'}^2)}{m_\chi(3m_\chi^2-m_{h'}^2)}\bigg)^2
\,,\\
\langle\sigma v\rangle_{\chi\chi^*\rightarrow h' h'}&=  \frac{ \lambda_{\phi\chi}^2}{64\pi m_\chi^2}\sqrt{1-\frac{m_{h'}^2}{m_\chi^2}} \bigg(1-\frac{2\lambda_{\phi\chi}v_\phi^2}{2m_\chi^2-m_{h'}^2} + \frac{6\lambda_\phi v_\phi^2}{4m_\chi^2-m_{h'}^2}\bigg)^2
\,,
\end{align}
with $R\equiv \zeta v_\phi/m_\chi$ and $m_{h'}\simeq m_{h_1}$,
and
\begin{align}
\langle\sigma v\rangle_{\chi\chi\rightarrow \chi^* Z'} &=
\frac{243g_X^2 R^2}{64\pi m_{Z'}^2}\bigg(1-\frac{m_{Z'}^2}{m_\chi^2}\bigg)^{3/2}\bigg(1-\frac{m_{Z'}^2}{3m_\chi^2}\bigg)^{-2}\bigg(1-\frac{m_{Z'}^2}{9m_\chi^2}\bigg)^{7/2}
\,, \\
\langle \sigma v \rangle_{\chi\chi^\ast \rightarrow Z' Z'} &= \frac{g_X^4}{16\pi m_\chi^2}\sqrt{1-\frac{m_{Z'}^2}{m_\chi^2}}\bigg( \frac{8m_\chi^4 - 8m_\chi^2 m_{Z'}^2 + 3m_{Z'}^4}{(2m_\chi^2-m_{Z'}^2)^2} + \frac{54\lambda_{\phi\chi}v_\phi^2}{4m_\chi^2-m_{h'}^2} \nonumber  \\
&\quad
+ \frac{81\lambda_{\phi\chi}^2v_\phi^4(4m_\chi^4-4m_\chi^2m_{Z'}^2+3m_{Z'}^4)}{m_{Z'}^4(4m_\chi^2-m_{h'}^2)^2} \bigg)\,.
\end{align}

The (semi-)annihilation cross sections for forbidden channels for $m_\chi<m_A$ with $A=h', Z'$ \cite{Choi:2016tkj} are
\begin{align}
\langle\sigma v\rangle_{\chi\chi\rightarrow \chi^* A}= \frac{n_{A}}{n_\chi} \, \langle\sigma v\rangle_{\chi^* A\rightarrow \chi\chi}
\,,
\end{align}
with
\begin{align}
\langle\sigma v\rangle_{\chi^* h'\rightarrow \chi\chi}&=
\frac{9R^2}{32\pi m_\chi(m_\chi+m_{h'})}\bigg(1-\frac{m_\chi}{m_{h'}}\bigg)^{1/2}\bigg(1+\frac{3m_\chi}{m_{h'}}\bigg)^{1/2} \nonumber \\
&\quad\times\bigg(\frac{m_\chi}{v_\phi}-\frac{\lambda_{\phi\chi}v_\phi(3m_\chi+2m_{h'})}{m_{h'}(2m_\chi+m_{h'})}\bigg)^2 \,,\\
\langle\sigma v\rangle_{\chi^* Z'\rightarrow \chi\chi}&=
\frac{3g_X^2R^2}{4\pi m_\chi m_{Z'} x}\bigg(1+\frac{m_\chi}{m_{Z'}}\bigg)^{-3}\bigg(1+2\frac{m_\chi}{m_{Z'}}\bigg)^{-2}\bigg(1+3\frac{m_\chi}{m_{Z'}}\bigg)^{5/2}\bigg(1-\frac{m_\chi}{m_{Z'}}\bigg)^{1/2} \nonumber \\
&\quad\times \bigg(1-4\frac{m_\chi^2}{m_{Z'}^2}+4\frac{m_\chi^3}{m_{Z'}^3}+11\frac{m_\chi^4}{m_{Z'}^4}\bigg)
\,,
\end{align}
and 
\begin{align}
\langle\sigma v\rangle_{\chi\chi^*\rightarrow A A}= \frac{(n_A)^2}{(n_\chi)^2} \, \langle\sigma v\rangle_{AA\rightarrow \chi\chi^*}
\,,
\end{align}
with
\begin{align}
\langle\sigma v\rangle_{h' h'\rightarrow \chi\chi^*}&=
\frac{\lambda_{\phi\chi}^2}{64\pi m_{h'}^2}\sqrt{1-\frac{m_\chi^2}{m_{h'}^2}} \bigg(1+\frac{2v_\phi^2(\lambda_\phi-\lambda_{\phi\chi})}{m_{h'}^2}\bigg)^2
\,, \\
\langle\sigma v\rangle_{Z' Z'\rightarrow \chi\chi^*}&=
\frac{g_X^4}{144\pi m_{Z'}^2}\sqrt{1-\frac{m_\chi^2}{m_{Z'}^2}}\bigg( 11 - \frac{24m_\chi^2}{m_{Z'}^2} + \frac{16m_\chi^4}{m_{Z'}^4} \nonumber \\
&\quad - \frac{18\lambda_{\phi\chi}v_\phi^2(4m_\chi^2-m_{Z'}^2)}{m_{Z'}^2(4m_{Z'}^2-m_{h'}^2)} + \frac{243\lambda_{\phi\chi}^2v_\phi^4}{(m_{h'}^2-4m_{Z'}^2)^2} \bigg)
\,.
\end{align}

\subsection*{\underline {$3\rightarrow 2$ processes}}

The $3\rightarrow 2$ annihilation cross sections involving only dark matter in the external states \footnote{We have corrected the symmetry factors for initial or final states as compared to Refs.~\cite{Choi:2015bya,Choi:2016tkj}} are
\begin{align}
\langle\sigma v^2\rangle_{\chi\chi\chi\rightarrow \chi\chi^\ast}&=
\frac{\sqrt{5}R^2}{64\pi m_\chi^5}\bigg( 2\lambda_\chi + 9R^2 + \frac{25g_X^2 m_\chi^2}{m_\chi^2+m_{Z'}^2} \nonumber \\
&\quad + \frac{2\lambda_{\phi\chi}m_\chi^2(13m_\chi^2-2m_{h'}^2)-\lambda_{\phi\chi}^2v_\phi^2(19m_\chi^2-m_{h'}^2)}{(9m_\chi^2-m_{h'}^2)(m_\chi^2+m_{h'}^2)} \bigg)^2
\,,
\end{align}
\begin{align}
\langle\sigma v^2\rangle_{\chi\chi\chi^\ast\rightarrow \chi^\ast\chi^\ast}&= 
\frac{\sqrt{5}R^2}{6144\pi m_\chi^5}\bigg( 74\lambda_\chi -117R^2 - \frac{200g_X^2 m_\chi^2}{m_\chi^2+m_{Z'}^2} \nonumber \\
&\quad + \frac{24\lambda_{\phi\chi}m_\chi^2(3m_\chi^2-2m_{h'}^2)-\lambda_{\phi\chi}^2v_\phi^2(43m_\chi^2-37m_{h'}^2)}{(4m_\chi^2-m_{h'}^2)(m_\chi^2+m_{h'}^2)} \bigg)^2
\,.
\end{align}

The $3\rightarrow 2$ cross sections with assisted annihilations for $m_\chi>m_A/2$ with $A=h', Z'$ are
\begin{align}
\langle\sigma v^2\rangle_{\chi\chi\chi^\ast\rightarrow \chi h'}&=
\frac{1}{36864\pi m_\chi^5}\bigg(1-\frac{m_{h'}^2}{16m_\chi^2}\bigg)^{1/2}\bigg(1-\frac{m_{h'}^2}{4m_\chi^2}\bigg)^{-3/2}  \nonumber \\
&\quad\times\bigg[\frac{192}{7} \frac{m^2_\chi}{v^2_\phi} R^2\bigg(1-\frac{m_{h'}^2}{4m_\chi^2}\bigg)\bigg( 1+\frac{m_{h'}^2}{2m_\chi^2} \bigg)\bigg( 1- \frac{m_{h'}^2}{7m_\chi^2}\bigg)^{-1}  \nonumber \\
&\quad \quad-\frac{192}{7}\lambda_{\phi \chi} R^2\bigg( 1-\frac{m_{h'}^2}{7m_\chi^2} \bigg)^{-1}\bigg( 1+ \frac{25m_{h'}^2}{64m_\chi^2} + \frac{m_{h'}^4}{64m_\chi^4} \bigg) \nonumber \\
&\quad\quad +64\lambda_\chi \lambda_{\phi\chi}\bigg(1+\frac{m_{h'}^2}{32m_\chi^2}\bigg) + 16\lambda_{\phi\chi}^2\bigg(1-\frac{5m_{h'}^2}{8m_\chi^2}\bigg)\bigg(1+\frac{m_{h'}^2}{2m_\chi^2}\bigg)^{-1}  \\
&\quad\quad +36\lambda_{\phi}\lambda_{\phi\chi}^2\frac{v_\phi^2}{m_\chi^2}\bigg(1+\frac{m_{h'}^2}{2m_\chi^2}\bigg)^{-1} \nonumber \\
&\quad\quad - 16\lambda_{\phi\chi}^3\frac{v_\phi^2}{m_\chi^2}\bigg(1-\frac{m_{h'}^2}{4m_\chi^2}\bigg)^{-1}\bigg(1+\frac{m_{h'}^2}{2m_\chi^2}\bigg)^{-1}\bigg(1-\frac{23m_{h'}^2}{64m_\chi^2}-\frac{m_{h'}^4}{128m_\chi^4}\bigg) \bigg]^2, \nonumber 
\end{align}
\begin{align}
\langle\sigma v^2\rangle_{\chi\chi\chi^\ast\rightarrow \chi Z'}&=
\frac{g_X^2}{1728\pi m_\chi^5}\frac{m_{Z'}^2}{m_\chi^2}\bigg(1-\frac{m_{Z'}^2}{16m_\chi^2}\bigg)^{3/2}\bigg(1-\frac{m_{Z'}^2}{4m_\chi^2}\bigg)^{-1/2} \nonumber  \\
&\quad\times\bigg( 2\lambda_\chi - \frac{12g_X^2 m_\chi^2}{2m_\chi^2+m_{Z'}^2} - \frac{3R^2(192 m_\chi^4 - 31m_\chi^2m_{Z'}^2+m_{Z'}^4)}{m_{Z'}^2(7m_\chi^2-m_{Z'}^2)} \nonumber \\
&\quad\quad + \frac{16\lambda_{\phi\chi} m_\chi^2(4m_\chi^2-m_{Z'}^2)}{(4m_\chi^2-m_{h'}^2)(2m_\chi^2+m_{Z'}^2)} + \frac{\lambda_{\phi\chi}^2 v_\phi^2}{4m_\chi^2-m_{h'}^2} \bigg)^2.
\end{align}

\subsection*{\underline {DM-SM scatterings }}

The effective Lagrangian for dark matter-quark elastic scattering is given by
\begin{align}
{\cal L}_{q,{\rm eff}}&= \frac{m_q}{v_{\rm ew}} \bigg(\frac{y_{h_2 \chi^*\chi}}{m^2_{h_2}}+\frac{y_{h_1 \chi^*\chi}}{m^2_{h_1}}\bigg) |\chi|^2 {\bar q}q +\frac{g_X q_\chi e\,\varepsilon Q_q}{m^2_{Z'}}\, i(\chi\partial_\mu\chi^*-\chi^*\partial_\mu \chi) {\bar q}\gamma^\mu q
\,.
\end{align}
Then, the relevant matching conditions between quark and nucleon operators are given by
\begin{align}
\langle N|{\bar q}q |N\rangle = \frac{m_N}{m_q}\,f^{(N)}_{Tq}
\,, 
\end{align}
for light quarks ($q=u,d,s$), and 
\begin{align}
\langle N|{\bar q}q |N\rangle &= \frac{2}{27}\,\frac{m_N}{m_q}\, f^{(N)}_{TG}, \quad f^{(N)}_{TG}=1-\sum_{q=u,d,s} f^{(N)}_{Tq}
\,,
\end{align}
for heavy quarks ($q=c,b,t$), and those for vector operators are given by
\begin{equation}\begin{aligned}
\langle N|{\bar u}\gamma^\mu u |N\rangle &= 2 {\bar N}\gamma^\mu N,
&
N&=p\,, \\
\langle N|{\bar d}\gamma^\mu d |N\rangle &= {\bar N}\gamma^\mu N, 
&
N&=p\,, \\
\langle N|{\bar u}\gamma^\mu u |N\rangle &= {\bar N}\gamma^\mu N, 
&
N&=n\,, \\
\langle N|{\bar d}\gamma^\mu d |N\rangle &= 2{\bar N}\gamma^\mu N, 
&
N&=n\,.
\end{aligned}\end{equation}
As a result, we get the effective Lagrangian for dark matter-nucleon elastic scattering as follows,
\begin{align}
{\cal L}_{N,{\rm eff}} &=\frac{m_N}{v_{\rm ew}}\bigg(\frac{y_{h_2 \chi^*\chi}}{m^2_{h_2}}+\frac{y_{h_1 \chi^*\chi}}{m^2_{h_1}}\bigg) \Big(\sum_{q=u,d,s}f^{(N)}_{Tq}+ \frac{2}{27}f^{(N)}_{TG}\times 3  \Big)  |\chi|^2 {\bar N} N \nonumber \\
&\quad+\frac{g_X q_\chi e\,\varepsilon}{m^2_{Z'}}\, i(\chi\partial_\mu\chi^*-\chi^*\partial_\mu \chi)\bigg((2Q_u+Q_d){\bar p}\gamma^\mu p+(Q_u+2Q_d) {\bar n}\gamma^\mu n\bigg)
\,,
\end{align}
where $y_{h_1 \chi^*\chi}$ and $y_{h_2 \chi^*\chi}$ are given in Eqs.~\eqref{hportal1} and \eqref{hportal2}, respectively.
Thus, from $2Q_u+Q_d=+1$ and $Q_u+2Q_d=0$, there is a nonzero interaction only for dark matter-proton scattering with $Z'$ portal.
Consequently, we obtained the $\chi$-nucleus scattering cross section as follows,
\begin{equation}
\begin{aligned}
& \sigma_{\chi -A}\ =\ \frac{\mu^2_A}{4\pi m^2_\chi}\Big[Z \Big(c_p f_p+g_p\Big)+(A-Z) \Big(c_n f_n+g_n\Big) \Big]^2\,,
\end{aligned}
\end{equation}
where $\mu_A=m_\chi m_A/(m_A+m_\chi)$ is the reduced mass of the dark matter-nucleus system with $m_A$ being the target nucleus mass, $Z, A$  are the number of protons and the atomic number, respectively, and the effective couplings and form factors \cite{GMDD} are given by
\begin{equation}
\begin{aligned}
& c_N\ \equiv\ \frac{m_N}{v_{\rm ew}}\bigg(\frac{y_{h_2 \chi^*\chi}}{m^2_{h_2}}+\frac{y_{h_1 \chi^*\chi}}{m^2_{h_1}}\bigg)
\,,\\
& f_p\ \equiv\ \sum_{q=u,d,s}f^{(p)}_{Tq}+ \frac{2}{9}f^{(p)}_{TG} \simeq 
0.28
\,,\\
& f_n\ \equiv\ \sum_{q=u,d,s}f^{(n)}_{Tq}+ \frac{2}{9}f^{(n)}_{TG}  \simeq  
0.28
\,,\\
& g_p\ =\ -\frac{2e q_\chi g_X \varepsilon m_\chi}{m_{Z'}^2},\\
& g_n\ \approx\ 0
\,.
\end{aligned}
\end{equation}
Similarly, the $\chi^*$-nucleus scattering cross section has the $Z'$ contributions flipped in sign, given by
\begin{equation}
\begin{aligned}
& \sigma_{\chi^* -A}\ =\ \frac{\mu^2_A}{4\pi m^2_\chi}\Big[Z \Big(c_p f_p-g_p\Big)+(A-Z) \Big(c_n f_n-g_n\Big) \Big]^2
\,.
\end{aligned}
\end{equation}
Then, the averaged dark matter-nucleus scattering cross section is given
\begin{align}
\sigma_{{\rm DM}-A}=\frac{1}{2} \Big( \sigma_{\chi -A}+ \sigma_{\chi^* -A} \Big)\,.
\end{align}

The above dark matter-nucleus scattering cross section is related to the normalized-to-proton scattering cross section \cite{GMDD}, $\sigma_{{\rm DM}-p}$, that is usually presented for experimental limits, by
\begin{align}
\sigma_{{\rm DM}-p}= \Big(\frac{\mu_N}{\mu_A}\Big)^2\, \frac{\sigma_{{\rm DM}-A}}{A^2}\,,
\end{align}
with $\mu_N=m_\chi m_N/(m_N+m_\chi)$.

The elastic scattering between dark matter and electron is mediated by CP-even scalars and $Z^\prime$, so the corresponding cross section is similarly given by 
\begin{align}
\sigma_{{\rm DM} -e}&= \frac{\mu^2_e}{8\pi m_\chi^2}\bigg[\frac{m_e}{v_{\rm ew}}\Big(\frac{y_{h_2 \chi^*\chi}}{m_{h_2}^2}+\frac{y_{h_1 \chi^*\chi}}{m_{h_1}^2}\Big)+\frac{2e q_\chi g_X \varepsilon m_\chi}{m_{Z'}^2}\bigg]^2 \nonumber \\
&\quad\quad+ \frac{\mu^2_e}{8\pi m_\chi^2}\bigg[\frac{m_e}{v_{\rm ew}}\Big(\frac{y_{h_2 \chi^*\chi}}{m_{h_2}^2}+\frac{y_{h_1 \chi^*\chi}}{m_{h_1}^2}\Big)-\frac{2e q_\chi g_X \varepsilon m_\chi}{m_{Z'}^2}\bigg]^2 \nonumber \\
&= \frac{\mu^2_e}{4\pi m_\chi^2}\bigg[\frac{m^2_e}{v^2_{\rm ew}}\Big(\frac{y_{h_2 \chi^*\chi}}{m_{h_2}^2}+\frac{y_{h_1 \chi^*\chi}}{m_{h_1}^2}\Big)^2+\frac{4e^2 q^2_\chi g^2_X \varepsilon^2 m^2_\chi}{m_{Z'}^4} \bigg]
\,,
\end{align}
where $\mu_e=m_\chi m_e/(m_e+m_\chi)$ is the reduced mass of the dark matter-electron system.

\subsection*{\underline {DM self-scattering }}

The self-interaction cross section for scalar dark matter in our model is given \cite{Choi:2015bya} by
\begin{align}
\sigma_{{\rm self}} = 
\frac{1}{64\pi m_\chi^2}\bigg(
|\mathcal{M}_{\chi\chi}|^2 
+ |\mathcal{M}_{\chi\chi^*}|^2
\bigg)\,,
\end{align}
where the squared amplitudes are given by
\begin{align}
|\mathcal{M}_{\chi\chi}|^2
&=
2\left(
2\lambda_\chi 
+\frac{3\zeta^2 m^2_{h'}}{\lambda_\phi m_\chi^2}
+\frac{8\lambda_\phi m_\chi^2}{9m_{h'}^2}
-\frac{\lambda_{\phi\chi}^2}{2\lambda_\phi }
\right)^2
\,, \label{M1} \\
|\mathcal{M}_{\chi\chi^*}|^2
&=
4\left(
2\lambda_\chi 
- \frac{9\zeta^2 m^2_{h'}}{2\lambda_\phi m_\chi^2}
-\frac{4\lambda_\phi m_\chi^2}{m_{h'}^2}
+\frac{\lambda_{\phi\chi}^2(m_{h^\prime}^2-2m_\chi^2)}{2\lambda_\phi (4m_\chi^2 - m_{h^\prime}^2)}
\right)^2 \label{M2}
\,.
\end{align}



\end{document}